\documentclass[fleqn,usenatbib]{mnras}

\DeclareUnicodeCharacter{0301}{\'{i}}

\usepackage[T1]{fontenc}
\usepackage{ae,aecompl}

\usepackage{graphicx}	
\usepackage{amsmath}	
\usepackage{amssymb}	
\usepackage{subfigure}
\usepackage{threeparttable}
\usepackage{adjustbox}
\usepackage{float}
\usepackage{gensymb}
\usepackage{siunitx}
\usepackage{float}
\usepackage{xspace}
\usepackage{ulem}

\newcommand{\hst}{\textit{HST}\xspace}

\newcommand{\2}{\,{\sc ii}}
\newcommand{\3}{\,{\sc iii}}

\newcommand{\msun}{\ensuremath{M_{\odot}}\xspace}			

\title[Synthetic photometry of OB star clusters with stochastically sampled IMFs]{Synthetic photometry of OB star clusters with stochastically sampled IMFs: analysis of models and \hst observations.}

\author[R. Orozco Duarte et al.]{\parbox{\textwidth}{
Rogelio Orozco-Duarte$^{1}$\thanks{E-mail: awofford@astro.unam.mx},
Aida Wofford$^{1}$,
Alba Vidal-Garc\'ia$^{2,3}$,
Gustavo Bruzual$^{4}$,
Stephane Charlot$^{4}$,
Mark R. Krumholz$^{5,6}$,
Stephen Hannon$^{7,8}$,
Janice Lee$^{9,7}$,
Timothy Wofford$^{10}$,
Michele Fumagalli$^{11}$,
Daniel Dale$^{12}$,
Matteo Messa$^{13,14}$,
Eva K. Grebel$^{15}$,
Linda Smith$^{16}$,
Kathryn Grasha$^{5}$,
David Cook$^{7}$
}\\\\
\parbox{\textwidth}{
$^{1}$Universidad Nacional Autónoma de México, Instituto de Astronomía, AP 106,  Ensenada 22800, BC,  México\\
$^{2}$Sorbonne Université, UPMC-CNRS, UMR7095, Institut d'Astrophysique de Paris, F-75014 Paris, France\\
$^{3}$LPENS, Ecole Normale Supérieure, Université PSL, CNRS, Sorbonne Université, Université Paris-Diderot, Paris, France\\
$^{4}$Instituto de Radioastronomía y Astrofísica, UNAM, Campus Morelia, Michoacán, Mé́xico, C.P. 58089, Mé́xico\\
$^{5}$Research School of Astronomy and Astrophysics, Australian National University, Canberra, ACT 2611, Australia\\
$^{6}$ARC Centre of Excellence for Astronomy in Three Dimensionns (ASTRO-3D), Canberra, ACT 2611, Australia\\
$^{7}$Caltech-IPAC, 1200 E. California Blvd. Pasadena, CA 91125, USA\\
$^{8}$Department of Physics $\And$ Astronomy, University of California, Riverside, CA, USA\\
$^{9}$Gemini Observatory/NSF’s NOIRLab, 950 N. Cherry Avenue, Tucson, AZ, 85719, USA\\
$^{10}$Facultad de Ciencias, Universidad Autónoma de Baja California, AP 1880, Ensenada 22800, BC, México\\
$^{11}$Dipartimento di Fisica G. Occhialini, Università degli Studi di Milano Bicocca, Piazza della Scienza 3, 20126 Milano, Italy\\
$^{12}$Department of Physics \& Astronomy, University of Wyoming, Laramie WY\\
$^{13}$Observatoire de Genève, Université de Genève, Chemin Pegasi 51, Versoix CH-1290, Switzerland\\
$^{14}$Department of Astronomy, Oscar Klein Centre, Stockholm University, AlbaNova, Stockholm SE-106 91, Sweden\\
$^{15}$Astronomisches Rechen-Institut, Zentrum für Astronomie der Universität Heidelberg, Mönchhofstraße 12-14, 69120 Heidelberg, Germany\\
$^{15}$Research School of Astronomy and Astrophysics, Australian National University, Canberra, ACT 2611, Australia\\
$^{16}$European Space Agency (ESA), ESA Office, Space Telescope  Science Institute, 3700 San Martin Drive, Baltimore, MD 21218, USA}}

\date{Accepted XXX. Received YYY; in original form ZZZ}

\pubyear{2015}

\begin{document}
\label{firstpage}
\pagerange{\pageref{firstpage}--\pageref{lastpage}}
\maketitle

\begin{abstract}
We present a pilot library of synthetic NUV, U, B, V, and I photometry of star clusters with stochastically sampled IMFs and ionized gas for initial masses, $M_i=10^3$, $10^4$, and $10^5$\,\msun; $t=1$, 3, 4, and 8 Myr; $Z=0.014$ and $Z=0.002$; and log(U$_{\rm S}$)\,=-2 and -3. We compare the library with predictions from deterministic models and observations of isolated low-mass ($<10^4$\,\msun) star clusters with co-spatial compact H\2~regions. The clusters are located in NGC~7793, one of the nearest galaxies observed as part of the \hst~LEGUS and H$\alpha$-LEGUS surveys. 1) For model magnitudes that only account for the stars: a) the residual |deterministic mag - median  stochastic mag| can be $\ge0.5$ mag, even for $M_i=10^5$\,\msun; and b) the largest spread in stochastic magnitudes occurs when Wolf-Rayet stars are present. 2) For $M_i=10^5$\,\msun: a) the median stochastic mag with gas can be $>$1.0 mag more luminous than the median stochastic magnitude without gas; and b) nebular emission lines can contribute with $>50\%$ and $>30\%$ to the total emission in the V and I bands, respectively. 3) Age-dating OB-star clusters via deterministic tracks in the U-B vs. V-I plane is highly uncertain at $Z=0.014$ for $M_i\sim10^3$\,\msun and $Z=0.002$ for $M_i\sim10^3-10^5$\,\msun. 4) For low-mass clusters, the V-band extinction derived with stochastic models significantly depends on the value of log(U$_{\rm S}$). 5) The youngest clusters tend to have higher extinction. 6) The majority of clusters have multi-peaked age PDFs. 7) Finally, we discuss the importance of characterising the true variance in the number of stars per mass bin in nature. 
\end{abstract}



\clearpage
\section{Introduction}\label{sec:introduction}

{\it Star clusters and cluster mass function.} Star clusters are groupings of stars that are born from the same molecular cloud and are gravitationally bound. They can contain anywhere between millions of stars to less than a few hundred members. Observations of star clusters with $<10^4\,$\msun\, exist for the Milky Way, M31, NGC~4214, and other comparatively-nearby galaxies. In particular, the PHAT survey (Panchromatic Hubble Andromeda Treasury, PI Dalcanton) covered approximately 1/3 of M31's star forming disk from the near ultraviolet (NUV) to the near infrared (NIR) at the high spatial resolution of the Hubble Space Telescope (\hst), and provided a robust distance measurement and high quality data for M31. Studies of the above three galaxies show that the cluster mass function (CMF) in the range $<10^4\,$\msun appears to be similar to the distribution at higher masses, down to the sensitivity limit, i.e., it is consistent with $dN/dM\sim M^{-2}$, where $N$ is the number of clusters and $M$ is the mass of the cluster \citep{Johnson2017, Krumholz2019}. The shape of the CMF has been confirmed by the studies of \cite{Adamo2017}, \cite{Messa2018}, and \cite{Cook2019} that target galaxies NGC~628, M51 and the dwarf galaxies from \hst's Legacy Extra Galactic Ultraviolet Survey (LEGUS, \citealt{Calzetti2015}), respectively. In their study of 25 LEGUS galaxies \citealt{Hannon2019} (hereafter H19) do not measure the CMF but find more clusters in the interval $10^3-10^4\,$\msun than with $>10^4\,$\msun.

{\it Importance of low-mass star clusters.}  Observations of the Small Magellanic Cloud analysed by \cite{Lamb2010} seem to suggest that a distribution of the type $dN/dM\sim M^{-2}$ describes systems with masses as low as $10\,$\msun. This means that the mass range between 10 and 1000\,\msun contains the same total mass in stars as the range between 1000 and $10^5\,$\msun. Thus, one cannot obtain a complete understanding of stellar populations without studying low-mass ($<10^4\,$\msun) star clusters. In addition, studying a statistically significant number of low-mass clusters in a diversity of environments is necessary in order to understand how cluster properties depend on their environment.  

\begin{figure}\label{fig:1000realizations}
\centering
\includegraphics[width=0.99\columnwidth]{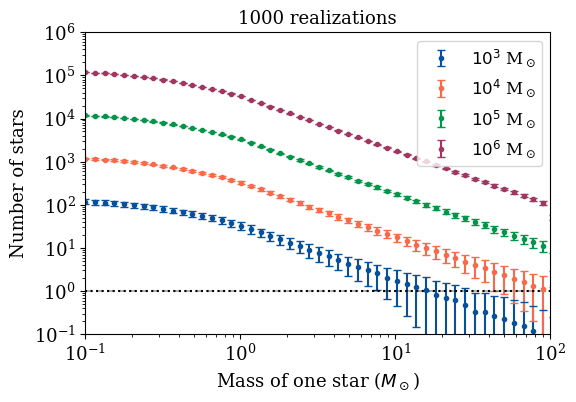}
\caption{Mean number of stars in each mass bin resulting from stochastically sampling the \citet{Chabrier2003}  IMF (his equation 1) 1000 times (filled circles), and standard deviation around the mean (error bars). We show this for clusters with initial masses of 10$^{3}$, 10$^{4}$, 10$^{5}$, and 10$^{6}$ M$_{\odot}$, as indicated by the legend. The horizontal dashed line corresponds to a number of stars equal to 1.}
\end{figure}

{\it The IMF of star clusters.} The  stellar initial mass function (IMF) describes the stellar mass distribution of the cluster at birth, when the cluster is still embedded. It is a main ingredient of population synthesis models (e.g. \citealt{Bruzual2003, Leitherer1999, Leitherer2014, Eldridge2017}), which aim to predict the radiative, mechanical, and chemical feedback of stellar populations; and are used as input to cosmological simulations (e.g., \citealt{Hirschmann2019}). The IMF has a universal form for conditions as they are found at the present time throughout galaxies of the Local Group \citep{Salpeter1955, Chabrier2003, Kroupa2012}; and it is stochastically sampled \citep{CervinoandLuridiana2003, Fouesneau2010, fumagalli2011, Krumholz2015a}. For star clusters with total initial masses ($M_i$) of 10$^{3}$, 10$^{4}$, 10$^{5}$, and 10$^{6}\,$\msun, and a stochastically-sampled IMF, Figure~\ref{fig:1000realizations} shows the mean and variation in the number of stars in each mass bin. For generating the figure, we assume a \cite{Chabrier2003} IMF shape and stop adding stars when the next star makes the total mass be $>M_i$. Since most stars have a low mass, the final total mass is rarely much different from $M_i$. The stars are collected in 49 bins with boundaries evenly spaced on a logarithmic scale from 0.1 to 100 $M_\odot$. The figure shows the following. 1) The standard deviation around the mean increases as the mass of the star increases, i.e., as the probability of a star belonging to the bin decreases. 2) In addition, as the value of $M_i$ decreases the standard deviation decreases while the relative standard deviation increases. These two results (1 and 2) are because the number of stars in each bin follows an approximately binomial distribution\footnote{For a fixed number of stars the collection of bin counts would follow a multinomial distribution. Fixing the total mass removes some of the independence, so a multinomial distribution is only an approximation.} with a probability given by the IMF. For example, if the IMF says that the probability in bin "j" is $P_j$ and we generate one realization of the IMF with N stars, then we expect to find $n_j=N P_j$  stars in that bin with a standard deviation of $\sigma_j=\sqrt{NP_j(1-P_j)}$ stars. The relative standard deviation $n_j/\sigma_j$ decreases as $N$ increases, i.e., as $M_i$ increases. 

{\it The need for improved models.} Beyond the Local Group of galaxies, star clusters can only be resolved into individual stars in comparatively nearby galaxies, and one requires model spectro-photometry generated with population synthesis codes to infer the extinction, mass, and age of a star cluster from its integrated light. Although in general, population synthesis codes do not account for the stochastic sampling of the IMF, {\texttt {SLUG}} \citep{daSilva2012} does. The latter code has been used to study broad-band \hst~NUV to NIR observations of large samples of star clusters. \cite{Krumholz2015a} find that the stochastic {\texttt {SLUG}}  models are generally a better fit to such observations compared to the deterministic {\texttt {Yggdrasil}} models of \cite{Zackrisson2011}, but that the overall properties of the star clusters recovered by both codes are qualitatively similar. \cite{Ashworth2017} find that including H$\alpha$ photometry in the {\texttt {SLUG}} stochastic models significantly improves the age determination of young clusters. However, in the latter two works, the stochastic models do not include the effect of the stochastic variation in the shape of the ionizing continuum, on the nebular emission. As we show in this work, properly accounting for the emission of the ionized gas is important when fitting models to observations that include the light from OB-star clusters and nebular emission, as is the case for compact H\2 regions, where the ionized gas is co-spatial with the stars. 

{\it This work.} In this work, we present a pilot library of synthetic photometry of young ($1-8$ Myr) star clusters that accounts for the stochastic sampling of the IMF and where the contribution of the ionized gas is fully modelled. The library includes the \hst F275W (UV), F336W (U), F438W (B), F555W (V), and F814W (I) broad band filters and is based on the model \textbf{\texttt{GALAXEV-C}} spectra that are presented in Vidal-García et al. (in prep.). Our specific objectives are: i) for clusters with different initial masses, ages, and metallicities, and for different values of the ionization parameter, to determine the spread in predicted magnitudes due to a) the stochastic sampling of the IMF and b) the inclusion of the ionized gas; ii) to quantify the relative contributions of the stellar continuum, nebular continuum, and emission lines to the total emission in the photometric bands; iii) to compare the location of the stochastic models relative to the deterministic predictions in the U - B versus V - I colour - colour magnitude diagram (this diagram is used for age-dating clusters); and iv) for a sample of observed star clusters with a) low masses according to deterministic models and b) compact H\2 regions, to obtain the probability distribution functions (PDFs) of the extinction, mass, and age corresponding to the independent stochastic {\texttt {GALAXEV-C}} and {\texttt {SLUG}}. We use cluster observations from the \hst~LEGUS \citep{Calzetti2015} and H$\alpha$-LEGUS (PID 13773, PI Chandar) surveys. The clusters are located in NGC~7793, which is one of the nearest galaxies in these surveys.

In Section~\ref{sec:Models}, we present our pilot library; in Section~\ref{sec:AnalysisModels}, we analyse the predictions of our pilot-library; in Section~\ref{sec:Observations}, we present the observations; in Section~\ref{sec:AnalysisObservations}, we describe \texttt{BAYESPHOT} \citep{Krumholz2015b}, which is the photometric interpretation tool that we use in this paper, and we discuss the derived cluster properties; in Appendix~\ref{app:enough}, we discuss what happens if we change the number of realizations of the IMF; in Appendices~\ref{app:more_ccds} and~\ref{app:cluster_pdfs} we present complementary colour-colour magnitude diagrams and PDF plots; finally, in Section~\ref{sec:Conclusions}, we summarise and conclude.  

\section{Models}\label{sec:Models}

In this Section, we present a pilot library of synthetic \hst-equivalent UV, U, B, V and I photometry, which accounts for the stochastic sampling of the IMF and the contribution of the ionized gas. The specific bands for which photometry was computed are: WFC3/UVIS F275W, F336W, and F438W; and ACS/WFC F555W and F814W, where WFC3 is the Wide Field Camera 3, ACS is the Advanced Camera for Surveys, and UVIS and WFC (Wide Field Channel) are the channels of the instruments. This is the filter set that was used for observing the west field of galaxy NGC 7793, whose star clusters are analysed in Section~\ref{sec:AnalysisObservations}. Our pilot library covers sufficient parameter space to i) determine if there is a significant difference between deterministic models and the median of the stochastic models; ii) quantify the impact of the nebular emission in the filters; and iii) show the limitations of colour-colour diagrams for age-dating low-mass star clusters. 

\subsection{Stars}\label{sub:stars}

The star clusters are assumed to be simple stellar populations (SSPs), i.e., populations where all stars are born simultaneously from the same molecular cloud. We compute models for four ages (1, 3, 4, and 8\,Myr); three initial cluster masses ($10^{3}$, $10^{4}$, and $10^{5}$\,\msun); and two metallicities ($Z=0.014$ and $Z=0.002$). The highest metallicity, $Z=0.014$, is the solar reference value of \cite{Asplund2009} and the closest metallicity of the star clusters (see \citealt{Pilyugin2014} and Section~\ref{sec:Observations}). However, we also compute models at $Z=0.002$ for comparison with the higher metallicity the models. The IMF of the star clusters is stochastically sampled and assumes a \cite{Chabrier2003} form and a mass range for the individual stars of 0.1 to 100\,\msun. When populating the IMF, we keep adding stars until we reach at least 0.05\,\msun above the required mass of the cluster. Since most of the stars are low mass, typically, our cluster masses are within 0.1 \msun of the target mass. For each combination of the above parameters, we generated 220 realizations. This number is set by \texttt{GALAXEV} \citep{Plat2019}, which is the code that we use to model the stellar population spectra. In its deterministic version, 220 corresponds to the maximum number of time steps and spectra that are generated in one run. In its stochastic version, the time steps are replaced by IMF realizations. In Appendix~\ref{app:enough}, we show that using a larger library of 1000 realisations does not significantly alter the mean to standard deviation or the main conclusions of this work. We compute deterministic models and corresponding stochastic models. In the deterministic models, the shape of the spectrum remains unchanged and the luminosity is simply scaled in proportion to the value of $M_i$, which effectively assumes that the IMF is fully sampled. The models were computed with the latest version of the population synthesis code {\texttt{GALAXEV}} (\citealt{Plat2019}; Charlot \& Bruzual, in prep.), which only includes the contribution from the stars. The models assume that the stars evolve as single stars with zero rotation. These models are useful for comparing with published results based on the same assumptions and more comprehensive models that account for the evolution of massive stars in close binary systems. In the latter systems the stars exchange mass and rotate.     

\subsection{Ionized gas}\label{sub:gas}

In order to compute the contribution of the gas ionized by the massive stars, we use the above stellar models as input to  photoionization models generated with  {\texttt{CLOUDY}} \citep{Ferland2017} and the approach presented in \cite{Vidal-Garcia2017}, i.e., we use: spherical geometry, a covering factor of 1, and a filling factor of 1. The H\2 regions are ionisation-bounded. The code used for this purpose is \texttt{GALAXEV-C} (C for Cloudy, Vidal-Garcia et al., in prep.). For the pilot library, we compute models for the following parameters: hydrogen number density, n(H)=100\,cm$^{-3}$; metallicities, $Z=0.002$ and $Z=0.014$; ionisation parameters at the Strömgren radius (as defined in \citealt{Gutkin2016}), log(U$_{\rm S}$) = -2 and -3; and C/O = (C/O)$_\odot$. 

\subsection{Extinction due to dust}\label{sub:extinction}

In order to account for the effect of dust mixed with the ionized gas, we include dust grains in {\texttt{CLOUDY}} and for consistency deplete the refractory elements in the ionized gas by using a dust-to-metal gas ratio of $\xi_d$=0.3. In order to account for dust in the intervening neutral medium when deriving the cluster properties, we apply an extinction law to the {\texttt{CLOUDY}} output. In the process of finding the extinction in the V-band, $A_{\rm V}$, of observed star clusters, we try $A_{\rm V}$ values in the range $0-3$ mag, in steps of 0.01 mag.


\section{Analysis of Models}\label{sec:AnalysisModels}

For all combinations of photometric band (NUV, U, B, V, and I ), initial mass ($10^3$, $10^4$, and $10^5\,$\msun), metallicity (Z=0.002 and 0.014), and log(U$_{\rm S}$) value (-2 and -3),  Figure~\ref{fig:magnitude_violin_plots} shows the magnitudes predicted by the deterministic (small squares) and  stochastic (violin plots) predictions. The photomtric bands are those used to observe field NGC 7793-W (see Section~\ref{sec:Observations}). The left side of the violin corresponds to stars only (\texttt{GALAXEV} output), while the right side corresponds to stars + ionized gas + dust mixed with the ionized gas (\texttt{GALAXEV-C} output).

\subsection{Models with just stars}\label{sub:stars_mag}

\begin{figure*}
    \centering
    \includegraphics[width=1.99\columnwidth]{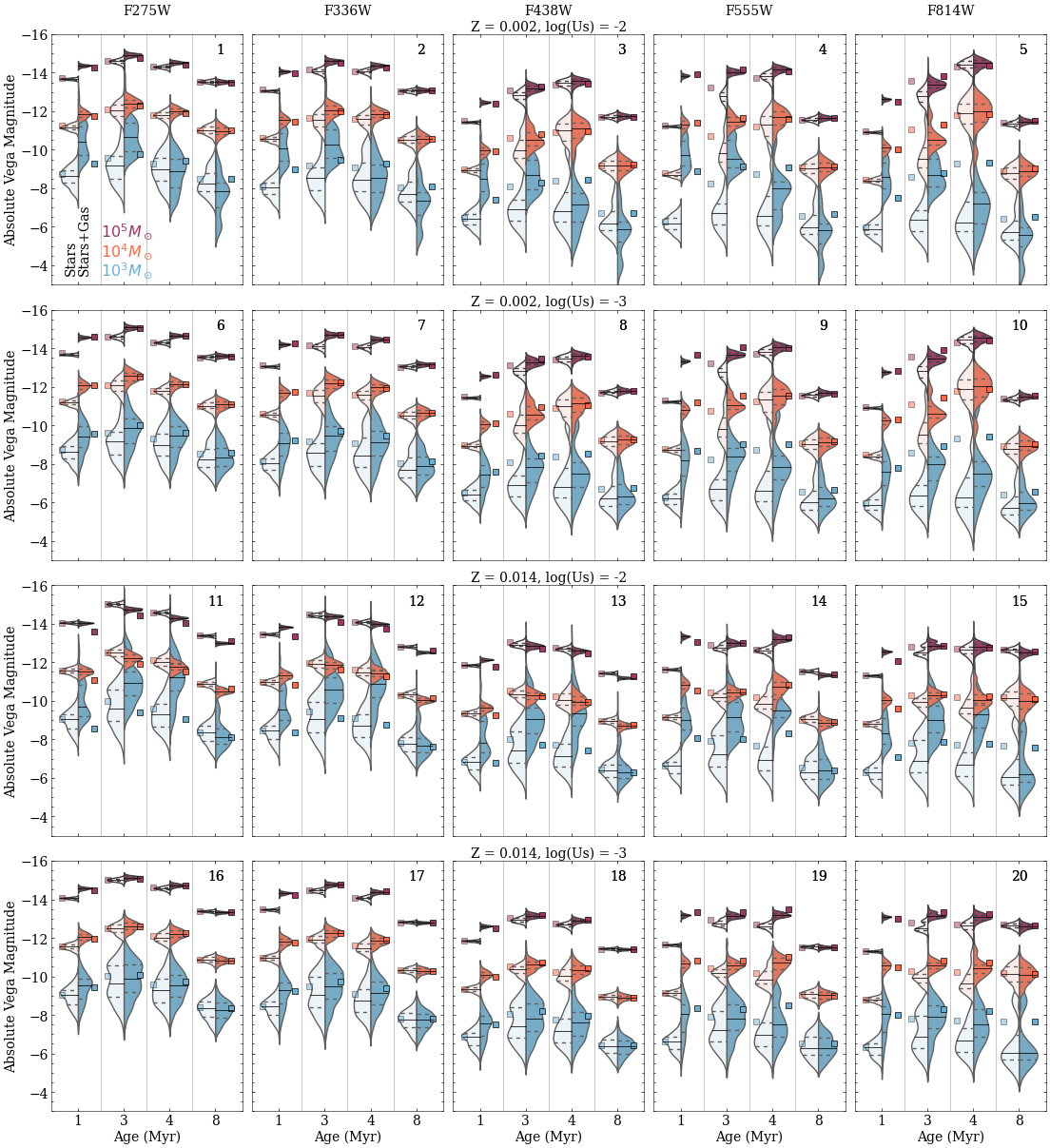}
    \caption{How accounting for the stochastic sampling of the IMF and including the contributions of the ionized gas + dust mixed with the ionized gas affects the magnitude predictions. Each violin diagram includes 220 realizations of the IMF. The left half of the violin represents the models with just stars and the right half the models that account for stars + ionized gas + dust. We show models for: five LEGUS bands (given by the column titles); initial cluster masses of $M_i=10^{3}$, 10$^{4}$, and 10$^{5}$\,\msun (bottom-blue, middle-red, and top-magenta violins, respectively, as given by the legend); ages of $t=1$, 3, 4, and 8 Myr (given by the x-axis label); and all combinations of metallicity (Z=0.002 or 0.014) and ionization parameter (log(U$_{\rm S}$)=-2 or -3, as given by the row titles). The shape of the half violin indicates the frequency of models at a given magnitude. The solid horizontal lines within the half violins give the median of the stochastic models while the 25th and 75th quartiles are given by the dashed lines. The squares on each side of the violin give the deterministic magnitude without gas (square on the left) and with gas (square on the right) at each value of $M_i$. We number the panels from 1 to 20 for facilitating the discussion in Table~\ref{tab:det_vs_soc}.}
    \label{fig:magnitude_violin_plots}
\end{figure*}

 In this subsection, we analyse the behaviour of the models that account for the stochastic sampling of the IMF and the contribution of the stars alone (left violin halves of Figure~\ref{fig:magnitude_violin_plots}).

{\it{Mass effect.}} As the initial cluster mass, $M_i$, increases from bottom to top in each panel, the spread in magnitudes decreases. This behaviour is observed in all LEGUS bands and at all ages and metallicities. This is because massive stars are significantly more luminous than lower mass stars. Thus,  the presence or absence of massive stars in the stellar population severely impacts its integrated luminosity. 

{\it{Age effect.}} In general, the spread in magnitudes is larger at 3 and 4 Myr than at 1 and 8 Myr, specially for $M_i=10^{3}$\,\msun. This is due to the presence of classical Wolf-Rayet stars at 3 and 4 Myr. These stars, which are in a phase of helium burning and have lost their hydrogen envelope, are the evolved descendants of the most massive stars ($\ge25\,M_\odot$, depending on metallicity, \citealt{Massey2000}). Their presence or absence in the population significant impacts the integrated luminosity. 

{\it{Metallicity effect.}} Stars of different chemical compositions evolve on different time-scales. For $M_i=10^{5}$\,\msun, and each broad band and age, Table~\ref{tab:stars_Z_effect} gives the difference between the median magnitude of the stochastic models at $Z=0.014$ and $Z=0.002$. At $t=1$\,Myr, the $Z=0.014$ models are more luminous in all bands, whereas at $t=8$\,Myr, the $Z=0.002$ models are more luminous in all bands except F814W. 

\begin{table}

\centering
\begin{tabular}{cccccc}
\hline
Age & F275W & F336W & F438W & F555W & F814W \\ 
(Myr) & \multicolumn{5}{c}{M(Z=0.002) - M(Z=0.014)} \\ 
\hline
1    & 0.38   & 0.40  & 0.40   & 0.40  & 0.39   \\
3    & 0.42   & 0.37   & 0.11   & -0.01   & -0.30   \\
4    & 0.28   & 0.00   & -0.77   & -1.19   & -1.80   \\
8    & -0.15  & -0.23  & -0.28  & -0.02  & 1.29  \\
\hline
\end{tabular}
\caption{For the \texttt{GALAXEV} models with $M_i=10^{5}$\,M$_{\odot}$, difference in median stochastic absolute magnitude, M(Z=0.002) - M(Z=0.014).}
\label{tab:stars_Z_effect}
\end{table}

\subsection{Models with stars, gas, and dust}\label{sub:stars_and_gas_mag}

 In this subsection, we analyse the behaviour of models that account for the stochastic sampling of the IMF and the added contributions of the ionized gas and dust mixed with the ionized gas (right violin halves of Figure~\ref{fig:magnitude_violin_plots}).
 
{\it{Gas effect.}} Adding the ionized gas significantly increases the luminosity for certain combinations of age, Z, and log(U$_{\rm S}$). This can be seen in Table~\ref{tab:median_mag_stars_stars_and_gas_difference}, which for models with $M_i=10^{5}$\,\msun shows the difference in median magnitude with and without gas. The effect of adding the gas is largest for the F555W (V) and F814W (I) bands, where at $t=1$ Myr, the difference is $>1$ mag for both values of $Z$ and log(U$_{\rm S}$). At $t=1$ Myr, significant differences also occur in other bands. At 4 Myr, for $Z=0.014$ and both values of log(U$_{\rm S}$), the difference is $\sim0.6$ mag in the F555W band. However, as expected, at 8 Myr, when the ionising flux from the most massive stars is greatly diminished, including the gas does not significantly change the synthetic magnitudes. The contributions of the nebular continuum and emission lines to the V and I bands are discussed next. 
 
\begin{table}

\centering
\begin{tabular}{cccccc}
\hline
Age & F275W & F336W & F438W & F555W & F814W \\ 
(Myr) & \multicolumn{5}{c}{M(stars) - M(stars+gas)} \\ 
\hline
\multicolumn{6}{c}{$Z=0.002$, log(U)=-2}\\
\hline
1 & 0.68 & 1.00 & 1.02 & 2.61 & 1.69 \\
3 & 0.33 & 0.50& 0.37 & 1.26 & 0.57 \\
4 & 0.22 & 0.29 & 0.09 & 0.35 & 0.06 \\
8 & -0.02 & 0.03 & 0.014 & 0.08 & 0.12 \\
\hline
\multicolumn{6}{c}{$Z=0.002$, log(U)=-3}\\
\hline
1    & 0.88   & 1.13   & 1.11   & 2.10   & 1.83   \\
3    & 0.48   & 0.62   & 0.45   & 0.90   & 0.65   \\
4    & 0.35  & 0.38   & 0.15   & 0.22   & 0.09   \\
8    & 0.06  & 0.09  & 0.06  & 0.10  & 0.14  \\
\hline
\multicolumn{6}{c}{$Z=0.014$, log(U)=-2}\\
\hline
1 & -0.03 & 0.35 & 0.28 & 1.70 & 1.23 \\
3 & -0.27 & -0.06 & -0.07 & 0.22 & 0.39 \\
4 & -0.27 & -0.08 & -0.09 & 0.58 & 0.18 \\
8 & -0.38 & -0.29 & -0.24 & -0.19 & -0.13 \\
\hline
\multicolumn{6}{c}{$Z=0.014$, log(U)=-3}\\
\hline
1    & 0.51  & 0.86  & 0.76  & 1.53  & 1.77  \\
3    & 0.11  & 0.31  & 0.23  & 0.37  & 0.63  \\
4    & 0.15  & 0.29  & 0.21  & 0.57  & 0.47  \\
8    & -0.06 & -0.03 & -0.04 & -0.03 & -0.03 \\
\hline
\end{tabular}
\caption{For the \texttt{GALAXEV} and \texttt{GALAXEV-C} models with $M_i=10^{5}$\,M$_{\odot}$, difference in median stochastic absolute magnitude, M(stars) - M(stars+gas).}
\label{tab:median_mag_stars_stars_and_gas_difference}
\end{table}

\subsection{Contributions of stellar continuum, nebular continuum, and emission lines to total in band.}\label{sub:stars_and_gas_mag}

\begin{figure}
    \centering
    \includegraphics[width=1\columnwidth]{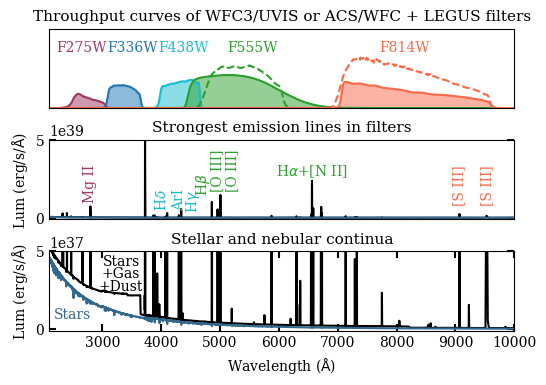}
    \caption{Top-panel--. Throughputs of the LEGUS filters + WFC3/UVIS (filled curves) or ACS/WFC (dashed curves) instrument/channel.  Middle-panel--. The strongest spectral features that contribute to the different LEGUS bands according to models that include the contributions of the stars the stars + ionized gas + dust mixed with the ionized gas. The stellar spectra used as input correspond to a cluster of initial mass = 10$^{5}$\,M$_{\odot}$, $Z=0.014$, and age = 1 Myr. The gas parameters are: $Z=0.014$ and log(U$_{\rm S}$)=-3. We use FWHM=130 km\,s$^{-1}$ for the width of the emission lines, which is the typical value obtained from VLT MUSE observations of H\2 regions in the galaxy under study (NGC~7793, \citealt{Wofford2020}).  Bottom-panel--. Enlargement of the middle panel to show the continua from the stars alone (blue curve) and stars + ionized gas + dust in the ionized gas (black curve).}\label{fig:throughputs}
\end{figure}

When gas is present in the models, the total emission includes the contributions from the nebular continuum and the emission lines. The importance of including nebular emission lines in spectral synthesis models has been shown by \cite{Bruzual2003}, \cite{Zackrisson2011}, and \cite{Schaerer2009}. For $M_i=10^5$\,\msun, $Z = 0.014$, and log(U$_{\rm S}$)=-3, Figure~\ref{fig:throughputs} shows the strongest spectral features that contribute to the LEGUS bands used to image two overlapping fields of galaxy NGC 7793. NGC 7793-E, was imaged with WFC3/UVIS in the \hst-equivalent NUV, U, B, V, and I bands; while NGC 7793-W used ACS/WFC for the V and I bands. The strongest emission lines in each filter are: Mg II $\lambda$2798 (F275W); H$\gamma$ $\lambda$4102, Ar I $\lambda$4300, and H$\delta$ $\lambda$4341 (F438W); [O\3]\,$\lambda\lambda$4859, 5007, H$\beta$ $\lambda4861$, and/or H$\alpha$ $\lambda6563$ + [N\2]\,$\lambda\lambda$6548, 6584 (F555W, depending on the instrument/channel combination, and [S\3] $\lambda$9069 and $\lambda$9532 (F814W). Note that WFC3/UVIS F555W contains H$\alpha$\,$\lambda6563$ + [N\2]\,$\lambda\lambda$6548, 6584 because it cuts out at 7000\,\AA~and even though the throughput at 6563 is low (0.05 compared to 0.28 at peak) the strength of H$\alpha$ can dominate depending on the strength of [O\3]. ACS/WFC F555W is different and cuts off at a shorter wavelength. The contribution of the nebular continuum  is strongest in the F275W and F336W bands. Note the presence of a strong Balmer break around $\sim$3800 \AA. Such Balmer breaks have been observed, see for example \cite{Guseva2006}. 

\begin{figure*}
    \centering
    \includegraphics[width=1.6\columnwidth]{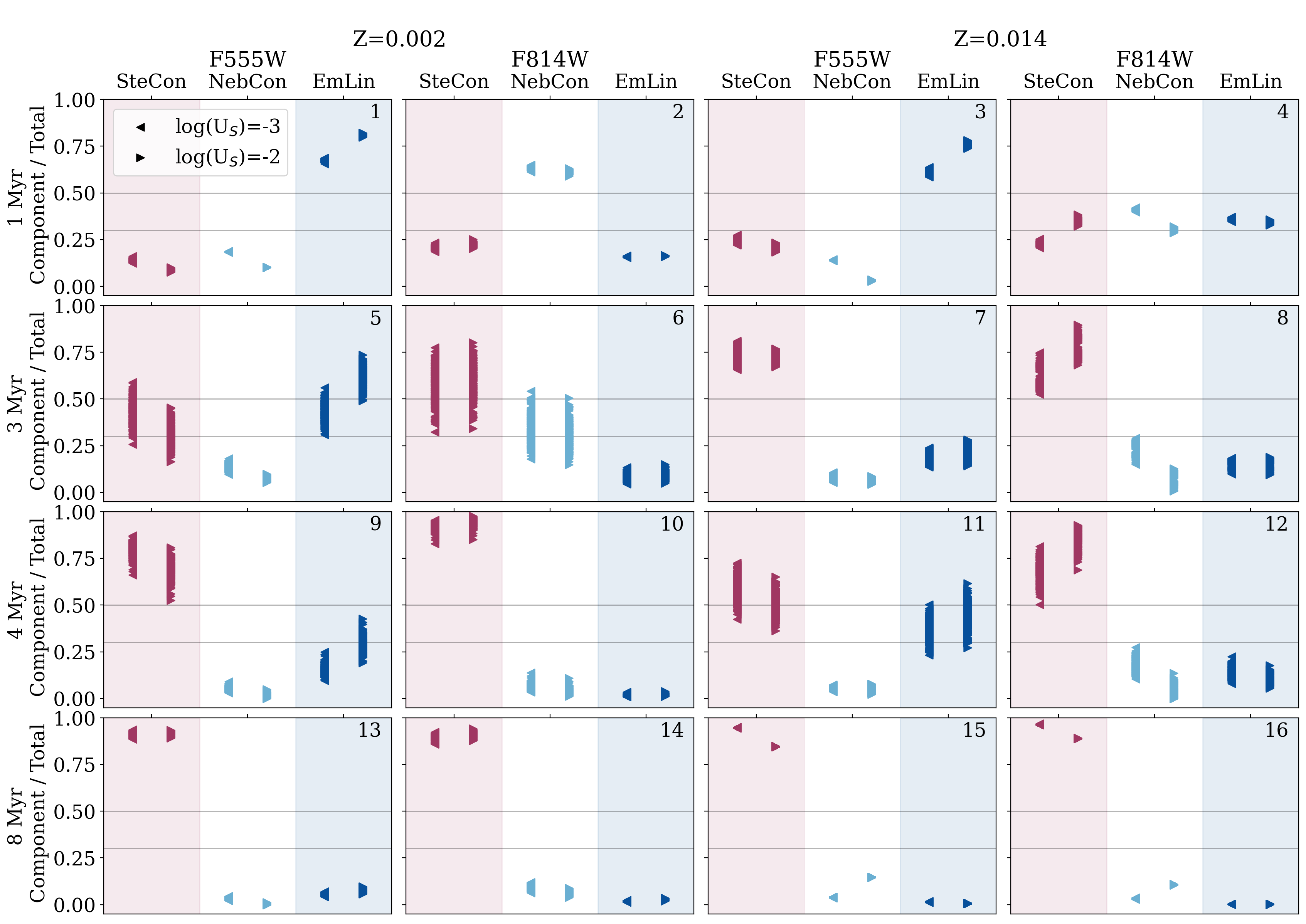}
    \caption{For the \texttt{GALAXEV} and \texttt{GALAXEV-C} models with $M_i=10^{5}$\,M$_{\odot}$, contributions to the total luminosities in the V and I bands of the stellar continuum (SteCon, magenta background), nebular continuum (NebCon, white background), and strongest emission lines (EmLin, blue background). The cluster age is given by the y-axis label the log(U$_{\rm S}$) value in the legend, and the metallicity at the top. The horizontal lines indicate contributions of 30 and 50\% to the total luminosity in the band. The panels are numbered from 1 to 16.}\label{fig:contributions}
\end{figure*}

For $M_i=10^{5}$\,\msun and the full range of parameters of our pilot library, Figure \ref{fig:contributions} shows the contributions to the total luminosity in the V and I bands of the stellar continuum, nebular continuum, and strongest emission lines. The 220 realizations of the IMF are included. We select the V and I bands because they have contributions from the emission lines alone of $\geq30$\%. The Figure~shows that the stellar continuum dominates at 8 Myr. This is because the ionizing flux from the massive stars in not significant at this age. The stellar continuum is also dominant in some cases at ages of 3 and 4 Myr (panels 5 to 12). The Figure also shows that the nebular continuum dominates in one case (panel 2). Finally, the Figure shows that the nebular emission lines can contribute with $>50\%$ to the total emission in the V-band (panels 1, 3, 5 and 11) and $>30\%$ to the I-bands (panel 4). The ranges corresponding to the y-axis of Figure \ref{fig:contributions} are given in Table~\ref{tab:contributions}. The Table and Figure illustrate the importance of accounting for the contribution of the ionized gas in the LEGUS bands. 

\begin{table*}
\begin{tabular}{@{\extracolsep{4pt}}lcccccc@{}}
\hline
\hfill & \multicolumn{3}{c}{F555W} & \multicolumn{3}{c}{F814W} \\
\cline{2-4} \cline{5-7}
Age	&	SteCon		&		NebCon			&	EmLin			&	SteCon		&		NebCon			&	EmLin			\\																										
\hline
\hfill	&	\multicolumn{6}{c}{Z=0.002, log(U$_{\rm S}$)=-2}\\
\hline
1	&	$0.08-0.10$	&	$0.10-0.10$	&	$0.80-0.82$	&	$0.20-0.25$	&	$0.59-0.63$	&	$0.16-0.17$\\																										
3	&	$0.17-0.45$	&	$0.05-0.10$	&	$0.49-0.74$	&	$0.34-0.80$	&	$0.15-0.51$	&	$0.05-0.15$\\																										
4	&	$0.53-0.81$	&	$0.00-0.05$	&	$0.19-0.43$	&	$0.85-0.98$	&	$0.01-0.11$	&	$0.01-0.04$\\																										
8	&	$0.89-0.93$	&	$0.00-0.01$	&	$0.06-0.10$	&	$0.88-0.94$	&	$0.04-0.09$	&	$0.02-0.03$\\																										
\hline
\hfill	&	\multicolumn{6}{c}{Z=0.002, log(U$_{\rm S}$)=-3}\\
\hline																																																
1	&	$0.12-0.16$	&	$0.18-0.19$	&	$0.66-0.69$ &	$0.19-0.23$	&	$0.61-0.65$	&	$0.15-0.16$\\																										
3	&	$0.26-0.59$	&	$0.10-0.18$	&	$0.31-0.56$ &	$0.32-0.78$	&	$0.18-0.54$	&	$0.05-0.13$\\																										
4	&	$0.66-0.87$	&	$0.03-0.09$	&	$0.10-0.25$ &	$0.83-0.95$	&	$0.04-0.14$	&	$0.01-0.03$\\																										
8	&	$0.89-0.94$	&	$0.02-0.04$	&	$0.04-0.07$ &	$0.86-0.92$	&	$0.06-0.12$	&	$0.01 -0.02$\\																										
\hline
\hfill	&	\multicolumn{6}{c}{Z=0.014, log(U$_{\rm S}$)=-2}\\
\hline																																																
1	&	$0.18-0.23$	&	$0.03-0.04$	&	$0.74-0.78$	&	$0.32-0.38$	&	$0.29-0.32$	&	$0.33- 0.36$\\																										
3	&	$ 0.67-0.77$	&	$0.04-0.09$	&	$0.14-0.28$	&	$0.68-0.90$	&	$0.01-0.13$	&	$0.10- 0.19$\\																										
4	&	$ 0.36-0.65$	&	$0.02-0.08$	&	$0.27-0.62$	&	$0.69-0.93$	&	$0.00-0.14$	&	$0.05- 0.18$\\																										
8	&	$ 0.84-0.85$	&	$0.15-0.15$	&	$0.01-0.01$	&	$0.89-0.89$	&	$0.11-0.11$	&	$0.00- 0.00$\\																										
\hline
\hfill	&	\multicolumn{6}{c}{Z=0.014, log(U$_{\rm S}$)=-3}\\
\hline																																																
1	&	$0.22-0.27$	&	$0.14-0.14$	&	$0.59-0.64$	&	$0.21-0.25$	&	$0.40-0.42$	&	$0.35- 0.37$\\																										
3	&	$0.66-0.81$	&	$0.05-0.11$	&	$0.14-0.24$	&	$0.53-0.75$	&	$0.15-0.29$	&	$0.10- 0.18$\\																										
4	&	$0.42-0.73$	&	$0.04-0.07$	&	$0.23-0.50$	&	$0.50-0.81$	&	$0.11-0.27$	&	$0.08- 0.22$\\																										
8	&	$0.94-0.95$	&	$0.04-0.04$	&	$0.01-0.02$	&	$0.96-0.97$	&	$0.03 - 0.04$	&	$0.00 - 0.00$\\
\hline
\end{tabular}
\caption{For the \texttt{GALAXEV} and \texttt{GALAXEV-C} models with $M_i=10^{5}$\,M$_{\odot}$, ranges of the contributions to the total luminosities in the V and I bands of the stellar continuum (SteCon), nebular continuum (NebCon), and the strongest emission lines in the band (EmLin, see Figure~\ref{fig:contributions}).}\label{tab:contributions}
\end{table*}

\subsection{Stochastic versus deterministic models.}\label{sec:det_vs_sto}

In Figure~\ref{fig:magnitude_violin_plots}, the filled squares represent the deterministic magnitudes for the cases with just stars (left square) and stars + gas + dust (right square) at each value of $M_i$. How the deterministic magnitude compares to the median of the stochastic models depends on whether the models include just stars or stars + gas + dust, and on the combination of $M_i$, age, $Z$, log(U$_{\rm S}$), and photometric band. Differences of less than 0.5 mag between the median and deterministic magnitude can be seen in the Figure for combinations of the parameters in the following cases: i) \{stars\} or \{stars + gas + dust\}, $M_i/M_\odot=10^4$ or $10^5$, $Z=0.014$, and log(U$_{\rm S}=-3$); and ii) \{stars\},  $M_i/M_\odot=10^3$, $t=1$\,Myr, both values of $Z$, and both values of log(U$_{\rm S}$). Examples of cases where $|$Det - <Sto>$|$> 0.5 mag are provided in Table~\ref{tab:det_vs_soc}. Note that $|$Det - <Sto>$|$ can be > 0.5 mag for $M_i/M_\odot$ as high as $10^5$, and that although the deterministic magnitude tends to be more luminous, this is not systematically the case.

\begin{table}
    \centering
    \begin{tabular}{lllll}
    \hline
        Gas? 	&	 M$_i$ &		Age 	& Panels & Most\\	
        \hfill	&	 (M$_\odot$) &	(Myr) 	& \hfill& Luminous\\		        
        \hline
        No      &	1E3\,-\,1E5	&	1	        &	none            & -\\
        No      &	1E3\,-\,1E5	&	3           &   5 \& 10         & Det \\
        No 	    &	1E4	        &	4	        &	15 \& 20        & Det \\        
        No 	    &	1E3	        &	4	        &	4, 5, 9 \& 10   & Det \\  
        No 	    &	1E3	        &	8	        &	15 \& 20        & Det \\        
        Yes	    &	1E3 	    &	1,\,3,\,4 	&   12              & <Sto>\\
        Yes 	&	1E3	        &   8 	        &	4 \& 5          & Det \\ 
        Yes  	&	1E4         &   3	        &   5 \& 10         & Det \\         
        \hline
    \end{tabular}
    \caption{Examples of cases where the absolute value of the residual (deterministic mag - median stochastic mag) is > 0.5 mag. We consider the \texttt{GALAXEV} and \texttt{GALAXEV-C} models. Column 1 indicates if gas is included in the models. Columns 2 and 3 give the initial mass and age of the cluster, respectively (or their ranges). Column 4 gives the IDs of the panels in Figure~\ref{fig:magnitude_violin_plots} where examples can be found. Column 5 says which of the two magnitudes is the most luminous (Det=deterministic, <Sto>=median stochastic mag).}    
    \label{tab:det_vs_soc}
\end{table}
For star-only broad-band magnitudes: a) the absolute value of the residual (deterministic prediction - median of  stochastic models) can be $\ge0.5$ mag, even for $M_i=10^5$\,\msun
\subsection{Models in the U - B versus V - I  diagram}\label{sub:analysis_models}

We now analyse the positions of the models in the U - B versus V - I diagram, which has been used to age-date star clusters by \citep{Chandar2004, Chandar2016}. Although \cite{Chandar2016} conclude that using UBVIH$\alpha$ photometry yields better agreement between photometric and spectroscopic ages, determining accurate H$\alpha$ photometry is very difficult in practice, particularly, when star clusters are not isolated and the H$\alpha$ morphology is complex, as we discuss in Section~\ref{sub:multiple_peaks}. We note that other diagrams such as (U - B) versus (B - V), which is not discussed here, have also been used to age-date star clusters (e.g., \citealt{Bica1991}).

\begin{figure*}
    \centering
    \includegraphics[width=1.69\columnwidth]{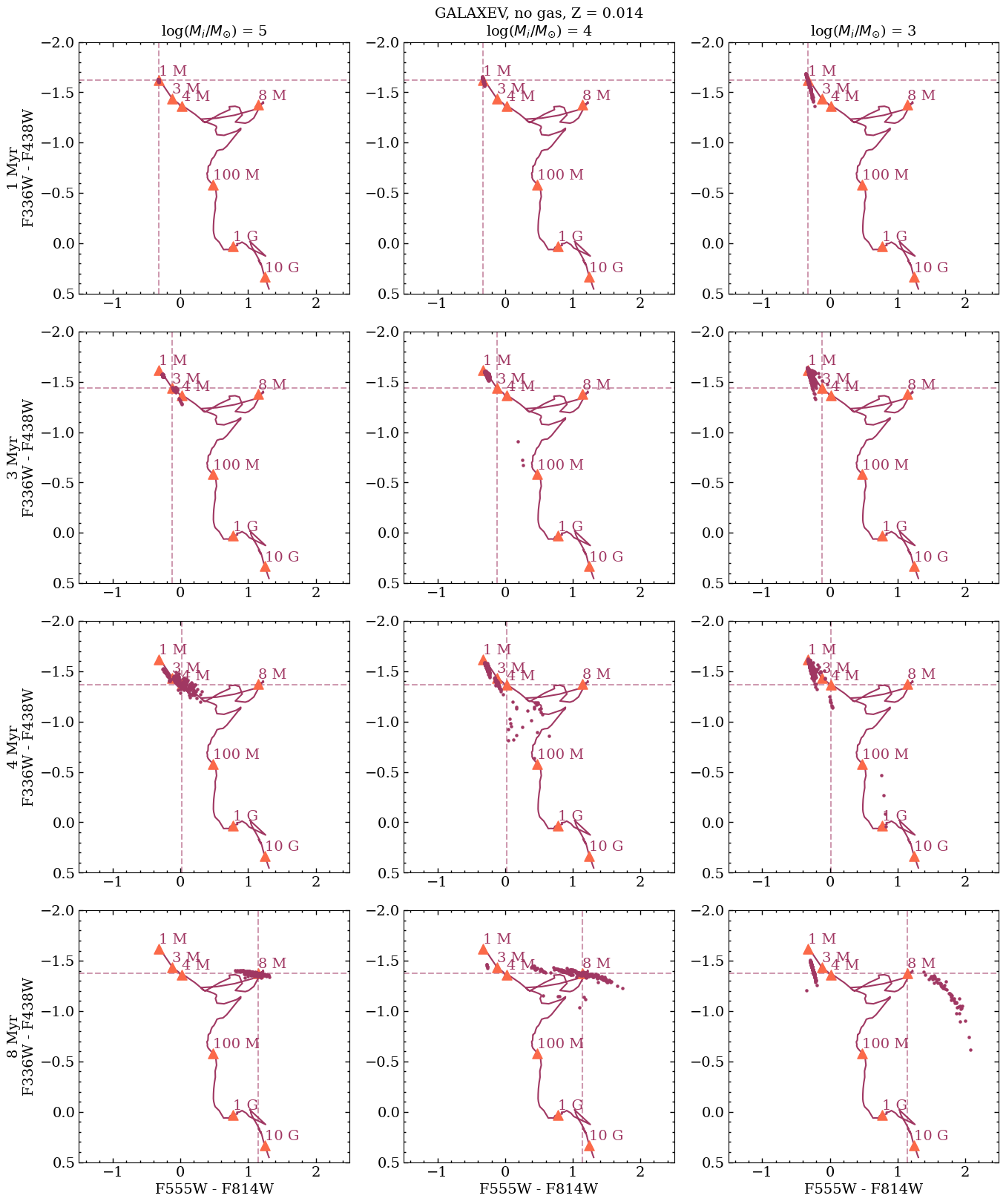}
    \caption{U-B vs. V-I diagrams with $Z=0.014$ star-only {\texttt GALAXEV} deterministic tracks (magenta solid curves) and stochastic models (magenta small filled symbols) overlaid. Ages along the D0 track are marked with orange-filled triangles and labelled using M=Myr and G=Gyr. The dashed vertical and horizontal lines mark the position of the deterministic prediction at the age given by the y-axis label. For clarity, for the stochastic models, each panel shows a different combination of cluster initial mass (given by the column title) and age (given by the y-axis label).}
    \label{fig:ccd_z014_u0}
\end{figure*}

\begin{figure*}
    \centering
    \includegraphics[width=1.69\columnwidth]{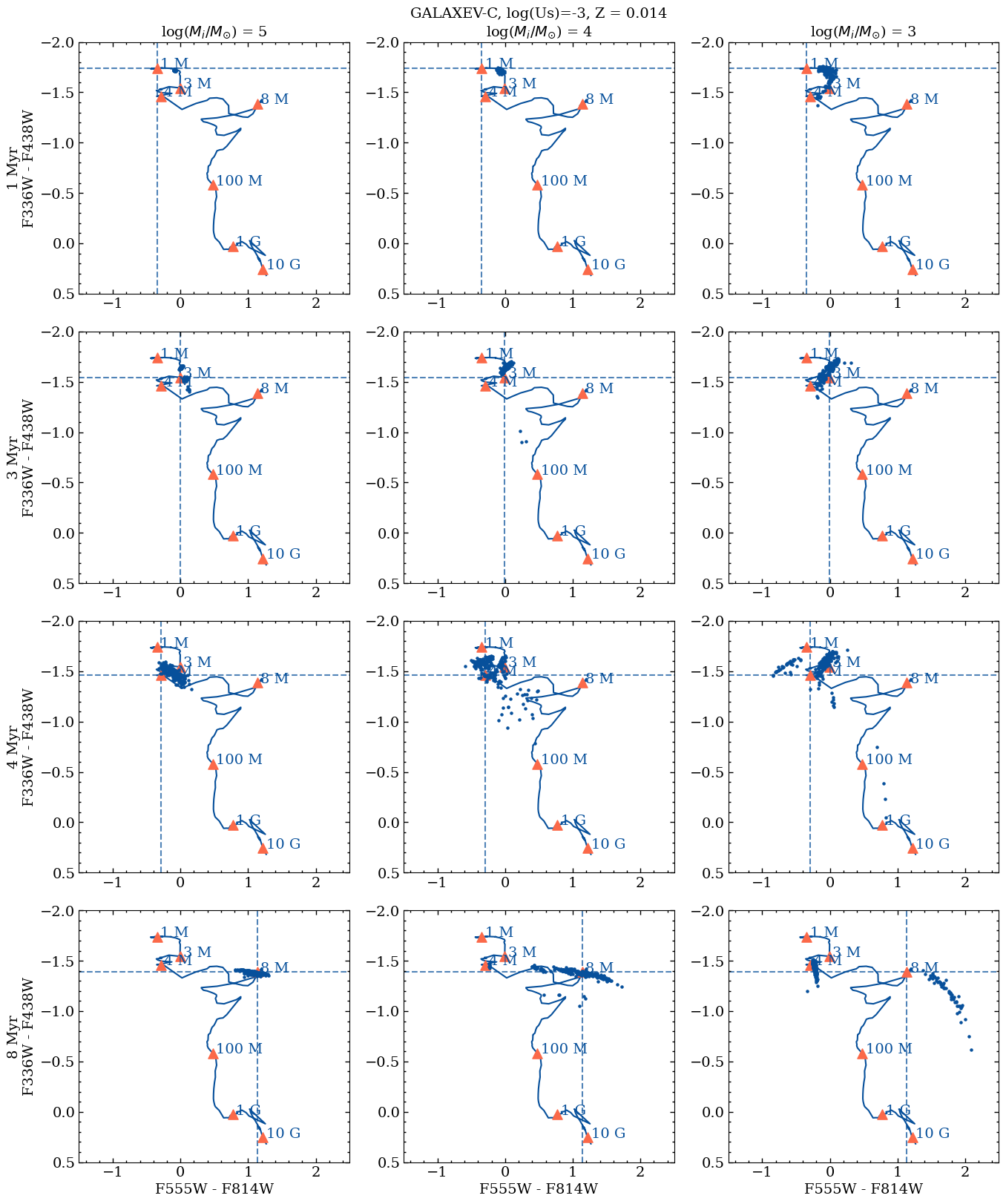}
    \caption{Similar to Figure~\ref{fig:ccd_z014_u0} but for log(Us)=-3 deterministic (dark blue curves) and stochastic (dark blue symbols) models that include the contributions of the ionized gas + dust mixed with the ionized gas.}
    \label{fig:ccd_z014_u3}
\end{figure*}

\begin{figure*}
    \centering
    \includegraphics[width=1.69\columnwidth]{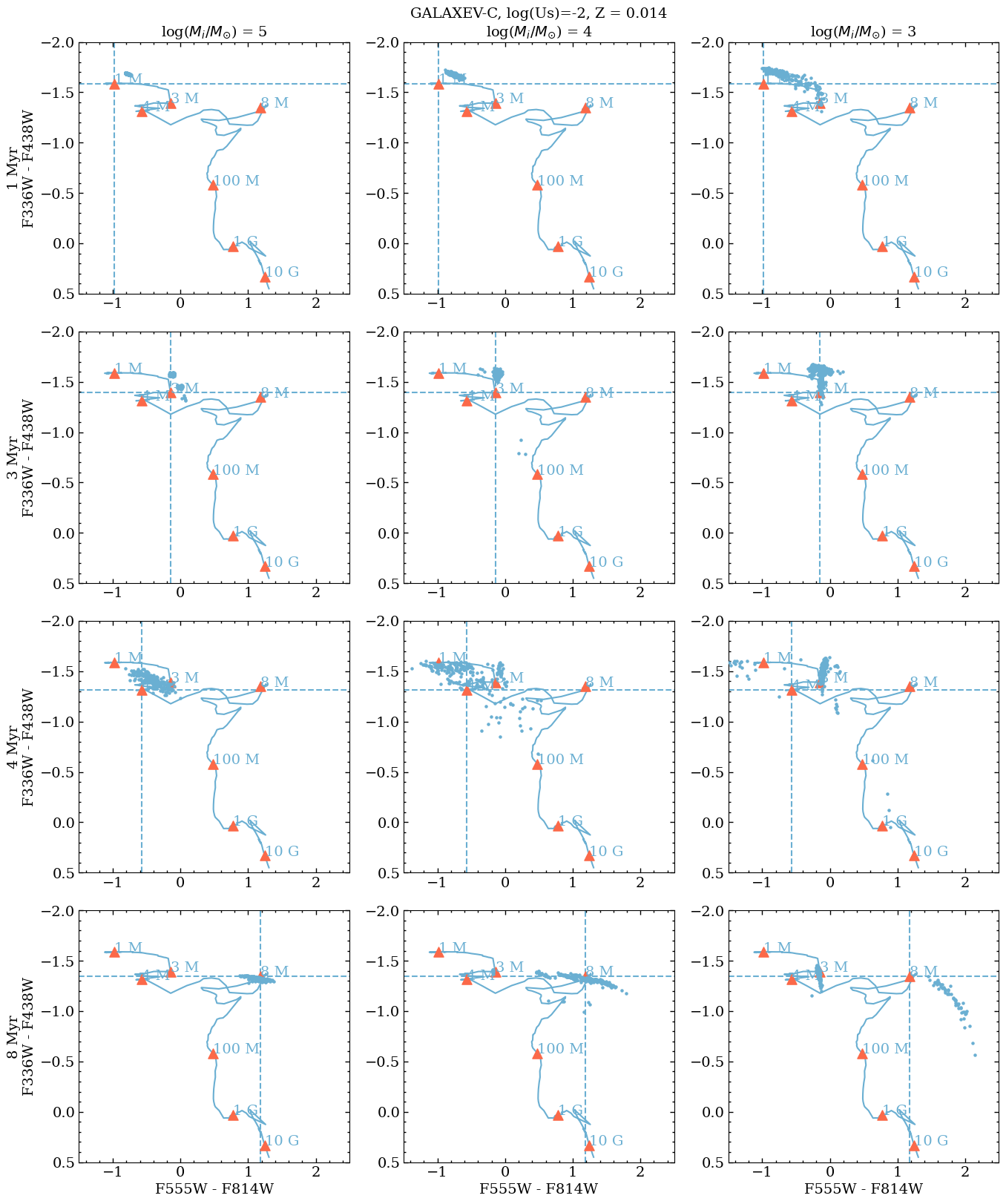}
    \caption{Similar to Figure~\ref{fig:ccd_z014_u3} but for log(Us)=-2.}
    \label{fig:ccd_z014_u2}
\end{figure*}

Let D0 and S0 be the deterministic and stochastic {\tt GALAXEV} models, respectively. Similarly, let (D2, S2) and (D3, S3) be the {\tt GALAXEV-C} pairs for log(Us)=-2 and -3, respectively. For $Z=0.014$, which is the closest metallicity to the observations (see Section~\ref{sec:Observations}), Figures~\ref{fig:ccd_z014_u0} -~\ref{fig:ccd_z014_u2} show comparisons of the D0 and S0, D3 and S3, and D2 and S2 predictions in the U - B vs. V - I diagram, respectively. Appendix Figures~\ref{fig:ccd_z002_u0} -~\ref{fig:ccd_z002_u3} are similar to Figures~\ref{fig:ccd_z014_u0} -~\ref{fig:ccd_z014_u2} but for $Z=0.002$.\\
Let us start by discussing Figure~\ref{fig:ccd_z014_u0}, which corresponds to the $Z=0.014$ models that only account for the stars. In this and similar figures, the age and initial mass of the cluster are given by the y-axis label and the column title, respectively. In Figure~\ref{fig:ccd_z014_u0}, the clouds of magenta filled-symbols are the stochastic S0 models and the magenta curves are the corresponding deterministic track. The comparison between the S0 and D0 predictions yields the following results, which are organised in order of increasing cluster age and decreasing cluster mass.

    \textit{1 Myr (top row of panels).--} For $M_i/M_\odot=10^5$, the S0 models are tightly grouped on top of the 1 Myr D0 point that is located at the intersection of the vertical and horizontal dashed lines. As mass decreases, the spread of the S0 models increases. In particular, for $M_i/M_\odot=10^3$, the spread is mostly in the vertical direction and towards the red.

   \textit{3 Myr (second row of panels from the top).--} For $M_i/M_\odot=10^5$, the S0 models break into three clouds, one tightly grouped on top of the 3 Myr D0 point, one bluer and closer to the 1 Myr D0 point, and the last one redder and closer to the 4 Myr D0 point. For $M_i/M_\odot=10^4$, a small fraction of the S0 models are located far away from the 3 Myr D0 point, towards the red, and approach the 100 Myr D0 point. Finally, for $M_i/M_\odot=10^4$ and $M_i/M_\odot=10^3$, most of the S0 models are bluer than the 3 Myr D0 point.
 
   \textit{4 Myr (third row of panels from the top).--} For $M_i/M_\odot=10^5$, the spread in S0 models is larger than at 1 Myr, and it is both towards the red and the blue. For  $M_i/M_\odot=10^4$, a larger but still small fraction of S0 models is found significantly offset to the red. For $M_i/M_\odot=10^3$, a small fraction of S0 models is significantly redder than the 4 Myr D0 point and is located between the 100 Myr and 1 Gyr markers along the deterministic track  (however, the main S0 cloud is spread closer to the 1-4 Myr D0 predictions.

   \textit{8 Myr (bottom row of panels.--} For $M_i/M_\odot=10^5$, the spread is larger than at 1 Myr, and it is both towards the red and the blue, similar to the behaviours at 3 and 4 Myr, but the spread is mostly horizontal. For $M_i/M_\odot=10^4$, the S0 models are spread almost horizontally towards the blue and the red and a few S0 models are significantly bluer than the D0 prediction. For $M_i/M_\odot=10^3$, the S0 models are divided into two clouds, one bluer in V - I and close to the 3 - 4 Myr D0 predictions and another starting at the D0 point and spreading towards significantly redder colours.

Let us now discuss Figures~\ref{fig:ccd_z014_u3} and~\ref{fig:ccd_z014_u2}, i.e., the $Z=0.014$ models with gas. The S2 and S3 models yield some of the same trends that are observed in the star-only case but also some additional behaviours. At 1 Myr, the S3 and S2 models are bluer in V - I relative to their respective deterministic predictions. At 3 and 4 Myr, for $M_i/M_\odot>10^5$, a significant fraction of the S2 and S3 models are offset towards bluer U - B relative to the corresponding deterministic models. Finally, a major conclusion from the figures is that the $Z=0.014$ D2 and D3 tracks are not useful for age-dating clusters with $M_i/M_\odot\sim10^3$. This is because at 8 Myr, for this mass, the number of stochastic models is similar in the $\sim$4 Myr cloud and the 8 M yr cloud.

We proceed with the analysis of the $Z=0.002$ models in the U - B vs. V - I plane. By looking at the star-only tracks of Figure~\ref{fig:ccd_z002_u0}, one can see that the 4 Myr D0 marker is almost coincident with the 100 Myr marker and that the 8 Myr marker is between the 1 and 3 Myr markers. When gas is included (Figures~\ref{fig:ccd_z002_u3} and~\ref{fig:ccd_z002_u2}), the 3 and 8 Myr deterministic markers are close to each other and they are almost coincident for log(Us)=-3. On the other hand, the 4 and 100 Myr deterministic markers become more separated than in the star-only case. In conclusion, age-dating star clusters with $Z=0.002$ using their positions in the U - B vs. V - I diagram along the \texttt{GALAXEV-C} deterministic tracks is highly uncertain at any mass.


\section{Sample and Observations}\label{sec:Observations}

In this work, we use \hst~images of two partially overlapping fields of galaxy NGC~7793, NGC~7793-E and NGC~7793-W. The images were obtained as part of the LEGUS and H$\alpha$ LEGUS programs. We fit models to the LEGUS broad-band photometry and investigate how close the H$\alpha$ equivalent widths of the clusters that were obtained by Hannon et al. from the H$\alpha$-LEGUS images are to the values predicted by our best-fitting models. In this section, we describe the observations, the galaxy, and how we selected the sample of star clusters. 

\subsection{\hst~LEGUS and H$\alpha$ LEGUS}\label{sub:galaxy}

LEGUS \citep[PID 13364]{Calzetti2015} is a Cycle 21 \hst~treasury program that obtained high spatial resolution ($\sim0.07"$) images of portions of 50 nearby ($\le16$ Mpc) galaxies, using the UVIS channel of the Wide Field Camera Three (WFC3), and the broad band filters F275W (2704 \AA), F336W (3355 \AA), F438W (4325 \AA), F555W (5308 \AA), and F814W (8024 \AA), which roughly correspond to the photometric bands NUV, U, B, V, and I, respectively. The survey includes galaxies of different morphological types and spans a factor of $\sim10^3$ in both star formation rate (SFR) and specific star formation rate (sSFR), $\sim10^4$ in stellar mass ($\sim10^7-10^{11}\,\rmn{M_\odot}$), and $\sim10^2$ in oxygen abundance ($12+\rmn{log\,O/H}=7.2-9.2$). Some of the targets in the survey have high quality archival images in bandpasses similar to those required by LEGUS, most of them from the Wide Field Channel of \hst's Advanced Camera for Surveys (ACS), and fewer of them from ACS's High Resolution Channel (HRC). For the latter targets, LEGUS completed the five band coverage. The choice of filters was dictated by the desire to distinguish young massive bright stars from faint star clusters, to derive accurate star formation histories for the stars in the field from their CMDs, and to obtain extinction-corrected estimates of age and mass for the star clusters. Star and star-cluster catalogues have been released for the LEGUS sample and are described in \cite{Sabbi2018} and \cite{Adamo2017} (hereafter A17), respectively.  

H$\alpha$ LEGUS (PI Chandar, PID 13773) is a Cycle 22 \hst program that obtained narrow-band, H$\alpha$ (F657N) and medium band, continuum (F547M) images for the 25 LEGUS galaxies with the highest star formation rates, using the WFC3. The corresponding H$\alpha$ observations reveal thousands of previously undetected H\2 regions, including those ionized by stellar clusters and ``single" massive stars. We note that the LEGUS data do not have the spatial resolution to visually resolve massive stars in close binary systems. 

\subsection{NGC~7793}\label{sub:galaxy}

We used the observations of NGC~7793 obtained by LEGUS and H$\alpha$ LEGUS which are summarised in Table~2 of \cite{Wofford2020}. NGC~7793 is a Southern SAd flocculent spiral  galaxy that is part of the Sculptor group and is located at a Cepheid distance of 3.44 Mpc \citep{Pietrzynsky2010}. It has a small bulge and a spiral filamentary morphology, and the following additional properties: an inclination of  47\degree; a colour excess due to the Galaxy of E(B - V) = 0.017 mag \citep{Schlafly2011}; a stellar mass determined from the extinction-corrected B-band luminosity and colour information of M$_{*}$ = $\num{3e9}$\,M$_{\odot}$ \citep{Bothwell2009}; lastly, a galaxy-wide star formation rate calculated from dust-attenuation corrected GALEX far-UV, adopting a distance of 3.44 Mpc, SFR = 0.52 M$_{\odot}$yr$^{-1}$ \citep{Lee2009}. According to \citet{Pilyugin2014}, the ionized-gas oxygen abundance at the centre of the galaxy is 12 + log(O/H)=$8.50\pm0.02$, and the O/H gradient is $-0.066\pm0.0104$ dex kpc$^{-1}$. 

\subsection{Sample of star clusters}\label{sub:star_clusters}

We select a sample of 17 isolated, low-mass ($<10^4$ \msun), young ($<10$ Myr) star clusters from the catalogue of clusters with compact H$\alpha$ morphologies of H19. H19 use the LEGUS datasets, which are aligned to the F438W image, and the LEGUS photometry, which uses an aperture with a radius of 5 pixels or 0.2” (3 pc), which was selected based on a curve of growth analysis. For H$\alpha$, H19 extracted their own photometry using apertures with different radii for each cluster, also based on a curve of growth analysis (see H21 in prep. for more details). The masses and ages used for the selection come from deterministic models, i.e., models where the luminosity is scaled in proportion to the initial cluster mass. 

\begin{figure*}
    \centering
    \includegraphics[width=1.99\columnwidth]{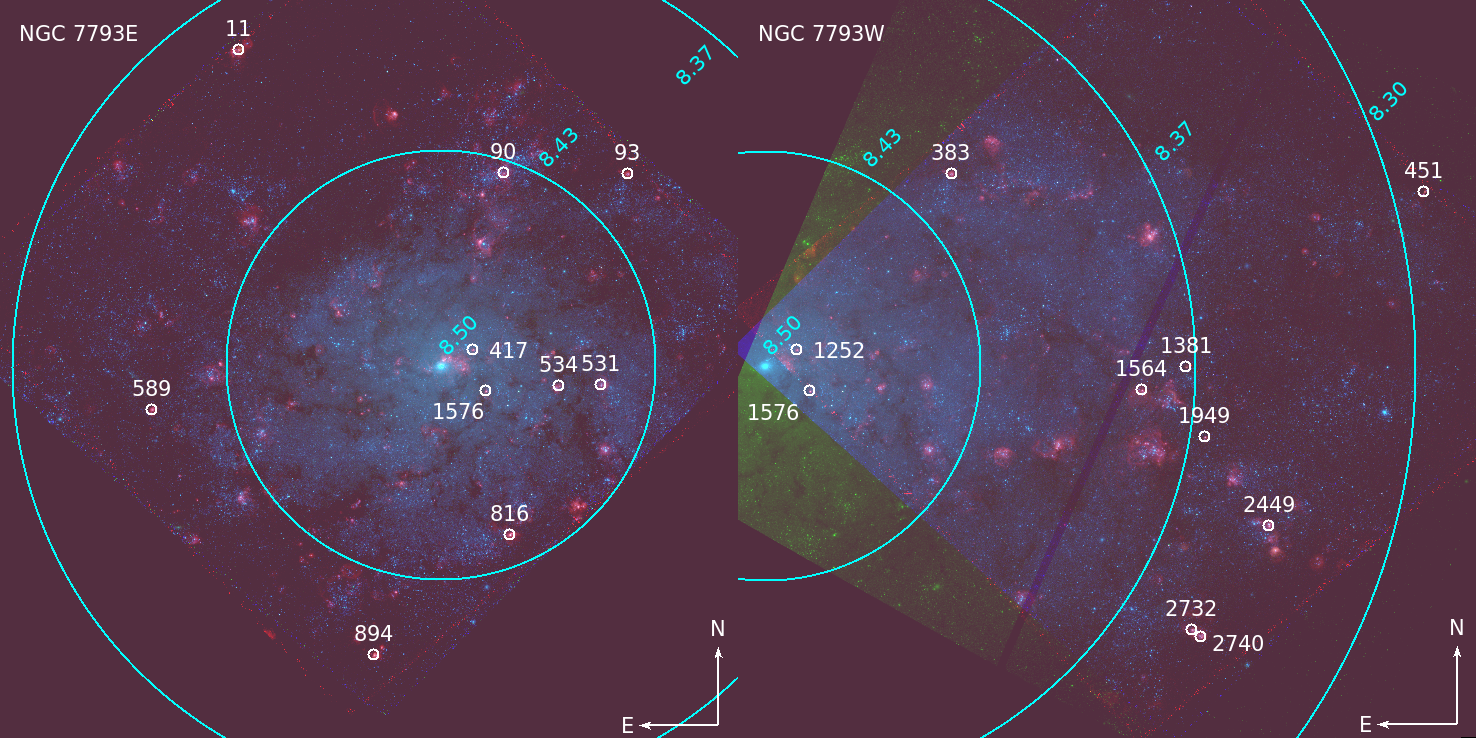}
    \caption{Composite RGB images of the partially-overlapping fields NGC~7793E (left panel) and NGC~7793W (righ panel), where red = continuum-subtracted H$\alpha$, green = F555W, and blue = F438W. We use white circles to indicate the positions of the star clusters in our sample and give their LEGUS ID. In both images, the centre of the galaxy appears as a cyan knot. We overlay cyan circles centered on this knot, of radii equal to 0, 1, 2, and 3 arcmin. The value of 12+log(O/H) in the ionized gas at each of these radii is shown with cyan characters. At 3.44 Mpc \citep{Pietrzynsky2010}, 1 armin is $\sim$ 1 kpc.  Note that clusters 93 and 383, and 417 and 1252 have similar coordinates and photometry but have different IDs (see Section~\ref{sub:star_clusters} for a discussion of these clusters). Also note that cluster 1576 appears in both fields and is at the edge of the H$\alpha$ image for field NGC~7793W.}\label{fig:cluster_locations}
\end{figure*}

Figure~\ref{fig:cluster_locations} shows the location of the star clusters within the galaxy. The figure shows that the star clusters are located within a radius of $\sim3$ arcmin from the centre of the galaxy. Adopting Z$_\odot=0.014$ as the reference solar metallicity \citep{Asplund2009} and using O/H as a gauge of metallicity, we find that the metallicity range of the clusters in our sample is Z=0.006 to 0.009. Thus, their metallicity is close to Z$_\odot$.

\begin{figure*}
    \centering
   \includegraphics[width=1.49\columnwidth]{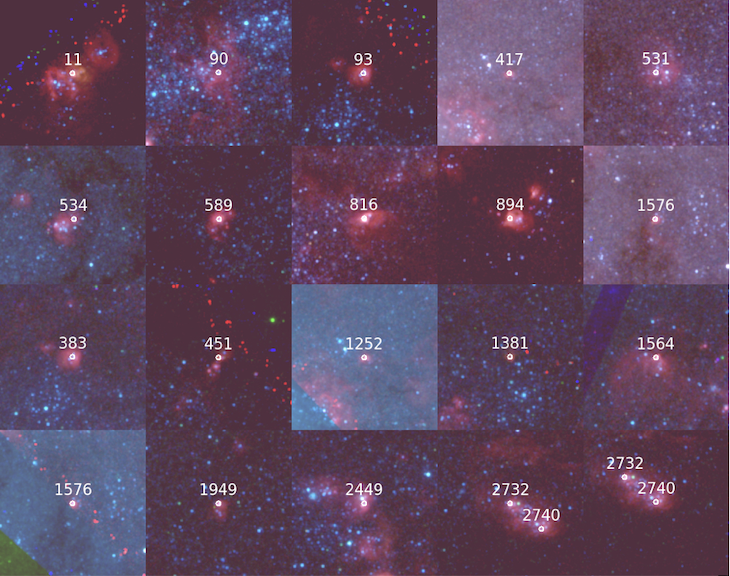}
    \caption{Postage stamps of the star clusters in our sample using the same colour scheme as in Figure~\ref{fig:cluster_locations}. The images are 12" (200 pc) on the side. The small white circle represents the aperture used for obtaining the photometry in the five LEGUS broad-bands. Star clusters from fields NGC~7793-E and NGC~7793-W are shown in the top-two and bottom-two rows show, respectively. Note that clusters 93 and 383, and 417 and 1252 have similar coordinates and photometry but have different IDs (see Section~\ref{sub:star_clusters} for a discussion of these clusters). Also note that cluster 1576 appears in both fields and at the edge of the H$\alpha$ image in field NGC~7793W.}\label{fig:stamps}
\end{figure*}

 Table~\ref{tab:observed_photometry} lists the J2000 coordinates and apparent Vega magnitudes of the star clusters in the LEGUS and H$\alpha$ LEGUS bands. Note that clusters 93 and 383, and 417 and 1252 have different IDs but very similar coordinates and photometry. This is because the clusters are the same, but their measurements come from field NGC~7793E in one case and NGC~7793W in the other. The clusters are in the overlapping region between these two fields. This constitutes a good check on repeatability of the cluster-finding process. When comparing models to observations, we keep the repeated clusters in order to check if small differences in the photometry due to the different pointings affect the derived properties of the clusters. Figure~\ref{fig:stamps} shows postage stamps of the clusters in our sample with the 5 pixel aperture overlaid.

\begingroup 
\renewcommand{\arraystretch}{1.0}
\begin{table*}
    \centering
\begin{tabular}{lcccccccccc}
\hline
ID & RA & Dec & F275W & F336W & F438W & F547M & F555W & F657N & F814W & Rad \\
\hfill & J2000 & J2000 & mag & mag & mag & mag & mag & mag & mag & pix \\ \hline
0011-E & 23:57:54.2424 & -32:33:59.148 & 20.97$\pm$0.09 & 20.81$\pm$0.09 & 22.16$\pm$0.08 & 20.70 & 21.07$\pm$0.07 & 17.39 & 20.71$\pm$0.06 & 30 \\
0090-E & 23:57:48.3768 & -32:34:33.672 & 18.80$\pm$0.09 & 19.32$\pm$0.08 & 20.82$\pm$0.07 & 21.06 & 20.99$\pm$0.05 & 19.80 & 21.06$\pm$0.05 & 10 \\
0093-E* & 23:57:45.6384 & -32:34:33.888 & 19.03$\pm$0.09 & 19.50$\pm$0.08 & 20.92$\pm$0.07 & 20.96 & 20.96$\pm$0.05 & 17.70 & 20.97$\pm$0.06 & 30 \\
0383-W* & 23:57:45.6384 & -32:34:33.852 & 18.95$\pm$0.07 & 19.40$\pm$0.07 & 20.88$\pm$0.06 & 21.06 & 20.97$\pm$0.05 & 17.62 & 21.06$\pm$0.07 & 40 \\
0417-E* & 23:57:49.0584 & -32:35:23.028 & 19.70$\pm$0.09 & 19.88$\pm$0.08 & 21.38$\pm$0.07 & 20.59 & 21.01$\pm$0.05 & 18.81 & 20.59$\pm$0.07 & 10 \\
0451-W & 23:57:35.1792 & -32:34:38.820 & 22.44$\pm$0.14 & 21.88$\pm$0.09 & 23.21$\pm$0.09 & 21.81 & 21.95$\pm$0.06 & 19.71 & 21.81$\pm$0.07 & 10 \\
0531-E & 23:57:46.2240 & -32:35:33.036 & 18.86$\pm$0.09 & 19.30$\pm$0.08 & 20.76$\pm$0.07 & 20.75 & 20.86$\pm$0.05 & 17.26 & 20.76$\pm$0.06 & 50 \\
0534-E & 23:57:47.1456 & -32:35:33.144 & 19.81$\pm$0.09 & 20.17$\pm$0.08 & 21.56$\pm$0.07 & 20.97 & 21.55$\pm$0.05 & 17.84 & 20.98$\pm$0.06 & 40 \\
0589-E & 23:57:56.1552 & -32:35:39.804 & 19.77$\pm$0.09 & 19.90$\pm$0.08 & 21.16$\pm$0.07 & 20.39 & 20.93$\pm$0.06 & 17.49 & 20.39$\pm$0.05 & 30 \\
0816-E & 23:57:48.2376 & -32:36:14.796 & 17.33$\pm$0.09 & 17.69$\pm$0.08 & 19.16$\pm$0.07 & 18.94 & 18.96$\pm$0.05 & 16.58 & 18.94$\pm$0.05 & 50 \\
0894-E & 23:57:51.2400 & -32:36:48.456 & 20.57$\pm$0.09 & 20.54$\pm$0.08 & 21.93$\pm$0.07 & 20.64 & 20.81$\pm$0.05 & 16.96 & 20.65$\pm$0.05 & 30 \\
1252-W* & 23:57:49.0656 & -32:35:23.028 & 19.69$\pm$0.07 & 19.91$\pm$0.07 & 21.43$\pm$0.07 & 20.70 & 20.96$\pm$0.05 & 18.83 & 20.71$\pm$0.07 & 10 \\
1381-W & 23:57:40.4448 & -32:35:27.888 & 21.77$\pm$0.09 & 21.79$\pm$0.08 & 23.27$\pm$0.09 & 22.49 & 22.02$\pm$0.06 & 20.15 & 22.50$\pm$0.07 & 10 \\
1564-W & 23:57:41.4264 & -32:35:34.368 & 20.10$\pm$0.07 & 20.18$\pm$0.07 & 21.53$\pm$0.07 & 20.42 & 20.80$\pm$0.05 & 18.84 & 20.43$\pm$0.06 & 10 \\
1576-W & 23:57:48.7728 & -32:35:34.656 & 18.73$\pm$0.07 & 19.24$\pm$0.06 & 20.80$\pm$0.06 & 20.74 & 20.73$\pm$0.05 & 18.85 & 20.75$\pm$0.07 & 10 \\
1949-W & 23:57:40.0344 & -32:35:47.436 & 18.93$\pm$0.07 & 19.38$\pm$0.06 & 20.97$\pm$0.06 & 20.96 & 20.71$\pm$0.05 & 19.22 & 20.96$\pm$0.06 & 10 \\
2449-W & 23:57:38.6064 & -32:36:12.312 & 19.67$\pm$0.07 & 19.84$\pm$0.07 & 21.19$\pm$0.07 & 20.47 & 20.89$\pm$0.06 & 17.08 & 20.47$\pm$0.06 & 50 \\
2732-W & 23:57:40.3200 & -32:36:41.328 & 19.32$\pm$0.07 & 19.61$\pm$0.07 & 21.06$\pm$0.06 & 20.68 & 20.59$\pm$0.06 & 18.80 & 20.69$\pm$0.06 & 10 \\
2740-W & 23:57:40.1136 & -32:36:43.452 & 20.39$\pm$0.08 & 20.78$\pm$0.08 & 22.10$\pm$0.07 & 21.77 & 21.83$\pm$0.06 & 19.15 & 21.77$\pm$0.06 & 10 \\
\hline
    \end{tabular}
    \caption{Column (1): ID of star cluster and field (NGC~7793-E or -W). Columns (2)-(3): Right Ascension and Declination. Columns (4)-(10): Apparent magnitudes from A17 (LEGUS-bands) and H19 (H$\alpha$-LEGUS bands), based on PSF-photometry. We use Vega and AB magnitudes for the LEGUS and H$\alpha$-LEGUS bands, respectively. The photometry is corrected for foreground extinction as explained in the text. Column (11): Radius in pixels used for the H$\alpha$ photometry. Note that clusters 93 and 383, and 417 and 1252, which are marked with an asterisk in the first column, have similar coordinates and photometry but have different IDs (see Section~\ref{sub:star_clusters} for a discussion of these clusters).}\label{tab:observed_photometry}
\end{table*}
\endgroup

\begin{figure*}
    \centering
    \includegraphics[width=1.69\columnwidth]{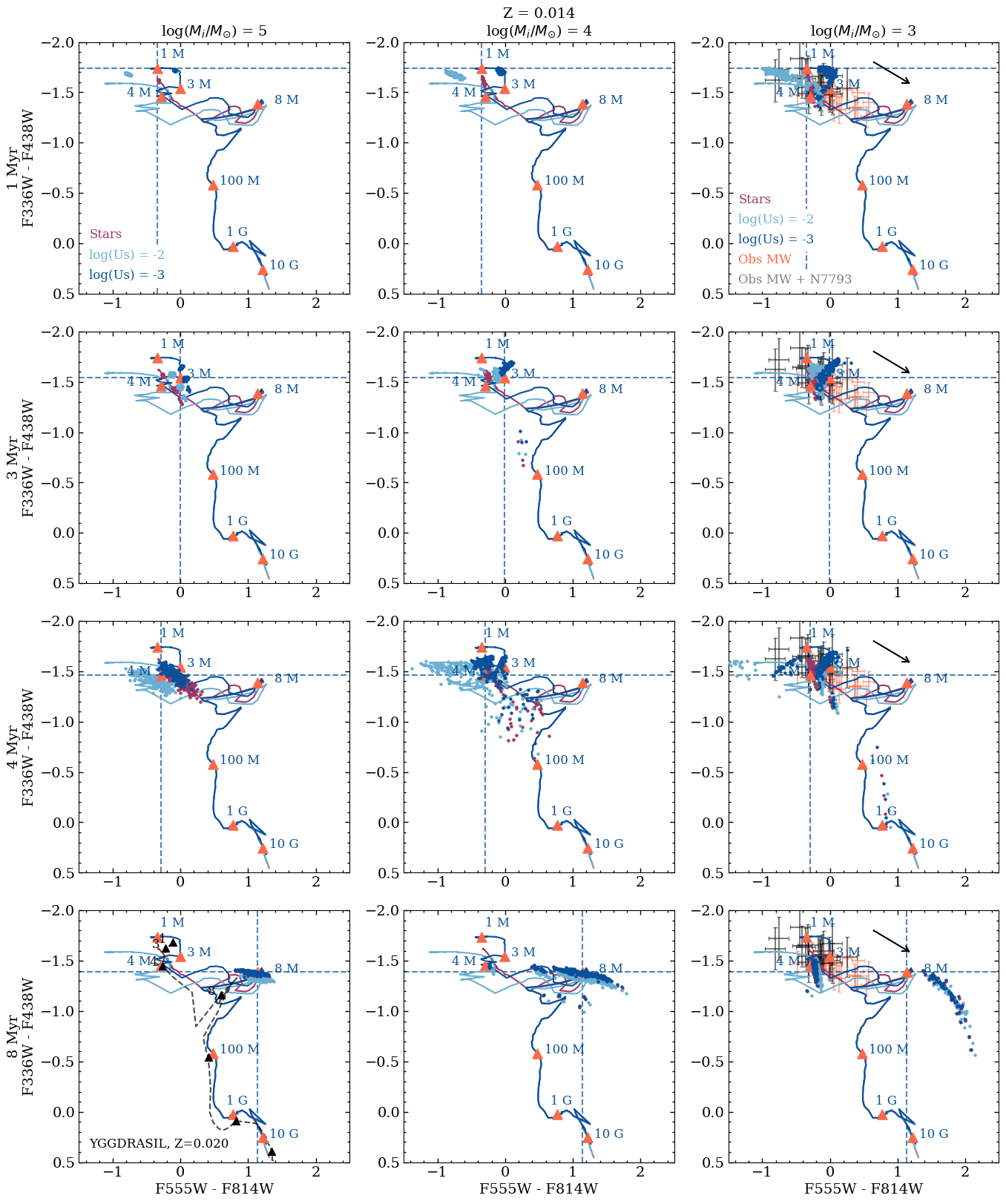}
    \caption{Combination of Figures~\ref{fig:ccd_z014_u0} -~\ref{fig:ccd_z014_u2}, as indicated by the legend on the top-left panel. Some ages along the log(Us)=-3 (dark-blue) track are marked with orange-filled triangles and labelled using M=Myr and G=Gyr. The bottom-left panel shows the {\texttt{Yggdrasil}} track corresponding to log(U)=-3 and $Z=0.020$ (dashed-black curve with black-filled triangles). The last column of panels includes: i) LEGUS observations corrected for reddening due to dust in the MW that are: uncorrected for dust in NGC 7793 (red error bars) and corrected for dust in NGC 7793 (black error bars), as given by the legend in the top-right panel; and ii) a reddening vector corresponding to an extinction of A$_{\rm V}$=1 mag.}\label{fig:ccd_z014}
\end{figure*}

Figure~\ref{fig:ccd_z014} combines Figures~\ref{fig:ccd_z014_u0},~\ref{fig:ccd_z014_u3}, and~\ref{fig:ccd_z014_u2} in one. This helps to see where the $Z=0.014$ S0, S2, and S3 models fall relative to the corresponding D0, D2, and D3 predictions. The corresponding figure for $Z=0.002$ is Figure~\ref{fig:ccd_z002}.

In addition, the panels in the right column of Figure~\ref{fig:ccd_z014} include the LEGUS observations. The red error bars represent the observations corrected for reddening due to  dust in the MW (using Av=0.053 mag) and uncorrected for intrinsic reddening while the black error bars are the observations also corrected for dust in NGC~7793, using the Av values of column 5 in Table~\ref{tab:extinctions}. The black error bars include the propagation of the uncertainties in the intrinsic Av values that are given in column 5 of Table~\ref{tab:extinctions}. For both the foreground and intrinsic reddening corrections, we use the MW-extinction law of \citet{Cardelli1989} for R(V)=3.1. As expected, after the full correction for reddening, the observations move towards bluer colours and the size of the error bars increases.

Note that none of our observations are found near the location of the ``outlier"  stochastic models, which are the few models located far from their corresponding deterministic prediction. This is expected since our observed sample is small and according to stochastic models, ``outlier" clusters have a low probability of being created in nature, and observed.

Several works of the LEGUS collaboration use deterministic {\texttt{Yggdrasil}} tracks of \cite{Zackrisson2011}. The bottom-left panel of Figure~\ref{fig:ccd_z014} shows the {\texttt{Yggdrasil}} track corresponding to log(U)=-3 and $Z=0.020$ (dashed-black curve with ages marked using black-filled triangles). $Z=0.020$ is the closest track to $Z=0.014$ that is available. The {\texttt{Yggdrasil}} track for log(U)=-3 and $Z=0.004$  is shown in Figure~\ref{fig:ccd_z002}. Note that along the $Z=0.004$ {\texttt{Yggdrasil}}, the 8 Myr marker is redder in U - B than the 4 Myr marker, contrary to what happens in the {\texttt{GALAXEV-C}} $Z=0.002$ track.

\section{Method for interpreting the observations}\label{sec:AnalysisObservations}

In this section, we use our models in order to derive the dust extinctions, masses, and ages of the star clusters in our sample. The derived properties are the physical parameters of the models that best fit the observations. In order to find the model that best fits the observations, one can use $\chi^{2}$  minimization \citep{Popescu2010}; construct probability maps in the parameter space to explore the nodes of the grid and then select the more probable solutions \citep{Fouesneau2010}; or use a Bayesian inference method \citep{Krumholz2015a, Wofford2016, Fouesneau2010_2}. We use the latter method, which is explained in the following section.

In order to find the probability distribution function of the physical properties, we use the method of conditional regression, coupled with a kernel density estimation, which is presented in \citep{Krumholz2015a}. In summary, if we let $p(x | y_{obs}; \sigma_{y})$ be the probability distribution of the physical parameters, $x$, given $N$ photometric observations, $y_{obs}$, with error $\sigma_{y}$, the probability distribution of the physical properties, given a set of photometric observations can be written as:

\begin{equation}\label{eqn:2.25}
p(x | y_{obs})  \equiv  \sum_{i} \omega_{i}G((x - x_{i}, y_{obs} - y_{i}), h^{'}),
\end{equation} 
where $h^{'}$ is a new bandwidth that depends on both, the bandwidth of the physical properties of the models $h_{x}$ and the photometric properties ($h_{y}$). This also allows us to find an expression for calculating the marginal probability distribution of each physical parameter, which we will call $x_{1}$.

\begin{equation}\label{eqn:2.26}
    p(x_{1} | y_{obs}) \propto \sum_{i} \omega_{i}G((x_{1} - x_{1, i}, h_{1}) G(y_{obs} - y_{i}),\sqrt{\sigma_{y}^{2} + h_{y}^{2}}).
\end{equation}
This procedure can be followed for each of the physical parameters. 
\\
In order to derive the physical properties of the clusters in our sample, we adapted our pilot-library based on the models presented in Vidal-García et al.(, in prep.) to the tool {\texttt{BAYESPHOT}}, which uses Bayesian inference to estimate joint and marginal PDFs by following the approach which was just described. In addition to the synthetic photometry, {\texttt{BAYESPHOT}} requires additional output from population synthesis, such as the time step, birth mass of the SSP, current mass of all stars in the cluster (accounting for the effects of mass loss and supernovae), number of living stars in the cluster at the present time, visual extinction, to name a few.  In this process, we used the python module {\texttt{SLUGPY}}, which is presented in \cite{Krumholz2015a}. This module is a series of functions that allows to handle spectro-photometric data generated with the population synthesis code {\texttt{SLUG}}. Since NGC~7793 is a spiral galaxy and the clusters have solar metallicity, in order to obtain the extinction due to dust mixed with the neutral gas, in the V-band, we adopted the Milky Way law of \cite{Mathis1990}, with $R(V) = 3.1$.  


\section{RESULTS}\label{sec:results}

In this section, we test our models using the observations which were presented in Section~\ref{sec:Observations}.

\subsection{S3 models versus LEGUS photometry}\label{sub:obs_vs_model_photometry}

In order to illustrate how well one can fit the LEGUS observations with our models, which are only available for three values of the initial mass, we use the S3 models, which have Z = 0.014 and log(U$_{\rm S}$) = -3.

\begin{figure*}
    \centering
    \includegraphics[width=1.99\columnwidth]{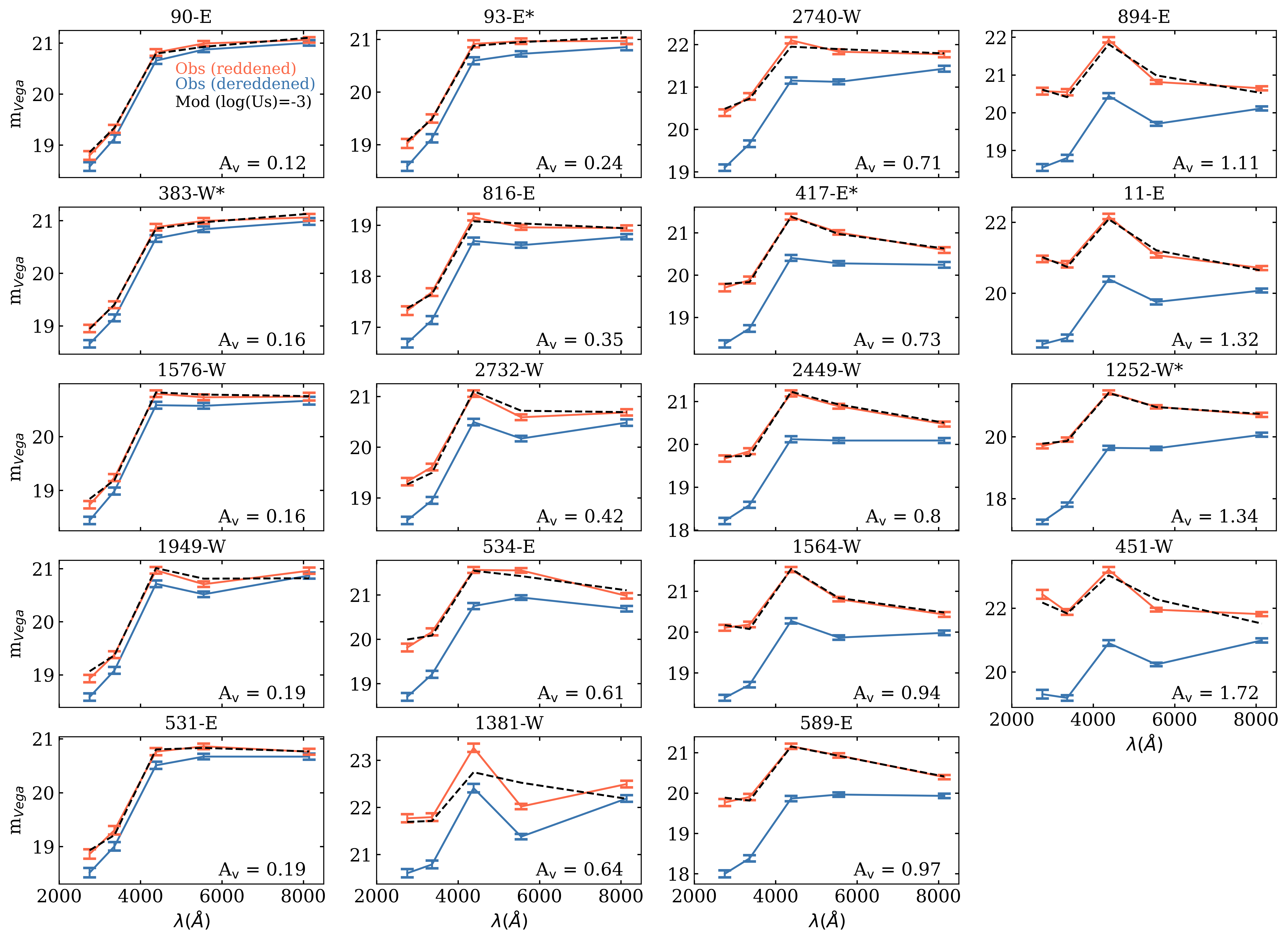}
    \caption{Observed apparent Vega magnitudes (red error bars, corrected for reddening in the Milky Way) versus best-fitting model with $Z=0.014$ and log($U_{\rm S}=-3$ (black dashed lines). For comparison, we also show the observations corrected for reddening in the Milky Way and NGC 7793 (blue error bars). The panels are arranged in order of increasing A$_{\rm V}$ value, which is why for the panels on the right, a larger offset between the red and blue curves can be observed. Note how the bluest magnitude (F275W) is more affected by the reddening correction compared to the reddest magnitude (F814W), which is expected from the shape of the extinction law. As discussed in Section~\ref{sub:star_clusters}, cluster 93 is a copy of 383, and cluster 417 is a copy of 1252. We mark the IDs of these clusters with asterisks. The clusters have different A$_{\rm V}$ values due their slightly different observed magnitudes (see Table~\ref{tab:observed_photometry}).}
    \label{fig:sed_shapes}
\end{figure*}

For each cluster in our sample, Figure~\ref{fig:sed_shapes} shows the comparison of the best-fitting S3 models (black-dashed curves) and the observations. In order to find the best-fitting model, we use equation (29) from \cite{Krumholz2015a}. The figure shows observations (apparent Vega magnitudes) with and without a correction for dust intrinsic to NGC~7793. The red error bars are the observations corrected for reddening due to dust in the Milky Way (using Av=0.017 mag), while the blue error bars include the additional correction for dust in NGC~7793 (using the median V-band extinctions of column 5 in Table~\ref{tab:extinctions}). For both corrections we use the Milky Way extinction law of \citet{Cardelli1989} for R(V)=3.1. Since in the figure, the models are uncorrected for reddening in the neutral gas, they should be compared to the red error bars.

In Figure~\ref{fig:sed_shapes}, the panels are arranged in order of increasing A$_{\rm V}$ value. This is why in the right-side panels there is a larger offset between the red and blue curves. As can be seen in the figure, the F275W magnitude is more affected by the reddening correction than the F814W magnitude, which is in agreement with the shape of the Milky Way extinction law. Note that there are differences in the $A_{\rm V}$ values of 0.08 and 0.61 mag between the clusters that are repeated and discussed above, i.e., 93 and 383; and 417 and 1252, respectively.

For each cluster, Table~\ref{tab:residuals} gives the A$_{\rm V}$ residual (observation - best-fitting model) in each LEGUS band. For the clusters in our sample, the median residual in each LEGUS band is within the observational error which is reported in Table~\ref{tab:observed_photometry}. The observations of cluster 1381-W are not well reproduced in the bands which are redder than U. This is likely because it is least massive cluster in our sample according to the K15 stochastic models which are presented in Table~\ref{tab:masses}. The mass of the latter cluster is 219 \msun while the minimum mass in our models is M=$10^3$ \msun. 

We find models that fit the observations reasonably well in spite of the poor sampling in cluster mass and age of our pilot library thanks to the fine sampling in V-band extinction values.

\begingroup 
\renewcommand{\arraystretch}{1.5}
\begin{table*}
\begin{tabular}{|c|ccccc|}
\hline
\hfill & \multicolumn{5}{c|}{Residual of observed - model magnitude}\\
ID & F275W & F336W & F438W & F555W & F814W \\
(1) & (2) & (3) & (4) & (5) & (6) \\
\hline
0011-E & -0.05 & 0.07 & 0.08 & -0.14 & 0.07 \\
0090-E & -0.06 & -0.02 & 0.02 & 0.07 & -0.04 \\
0093-E & -0.04 & 0.01 & 0.04 & 0.02 & -0.07 \\
0383-W & 0.01 & -0.00 & 0.03 & 0.03 & -0.07 \\
0417-E & -0.09 & 0.05 & 0.00 & 0.04 & -0.04 \\
0451-W & 0.26 & 0.05 & 0.18 & -0.32 & 0.29 \\
0531-E & -0.07 & 0.09 & -0.04 & 0.03 & -0.01 \\
0534-E & -0.18 & 0.08 & 0.02 & 0.13 & -0.13 \\
0589-E & -0.12 & 0.09 & 0.01 & 0.01 & -0.02 \\
0816-E & -0.05 & 0.04 & 0.08 & -0.08 & 0.01 \\
0894-E & -0.04 & 0.13 & 0.12 & -0.17 & 0.12 \\
01252-W & -0.09 & 0.05 & 0.02 & 0.01 & -0.03 \\
01381-W & 0.08 & 0.08 & 0.52 & -0.51 & 0.31 \\
01564-W & -0.07 & 0.11 & -0.02 & -0.03 & -0.05 \\
01576-W & -0.11 & 0.05 & -0.02 & -0.05 & -0.01 \\
01949-W & -0.14 & 0.01 & -0.04 & -0.11 & 0.14 \\
02449-W & -0.05 & 0.11 & -0.04 & -0.04 & -0.03 \\
02732-W & 0.05 & 0.11 & -0.05 & -0.13 & -0.01 \\
02740-W & -0.09 & 0.05 & 0.15 & -0.06 & -0.02 \\ \hline
RANGE & -0.18 - 0.26 & -0.02 - 0.13 & -0.05 - 0.52 & -0.51 - 0.13 & -0.13 - 0.31 \\
MEDIAN & -0.04 & 0.06 & 0.06 & -0.07 & 0.02 \\ \hline
\end{tabular}%
\caption{Residuals in each photometric band. Column (1) shows the observed cluster ID. Columns (2) - (6) shows the difference between the observed magnitudes of the clusters in NGC~7793 and the best-fitting models corresponding to $Z = 0.014$, log(U$_{s}$) = -3, and cluster mass = 10$^{3}$~\msun. For each LEGUS filter, the last two rows give the range and median of the values in the column.}\label{tab:residuals}
\end{table*}
\endgroup
\subsection{Equivalent width of H$\alpha$}\label{sub:ew_ha}

The F656N filter includes the H$\alpha$ and [N\2] lines. We compare the equivalent width of the combined H$\alpha$ + [N\2] emission from the best-fitting S2 and S3 models with $Z = 0.014$, against the value measured by Hannon et al. (in prep., hereafter H21) using the H$\alpha$-LEGUS observations. In order to catch all of the H$\alpha$ emission due to ionisation by the cluster, the size of the aperture used by H21 to obtain EW(H$\alpha$+ [N\2]) was selected based on the curve of growth of each cluster. The radius of the aperture is provided in the last column of Table~\ref{tab:observed_photometry}. We present the observed and best-fit model EWs in Table~\ref{tab:EW}. We find that for two clusters (816-E and 1564-W) the S2 models are within $100$\,\AA of the observed value, while the S3 models are within $100$\,\AA of the observed value only in one case (531-E). Note that the observed EW(H$\alpha$+ [N\2]) value is highly uncertain. We also find that for the S2 models, the mean of EW(H$\alpha$+ [N\2]) is a factor of $\sim4$ lower than the mean of the observations, and that the mean of the S3 models is closer to the mean of the observations. Finally, we find that in general, the youngest clusters have the largest model value of EW(H$\alpha$+ [N\2]) (see Table~\ref{tab:ages}).

\begingroup 
\renewcommand{\arraystretch}{1.5}
\begin{table*}
\begin{tabular}{|l|ccc|cc|}
\hline
ID & \multicolumn{3}{c|}{\begin{tabular}[c]{@{}c@{}}EW(H$\alpha$)\\ \AA\end{tabular}} & \multicolumn{2}{c|}{\begin{tabular}[c]{@{}c@{}}$\Delta$EW(H$\alpha$)\\ \AA\end{tabular}} \\
\begin{tabular}[c]{@{}l@{}}\hfill\\ (1)\end{tabular} & \begin{tabular}[c]{@{}c@{}}H21\\ (2)\end{tabular} & \begin{tabular}[c]{@{}c@{}}S2\\ (3)\end{tabular} & \multicolumn{1}{c|}{\begin{tabular}[c]{@{}c@{}}S3\\ (4)\end{tabular}} & \begin{tabular}[c]{@{}c@{}}(3) - (2)\\ (5)\end{tabular} & \begin{tabular}[c]{@{}c@{}}(4) - (2)\\ (6)\end{tabular} \\ \hline
0011-E & 3038 & 1919 & 2097 & -1119  &  -841\\
0090-E & 1489  & 171 & 235 &  -1318&  -1254 \\
0093-E & 1232 & 128 & 1987 &  -1104&  755\\
0383-W & 1043 & 128 & 325 & -915 & -718 \\
0417-E & 1640  & 102 & 2365 &  -1538&  725\\
0451-W & 3288 & 102 & 2299 & -3186 & -989 \\
0531-E & 364  & 54 & 293 &  -310&  -71\\
0534-E & 444  & 143 & 2080 &  -301&  1636\\
0589-E & 2423  & 41 & 2080 &  -2382&  -415\\
0816-E & 1988  & 1919 & 803 &  -69&  -1185\\
0894-E & 2043  & 1919 & 1913 &  -124&  -130\\
1252-W & 1823 & 41 & 3121 & -1782 & 1298 \\
1381-W & 789 & 63 & 14 & -726 &  -775 \\
1564-W & 1197 & 1290 & 3852 & 93 & 2655 \\
1576-W & 2472 & 91 & 2177 & -2381 & -295 \\
1949-W & 1053 & 41 & 2048 & -1012 & 995 \\
2449-W & 358 & 80 & 3187 & -278 & 2829 \\
2732-W & 796 & 41 & 577 & -755 & -219 \\
2740-W & 826 & 153 & 325 & -673 & -501 \\ \hline
RANGE & 358 - 3288 & 41 - 1919 & 14 - 3852 & \multicolumn{1}{l}{- 3186 - 93} & \multicolumn{1}{l|}{-1254 - 2829} \\
MEAN & 1489 & 387 & 1674 & -1046 & -184 \\ \hline
\end{tabular}%
\caption{Observed versus predicted H$\alpha$ equivalent widths. Column (1) - Cluster ID. Column (2) - Observed EW(H$\alpha$) from H21. Columns (3) and (4) - value from best-fitting S2 and S3 models, respectively ($Z=0.014$). Columns (5) and (6) give the differences between values in the columns which are indicated in the header of the table.}\label{tab:EW}
\end{table*}
\endgroup

\subsection{Physical properties of the star clusters}\label{sub:physical_properties}
We use the Bayesian inference tool \texttt{BAYESPHOT} and the formalism presented in Section~\ref{sec:AnalysisObservations} in order to determine the extinction, mass and age of the star clusters in our sample. 

{\it V-band extinction}. Table~\ref{tab:extinctions} gives V-band extinctions (A$_{\rm V}$) from the literature and our work. Column (2) gives the value from A17, which is derived via deterministic models and $\chi^{2}$ minimization. Column (3) gives the median value that is derived with \texttt{BAYESPHOT} and the Z=0.020 stochastic models of \citet[hereafter K15]{Krumholz2015a}. Columns (4) and (5) give the median values that are derived with \texttt{BAYESPHOT} and the Z=0.014 S2 and S3 \texttt{GALAXEV-C} models, respectively. We find that the S2 models yield lower extinctions than the S3 models. Thus, the extinction depends on the log(U$_{\rm S}$) value of the models. We also find good general agreement between the A17, K15 and S3 extinctions (within the error bars), and that the A17 extinctions tend to be the largest. 

{\it Mass}. Table~\ref{tab:masses} gives masses from the literature and our work. Although the models in our pilot library use a coarse grid of cluster masses, we find that the observed clusters are low mass, in agreement with previous results. The mean cluster mass is 10$^3$ \msun using the A17, K15 S2, and S3 models. Thus, the value of log(U$_{\rm S}$) does not affect the estimated mass value. 

{\it Age.} Table~\ref{tab:ages} gives ages from the literature and our work. We find that the S2 models yield an older mean cluster age relative to the S3 models and that the A17 models yield the youngest mean age (2 Myr). We also find that the oldest/youngest clusters using K15 are the oldest/youngest clusters from S3 as well. Finally, according to the S3 models, four clusters in our sample are 1 Myr. This is an age when nebular emission lines contribute significantly to the V-band luminosity and the nebular continuum to the I-band luminosity.

\begingroup 
\renewcommand{\arraystretch}{1.5}
\begin{table*}\label{tab:extinction}
\begin{tabular}{|l|cccc|ccc|}
\hline
ID & \multicolumn{4}{c|}{Av} & \multicolumn{3}{c|}{$\Delta$Av}  \\ 
\hfill & A17 & K15 & S2 & S3 & (3)-(2) & (3)-(5) & (5)-(4) \\
(1) & (2) & (3) & (4) & (5) & (6) & (7) & (8) \\ \hline
0011-E	&	1.89	$^{+1.80}_{-1.98}$	&	1.25	$^{+0.24}_{-0.19}$	&	0.54	$^{+0.10}_{-0.09}$	&	1.32	$^{+0.16}_{-0.14}$	&	-0.64	&	-0.07	&	0.78	\\
0090-E	&	0.12	$^{+0.00}_{-0.19}$	&	0.17	$^{+0.11}_{-0.12}$	&	0.07	$^{+0.07}_{-0.05}$	&	0.12	$^{+0.14}_{-0.07}$	&	0.05	&	0.05	&	0.05	\\
0093-E	&	0.37	$^{+0.00}_{-0.46}$	&	0.31	$^{+0.16}_{-0.17}$	&	0.09	$^{+0.1}_{-0.07}$	&	0.24	$^{+0.16}_{-0.15}$	&	-0.06	&	0.07	&	0.15	\\
0383-W	&	0.25	$^{+0.12}_{-0.31}$	&	0.21	$^{+0.14}_{-0.14}$	&	0.07	$^{+0.09}_{-0.05}$	&	0.16	$^{+0.15}_{-0.11}$	&	-0.04	&	0.05	&	0.09	\\
0417-E	&	1.05	$^{+0.93}_{-1.15}$	&	0.83	$^{+0.23}_{-0.26}$	&	0.28	$^{+0.10}_{-0.12}$	&	0.73	$^{+0.19}_{-0.19}$	&	-0.22	&	0.10	&	0.45	\\
0451-W	&	2.08	$^{+1.98}_{-2.17}$	&	1.39	$^{+0.26}_{-0.26}$	&	0.64	$^{+0.11}_{-0.12}$	&	1.72	$^{+0.28}_{-0.17}$	&	-0.69	&	-0.33	&	1.08	\\
0531-E	&	0.37	$^{+0.22}_{-0.46}$	&	0.33	$^{+0.17}_{-0.19}$	&	0.09	$^{+0.10}_{-0.07}$	&	0.19	$^{+0.16}_{-0.12}$	&	-0.04	&	0.14	&	0.10	\\
0534-E	&	0.81	$^{+0.00}_{-0.90}$	&	0.71	$^{+0.21}_{-0.21}$	&	0.21	$^{+0.12}_{-0.09}$	&	0.61	$^{+0.17}_{-0.16}$	&	-0.10	&	0.10	&	0.40	\\
0589-E	&	1.21	$^{+1.12}_{-1.36}$	&	0.99	$^{+0.24}_{-0.64}$	&	0.31	$^{+0.09}_{-0.12}$	&	0.97	$^{+0.23}_{-0.22}$	&	-0.22	&	0.02	&	0.66	\\
0816-E	&	0.62	$^{+0.50}_{-0.71}$	&	0.38	$^{+0.19}_{-0.21}$	&	0.12	$^{+0.09}_{-0.07}$	&	0.35	$^{+0.17}_{-0.19}$	&	-0.24	&	0.03	&	0.23	\\
0894-E	&	1.71	$^{+1.61}_{-1.86}$	&	1.02	$^{+0.18}_{-0.19}$	&	0.47	$^{+0.12}_{-0.09}$	&	1.11	$^{+0.16}_{-0.14}$	&	-0.69	&	-0.09	&	0.64	\\
1381-W	&	1.21	$^{+1.15}_{-1.30}$	&	0.33	$^{+0.40}_{-0.24}$	&	0.49	$^{+1.58}_{-0.18}$	&	1.34	$^{+0.21}_{-0.21}$	&	-0.88	&	-1.01	&	0.85	\\
1252-W	&	0.96	$^{+0.87}_{-1.05}$	&	0.66	$^{+0.28}_{-0.23}$	&	0.26	$^{+0.12}_{-0.10}$	&	0.64	$^{+0.21}_{-0.17}$	&	-0.30	&	0.02	&	0.38	\\
1564-W	&	1.46	$^{+1.40}_{-1.61}$	&	1.04	$^{+0.21}_{-0.24}$	&	0.42	$^{+0.10}_{-0.09}$	&	0.94	$^{+0.24}_{-0.16}$	&	-0.42	&	0.10	&	0.52	\\
1576-W	&	0.34	$^{+0.19}_{-0.40}$	&	0.24	$^{+0.16}_{-0.15}$	&	0.09	$^{+0.07}_{-0.07}$	&	0.16	$^{+0.15}_{-0.11}$	&	-0.10	&	0.08	&	0.07	\\
1949-W	&	0.37	$^{+0.03}_{-0.43}$	&	0.24	$^{+0.16}_{-0.15}$	&	0.09	$^{+0.10}_{-0.07}$	&	0.19	$^{+0.16}_{-0.12}$	&	-0.13	&	0.05	&	0.10	\\
2449-W	&	1.08	$^{+0.99}_{-1.24}$	&	0.94	$^{+0.19}_{-0.25}$	&	0.28	$^{+0.10}_{-0.12}$	&	0.80	$^{+0.24}_{-0.19}$	&	-0.14	&	0.14	&	0.52	\\
2732-W	&	0.74	$^{+0.46}_{-0.90}$	&	0.54	$^{+0.26}_{-0.28}$	&	0.21	$^{+0.10}_{-0.12}$	&	0.42	$^{+0.29}_{-0.18}$	&	-0.20	&	0.12	&	0.21	\\
2740-W	&	0.50	$^{+0.00}_{-0.78}$	&	0.66	$^{+0.21}_{-0.21}$	&	0.19	$^{+0.12}_{-0.10}$	&	0.71	$^{+0.16}_{-0.17}$	&	0.16	&	-0.05	&	0.52	\\
\hline																			
RANGE	&	0.1 - 2.1		&	0.2 - 1.4		&	0.1 - 0.6		&	0.1 - 1.7		&	-0.9 - 0.2	&	-1.0 - 0.1	&	0.1 - 1.1	\\
MEAN	&	0.9		&	0.6		&	0.3		&	0.7		&	-0.3	&	-0.0	&	0.4		\\
\hline
\end{tabular}
\caption{V-band extinctions of star-clusters derived with different models, all of which include gas. The difference between values given by the different models is also given. Column (1) - Cluster ID. Column (2) - Deterministic models with Z=0.020 of A17. Column (3) - Stochastic models with Z=0.020 of \citet{Krumholz2015a}. Columns (4) and (5) - Stochastic models with Z=0.014 and log(U$_{\rm S}$)=-2 and log(U$_{\rm S}$)=-3, respectively, presented in this work. Columns (6) to (8) - Differences between values in the columns which are indicated. Column (9) Comment.}\label{tab:extinctions}
\end{table*}
\endgroup

\begingroup
\renewcommand{\arraystretch}{1.5}
\begin{table*}
\begin{tabular}{|l|cccc|c|}
\hline
ID & \multicolumn{4}{c|}{log(M / M$_\odot$)} & $\Delta$log(M / M$_\odot$) \\
\hfill & A17 & K15 & S2 & S3 & (3)-(2) \\
(1) & (2) & (3) & (4) & (5) & (6) \\ \hline
0011-E & 3.43$^{+3.47}_{-3.39}$ & 2.59$^{+0.40}_{-0.35}$ & 3.0$^{+0.02}_{-0.00}$ & 3.0$^{+0.02}_{-0.00}$ & -0.44 \\
0090-E & 2.77$^{+2.86}_{-2.63}$ & 2.89$^{+0.35}_{-0.55}$ & 3.0$^{+0.02}_{-0.00}$ & 3.0$^{+0.02}_{-0.00}$ & 0.47 \\
0093-E & 2.88$^{+2.95}_{-2.72}$ & 2.94$^{+0.40}_{-0.55}$ & 3.0$^{+0.02}_{-0.00}$ & 3.0$^{+0.02}_{-0.00}$ & 0.46 \\
0383-W & 2.71$^{+2.85}_{-2.68}$ & 2.94$^{+0.35}_{-0.55}$ & 3.0$^{+0.02}_{-0.00}$ & 3.0$^{+0.02}_{-0.00}$ & 0.58 \\
0417-E & 3.25$^{+3.30}_{-3.19}$ & 2.69$^{+0.40}_{-0.40}$ & 3.0$^{+0.02}_{-0.00}$ & 3.0$^{+0.02}_{-0.00}$ & -0.16 \\
0451-W & 2.96$^{+3.00}_{-2.92}$ & 2.54$^{+0.45}_{-0.35}$ & 3.0$^{+0.02}_{-0.00}$ & 3.0$^{+0.02}_{-0.00}$ & 0.03 \\
0531-E & 2.94$^{+3.04}_{-2.82}$ & 2.89$^{+0.45}_{-0.55}$ & 3.0$^{+0.02}_{-0.00}$ & 3.0$^{+0.02}_{-0.00}$ & 0.40 \\
0534-E & 2.99$^{+3.03}_{-2.90}$ & 2.64$^{+0.40}_{-0.40}$ & 3.0$^{+0.02}_{-0.00}$ & 3.0$^{+0.02}_{-0.00}$ & 0.05 \\
0589-E & 3.37$^{+3.41}_{-3.28}$ & 3.04$^{+0.50}_{-0.60}$ & 3.0$^{+0.02}_{-0.00}$ & 3.0$^{+0.02}_{-0.00}$ & 0.17 \\
0816-E & 3.65$^{+3.83}_{-3.62}$ & 3.49$^{+0.25}_{-0.25}$ & 3.0$^{+0.02}_{-0.00}$ & 3.0$^{+0.02}_{-0.00}$ & 0.09 \\
0894-E & 3.43$^{+3.46}_{-3.25}$ & 2.54$^{+0.40}_{-0.35}$ & 3.0$^{+0.02}_{-0.00}$ & 3.0$^{+0.02}_{-0.00}$ & -0.49 \\
1252-W & 3.19$^{+3.24}_{-3.14}$ & 2.69$^{+0.35}_{-0.40}$ & 3.0$^{+0.02}_{-0.00}$ & 3.0$^{+0.02}_{-0.00}$ & -0.15 \\
1381-W & 2.48$^{+2.52}_{-2.45}$ & 2.34$^{+0.30}_{-0.20}$ & 3.02$^{+1.27}_{-0.02}$ & 3.0$^{+0.16}_{-0.00}$ & 0.16 \\
1564-W & 3.43$^{+3.47}_{-3.28}$ & 2.74$^{+0.40}_{-0.40}$ & 3.0$^{+0.02}_{-0.00}$ & 3.0$^{+0.02}_{-0.00}$ & -0.29 \\
1576-W & 2.95$^{+3.06}_{-2.81}$ & 2.79$^{+0.40}_{-0.50}$ & 3.0$^{+0.02}_{-0.00}$ & 3.0$^{+0.02}_{-0.00}$ & 0.24 \\
1949-W & 2.81$^{+2.91}_{-2.76}$ & 2.79$^{+0.35}_{-0.45}$ & 3.0$^{+0.02}_{-0.00}$ & 3.0$^{+0.02}_{-0.00}$ & 0.33 \\
2449-W & 3.31$^{+3.34}_{-3.21}$ & 2.84$^{+0.45}_{-0.50}$ & 3.0$^{+0.02}_{-0.00}$ & 3.0$^{+0.02}_{-0.00}$ & -0.02 \\
2732-W & 3.05$^{+3.13}_{-3.00}$ & 2.89$^{+0.35}_{-0.45}$ & 3.0$^{+0.02}_{-0.00}$ & 3.0$^{+0.02}_{-0.00}$ & 0.19 \\
2740-W & 2.58$^{+2.62}_{-2.26}$ & 2.69$^{+0.40}_{-0.40}$ & 3.0$^{+0.02}_{-0.00}$ & 3.0$^{+0.02}_{-0.00}$ & 0.51 \\ \hline
RANGE & 2.48 - 3.65 & 2.34 - 3.49 & 3.00 - 3.00 & 3.00 - 3.00 & -0.49 - 0.58 \\
MEAN & 3.06 & 2.79 & 3.00 & 3.00 & 0.11 \\ \hline
\end{tabular}
\caption{Total mass in stars of star-cluster derived with different models, all of which include gas. The difference between values given by the different models is also given. Column (1) - Cluster ID. Column (2) - deterministic models with Z=0.020 A17. Column (3) - Stochastic models with Z=0.020 of \citet{Krumholz2015a}. Columns (4) and (5) - stochastic models with (Z=0.014, log(U$_{\rm S}$)=-2) and (Z=0.014, log(U$_{\rm S}$)=-3) from this work, respectively. Columns (6) Differences between values in the columns which are indicated.}\label{tab:masses}
\end{table*}
\endgroup

\begingroup
\renewcommand{\arraystretch}{1.5}
\begin{table*}
\begin{tabular}{|l|cccc|ccc|}
\hline
ID & \multicolumn{4}{c|}{t / Myr} & \multicolumn{3}{c|}{$\Delta$t / Myr} \\
\hfill & A17 & K15 & S2 & S3 & (3)-(2) & (3)-(5) & (5)-(4) \\
(1) & (2) & (3) & (4) & (5) & (6) & (7) & (8) \\ \hline
0011-E & 1.0$^{+1.0}_{-1.0}$ & 0.7$^{+1.9}_{-0.5}$ & 3.0$^{+0.1}_{-0.0}$ & 1.0$^{+0.0}_{-0.0}$ & -0.3 & -0.4 & -2.0 \\
0090-E & 2.0$^{+3.0}_{-1.0}$ & 5.6$^{+1.6}_{-0.6}$ & 7.8$^{+0.1}_{-0.1}$ & 7.8$^{+0.1}_{-3.8}$ & 3.7 & -2.2 & 0.0 \\
0093-E & 2.0$^{+5.0}_{-1.0}$ & 5.1$^{+1.9}_{-0.5}$ & 7.8$^{+0.1}_{-0.1}$ & 4.0$^{+3.8}_{-0.2}$ & 3.2 & 1.1 & -3.8 \\
0383-W & 3.0$^{+4.0}_{-2.0}$ & 5.6$^{+1.8}_{-0.5}$ & 7.8$^{+0.1}_{-0.1}$ & 7.6$^{+0.4}_{-3.6}$ & 2.6 & -1.9 & -0.3 \\
0417-E & 1.0$^{+1.0}_{-1.0}$ & 2.0$^{+1.5}_{-0.6}$ & 7.8$^{+0.1}_{-0.1}$ & 3.0$^{+0.9}_{-0.1}$ & 1.0 & -1.0 & -4.8 \\
0451-W & 4.0$^{+4.0}_{-4.0}$ & 0.7$^{+7.1}_{-0.3}$ & 7.8$^{+0.1}_{-0.1}$ & 1.0$^{+3.0}_{-0.0}$ & -3.3 & -0.4 & -6.8 \\
0531-E & 2.0$^{+3.0}_{-1.0}$ & 4.3$^{+1.9}_{-0.5}$ & 7.8$^{+0.1}_{-0.1}$ & 3.9$^{+0.1}_{-0.2}$ & 2.3 & 0.3 & -3.9 \\
0534-E & 1.0$^{+15.0}_{-1.0}$ & 2.2$^{+1.6}_{-0.8}$ & 7.8$^{+0.1}_{-0.1}$ & 3.9$^{+0.1}_{-1.0}$ & 1.2 & -1.7 & -3.9 \\
0589-E & 1.0$^{+2.0}_{-1.0}$ & 2.7$^{+8.5}_{-0.6}$ & 7.8$^{+0.1}_{-0.1}$ & 3.0$^{+1.0}_{-0.1}$ & 1.7 & -0.4 & -4.8 \\
0816-E & 3.0$^{+3.0}_{-1.0}$ & 2.5$^{+1.6}_{-0.6}$ & 3.0$^{+0.8}_{-0.1}$ & 3.0$^{+0.1}_{-0.1}$ & -0.6 & -0.6 & 0.0 \\
0894-E & 1.0$^{+3.0}_{-1.0}$ & 0.5$^{+1.9}_{-0.5}$ & 3.0$^{+0.1}_{-0.1}$ & 1.0$^{+0.0}_{-0.0}$ & -0.5 & -0.5 & -2.0 \\
1252-W & 1.0$^{+1.0}_{-1.0}$ & 1.7$^{+1.8}_{-0.6}$ & 7.8$^{+0.1}_{-0.1}$ & 3.0$^{+0.9}_{-2.0}$ & 0.7 & -1.4 & -4.8 \\
1381-W & 4.0$^{+4.0}_{-4.0}$ & 4.3$^{+1.3}_{-0.8}$ & 7.8$^{+0.1}_{-5.1}$ & 7.8$^{+0.1}_{-0.3}$ & 0.3 & -3.5 & 0.0 \\
1564-W & 1.0$^{+3.0}_{-1.0}$ & 1.3$^{+2.1}_{-0.6}$ & 3.0$^{+0.1}_{-0.0}$ & 1.0$^{+2.0}_{-0.0}$ & 0.3 & 0.2 & -2.0 \\
1576-W & 2.0$^{+3.0}_{-1.0}$ & 3.0$^{+1.9}_{-0.6}$ & 7.8$^{+0.1}_{-0.1}$ & 3.9$^{+0.1}_{-0.9}$ & 0.9 & -1.0 & -3.9 \\
1949-W & 3.0$^{+5.0}_{-2.0}$ & 3.5$^{+1.6}_{-0.6}$ & 7.8$^{+0.1}_{-0.1}$ & 4.0$^{+3.8}_{-1.0}$ & 0.5 & -0.5 & -3.8 \\
2449-W & 1.0$^{+2.0}_{-1.0}$ & 2.2$^{+1.8}_{-0.7}$ & 7.8$^{+0.1}_{-0.1}$ & 3.0$^{+1.0}_{-0.1}$ & 1.2 & -0.8 & -4.8 \\
2732-W & 4.0$^{+5.0}_{-2.0}$ & 2.2$^{+1.9}_{-0.4}$ & 7.8$^{+0.1}_{-0.1}$ & 3.0$^{+1.0}_{-2.0}$ & -1.8 & -0.8 & -4.8 \\
2740-W & 5.0$^{+6.0}_{-4.0}$ & 3.9$^{+2.3}_{-0.4}$ & 7.8$^{+0.1}_{-0.1}$ & 7.8$^{+0.1}_{-3.9}$ & -1.1 & -3.9 & 0.0 \\ \hline
RANGE & 1 - 5 & 0.5 - 5.6 & 2.9 - 7.8 & 1.0 - 7.8 & -3.3 - 3.6 & -3.9 - 1.0 & -6.8 - 0.0 \\
MEAN & 2.2 & 2.8 & 6.79 & 3.8 & 0.6 & -1.0 & -3.0 \\ \hline
\end{tabular}
\caption{Ages of star-cluster derived with different models, all of which include gas. The difference between values given by the different models is also given. Column (1) - Cluster ID. Column (2) Deterministic models with Z=0.020 (A17). Column (2) - Stochastic models with Z=0.020 of \citet{Krumholz2015a}. Columns (4) and (5) - Stochastic models with Z=0.014 and log(U$_{\rm S}$)=-2 and log(U$_{\rm S}$)=-3, respectively, presented in this work. Columns (6) to (8) - Differences between values in the columns which are indicated.}\label{tab:ages}
\end{table*}
\endgroup

\subsection{Well versus poorly constrained solutions}\label{sub:multiple_peaks}

\begin{figure*}
    \centering
   \includegraphics[width=1.6\columnwidth, height = 1.1\columnwidth]{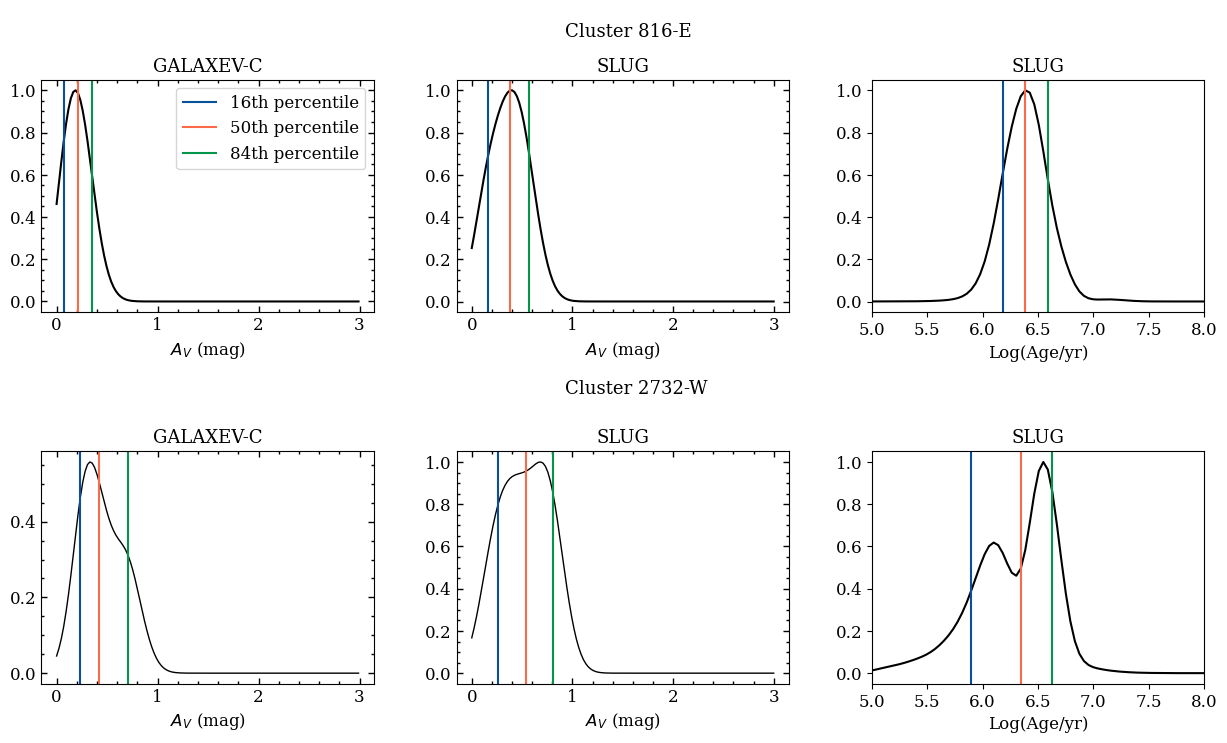}
    \caption{Left column--. The Z=0.014/log(Us)=-3/\texttt{GALAXEV-C} V-band extinction PDFs of clusters 816-E (top panel) and 2732-W (bottom panel). The blue, red, and green lines give the 16th, 50th and 84th percentiles, respectively.  Middle column--. Similar to the left column but we show the Z=0.020/log(Us)=-3/{\texttt{SLUG}} V-band extinction PDFs. Right column--. Z=0.020/log(Us)=-3/{\texttt{SLUG}} age PDFs.}\label{fig:PDF}
\end{figure*}

For clusters 816-W (top row) and 2732-W (bottom row), Figure~\ref{fig:PDF} shows the extinction PDFs corresponding to the Z = 0.014 / S3 / \texttt{GALAXEV-C} models (left column) and the Z=0.020 / log(Us)=-3 / \texttt{SLUG} models (middle column). For 2732-W, note that \texttt{GALAXEV-C} and \texttt{SLUG} yield single- and multiple-valued extinction PDFs, respectively. That one PDF is single-peaked and the other one is not is attributed to differences between the \texttt{GALAXEV-C} and \texttt{SLUG} libraries. In particular, as explained in the introduction, the \texttt{SLUG} models do not include the effect of the stochastic variation in the shape of the ionizing continuum, on the nebular emission.\\

The right column of Figure~\ref{fig:PDF} shows the \texttt{SLUG} age PDFs corresponding the the above two clusters. We only show the \texttt{SLUG} results because the pilot \texttt{GALAXEV-C} age grid is very coarse. The age PDF is single-peaked for cluster 816-W and multi-peaked for cluster 2732-W. A discussion of the PDFs for the whole sample of clusters is provided in  Appendix~\ref{app:cluster_pdfs}.


\section{Summary and Conclusions}\label{sec:Conclusions}

\begin{enumerate}
    \item We present a pilot \textbf{\texttt{GALAXEV-C}} library of synthetic \hst-equivalent NUV, U, B, V, and I photometry of star clusters that accounts for the stochastic sampling of the stellar IMF and the contribution of the ionized gas and dust mixed with the ionized gas (Section~\ref{sec:Models}). The library uses the spectra that are presented in Vidal-García et al. (in prep.) and includes models for clusters with initial masses, $M_i=10^3$, $10^4$, and $10^5$ \msun; ages,  $t=1$, 3, 4, and 8 Myr; metallicities, $Z=0.002$ and $Z=0.014$ (solar); and ionisation parameters, log(U$_{\rm S})=-2$ (S2 models) and -3 (S3 models). We compare the stochastic models to corresponding deterministic models (Section~\ref{sec:det_vs_sto}); and to \hst LEGUS and H$\alpha$-LEGUS observations of star clusters in galaxy NGC~7793 that are isolated, have compact H$\alpha$ morphologies, $Z\sim0.014$ (Figure~\ref{fig:cluster_locations}), and deterministic masses and ages of $<10^4$ \msun and $\leq10$ Myr, respectively. We determine the V-band extinctions, masses, and ages of these clusters using the stochastic models (Section~\ref{sec:AnalysisObservations}). We compare the cluster properties derived with deterministic models that are published in A17, and derived with independent \texttt{GALAXEV-C} and \texttt{SLUG} (K15) stochastic models (Section~\ref{sub:physical_properties}).
   \item For \textbf{\texttt{GALAXEV}} magnitudes that only account for the stars: a) the absolute value of the residual, deterministic mag - median stochastic mag, can be $\ge0.5$ mag, even for $M_i=10^5$\,\msun (Table~\ref{tab:det_vs_soc}); and b) the largest spread of the stochastic models occurs at 3 and 4\,Myr, when Wolf-Rayet stars are present (Figure~\ref{fig:magnitude_violin_plots}).
   \item For $M_i=10^5$\,\msun: a) the median stochastic mag with gas can be $>$1.0 mag more luminous than the median stochastic mag without gas  (Table~\ref{tab:median_mag_stars_stars_and_gas_difference}); and b) the nebular emission lines can contribute with $>50\%$ and $>30\%$ to the total emission in the V and I bands, respectively (Figure~\ref{fig:contributions}). 
   \item Regarding age-dating OB clusters via deterministic tracks in the U - B vs. V - I diagram, we find that this method leads to highly uncertain ages at $Z=0.014$ for $M_i\sim10^3$\,\msun (Figures~\ref{fig:ccd_z014_u0} - \ref{fig:ccd_z014_u2}) and $Z=0.002$ for all masses in the stochastic library  (Figures~\ref{fig:ccd_z002_u0} - \ref{fig:ccd_z002_u2}).
    \item Also regarding the U - B vs. V - I diagram, we find that at $Z=0.014$, a small fraction of models with $M_i\sim10^3$ and $M_i\sim10^4$~\msun are located far from their corresponding deterministic predictions and none of our observations are found near these outlier models. This is expected given that our observational sample is small and according to stochastic models the corresponding outlier clusters have a low probability of being created.
    \item Regarding the SED fitting, we find good agreement between the best-fitting S3 model and the observations (Figure~\ref{fig:sed_shapes}, Table~\ref{tab:residuals}).
    \item We derive the extinctions in the V-band, masses, and ages of the star clusters in our sample using two independent libraries of stochastic models with gas, the K15 library and our pilot library. We compare the results with those of A17, which are based on deterministic models (Tables~\ref{tab:extinctions} to~\ref{tab:ages}).
    \item Regarding the extinction, we find that the \texttt{GALAXEV-C} A$_{\rm V}$ value is systematically lower for log(U$_{\rm S}$)=-2 than for log(U$_{\rm S}$)=-3. We also find that the A17, K15, and S3 extinctions are in general agreement (within the errors), and that the A17 extinctions tend to be the largest. Finally, we find that for a given cluster, the extinction PDF can be single-peaked for \texttt{GALAXEV-C} and multi-peaked for \texttt{SLUG} and vise versa, which is attributed to differences in the stochastic libraries. 
    \item Regarding the masses, we find that the observed clusters are low mass, in agreement with the deterministic predictions (Table~\ref{tab:masses}).
    \item Regarding the ages, we find that the S2 models tend to yield ages relative to the S3 models, and that the A17 models yield the youngest mean age (2 Myr; Table~\ref{tab:ages}). We also find that in several cases, the age PDF presents multiple peaks. 
    \item Regarding models versus nature, we recall that for a multinomial distribution, the standard deviations of the different mass bins are not independent from each other, whereas we have no reason to believe that the same correlation between the standard deviations is true in nature. This is why it is important to observationally characterize the true variance in nature.
    \item An extension of the pilot library to other metallicities and ages is near completion and can be made available upon request to AVG, who is a co-author of this work.
\end{enumerate}

\appendix

\section{Is 220 realizations enough?}\label{app:enough}

\begin{figure*}
    \centering
    \includegraphics[width=0.89\columnwidth]{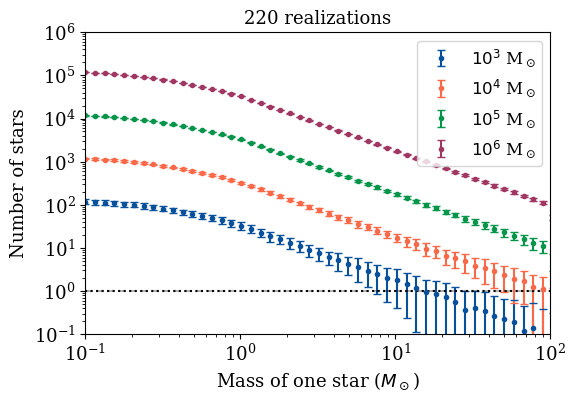}%
    \includegraphics[width=0.89\columnwidth]{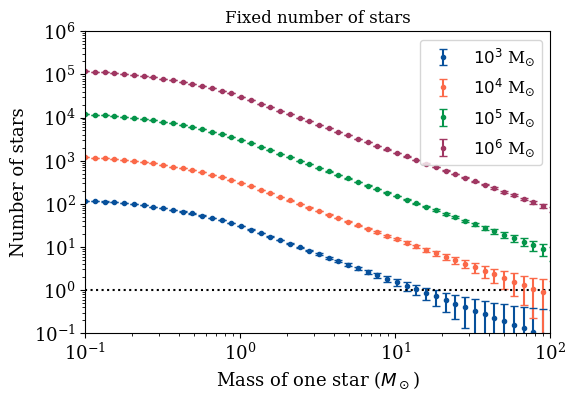}
    \caption{Left-panel---Same as Figure~\ref{fig:1000realizations} but for 220 realizations. Right-panel---Expected mean and standard deviations for a number of stars that is fixed such that the expected mass is $M_i$.}
    \label{fig:220realizations}
\end{figure*}

The left-panel of Figure~\ref{fig:220realizations} is similar to Figure~\ref{fig:1000realizations} but includes only the number of realizations that are used in this work. These Figures show that reducing the number of realizations from 1000 to 220 has no significant effect on the estimated mean and standard deviation of the number of stars in each mass bin. Increasing the number of realizations only changes the uncertainty of the standard deviation estimates. 

If we fix the total number of stars rather than the total mass, then the distribution of bin counts will follow a multinomial distribution with the desired expected total mass, as shown in the right panel of Figure~\ref{fig:220realizations}. We can use the known statistics of this multinomial distribution to approximate the standard deviations for the case in which the total mass is strictly fixed (compare the standard deviations in the left and right panels). The standard deviations estimated from the stochastically-generated distributions (left panel) show a somewhat larger standard deviation due to having a variable total number of stars. An in depth discussion about mass-limited sampling versus other sampling procedures can be found in \cite{Cervino2013}. 

The importance of computing 220 realizations of the IMF is thus, not to characterize the spread in observables predicted by the stochastic models (because in principle, that can be calculated analytically) but to fill in the space between random realizations in diagrams such as the colour-colour diagram, which is useful for comparison to the observations. 


\section{Additional colour-colour diagrams.}\label{app:more_ccds}

\begin{figure*}
    \centering
    \includegraphics[width=1.69\columnwidth]{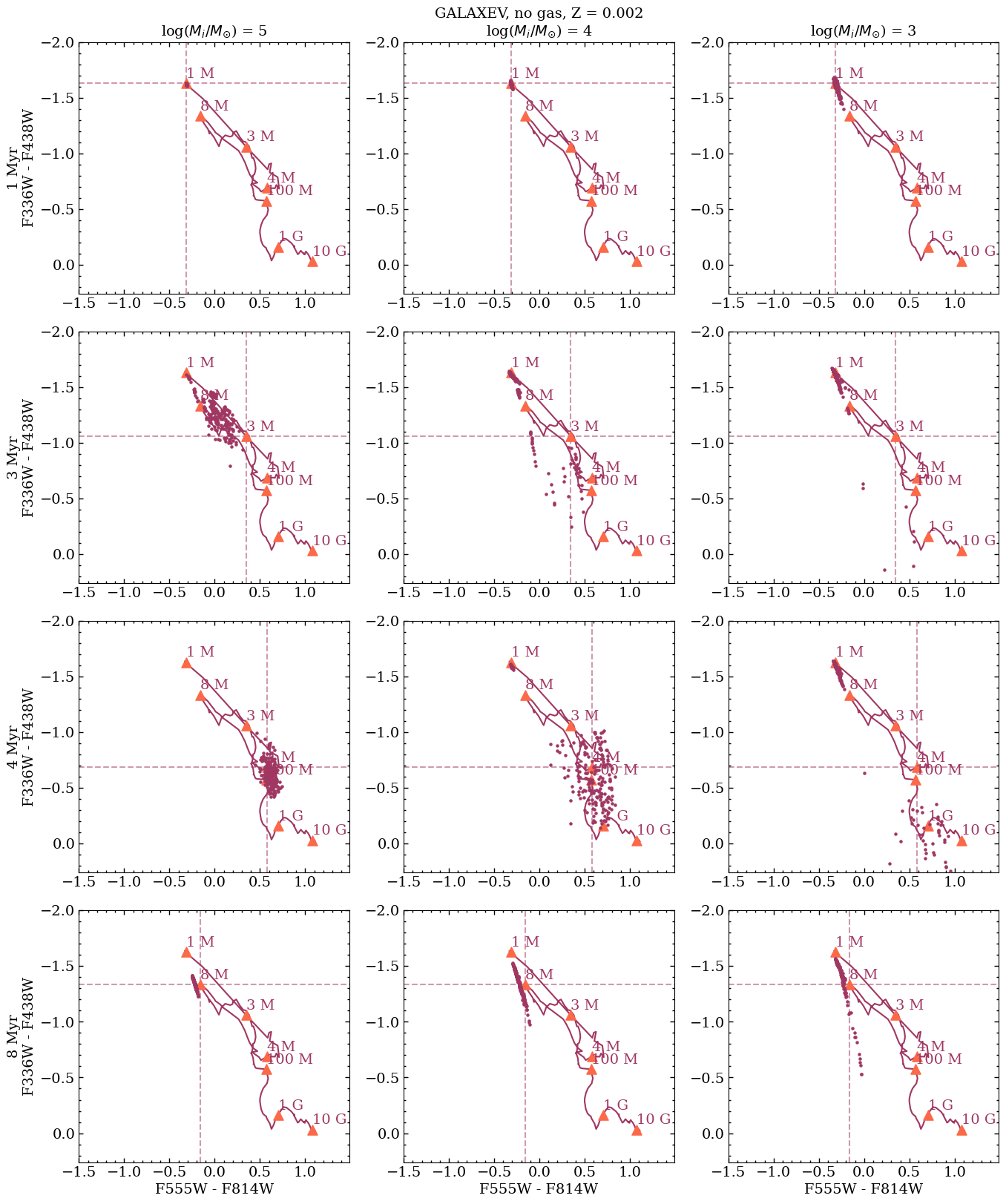}
    \caption{Similar to Figure~\ref{fig:ccd_z014_u0} but for $Z=0.002$.}
    \label{fig:ccd_z002_u0}
\end{figure*}

\begin{figure*}
    \centering
    \includegraphics[width=1.69\columnwidth]{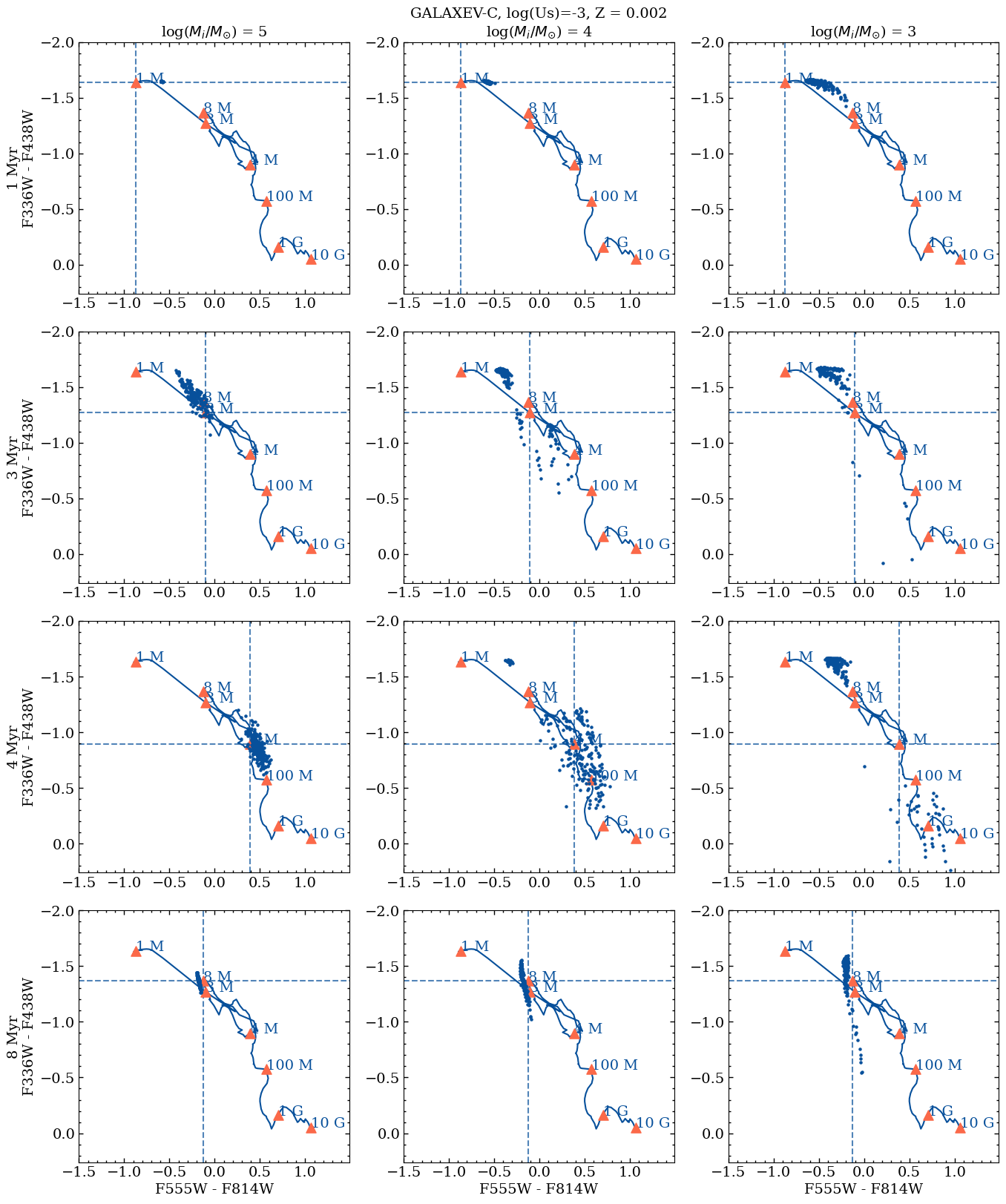}
    \caption{Similar to Figure~\ref{fig:ccd_z014_u3} but for $Z=0.002$.}
    \label{fig:ccd_z002_u3}
\end{figure*}

\begin{figure*}
    \centering
    \includegraphics[width=1.69\columnwidth]{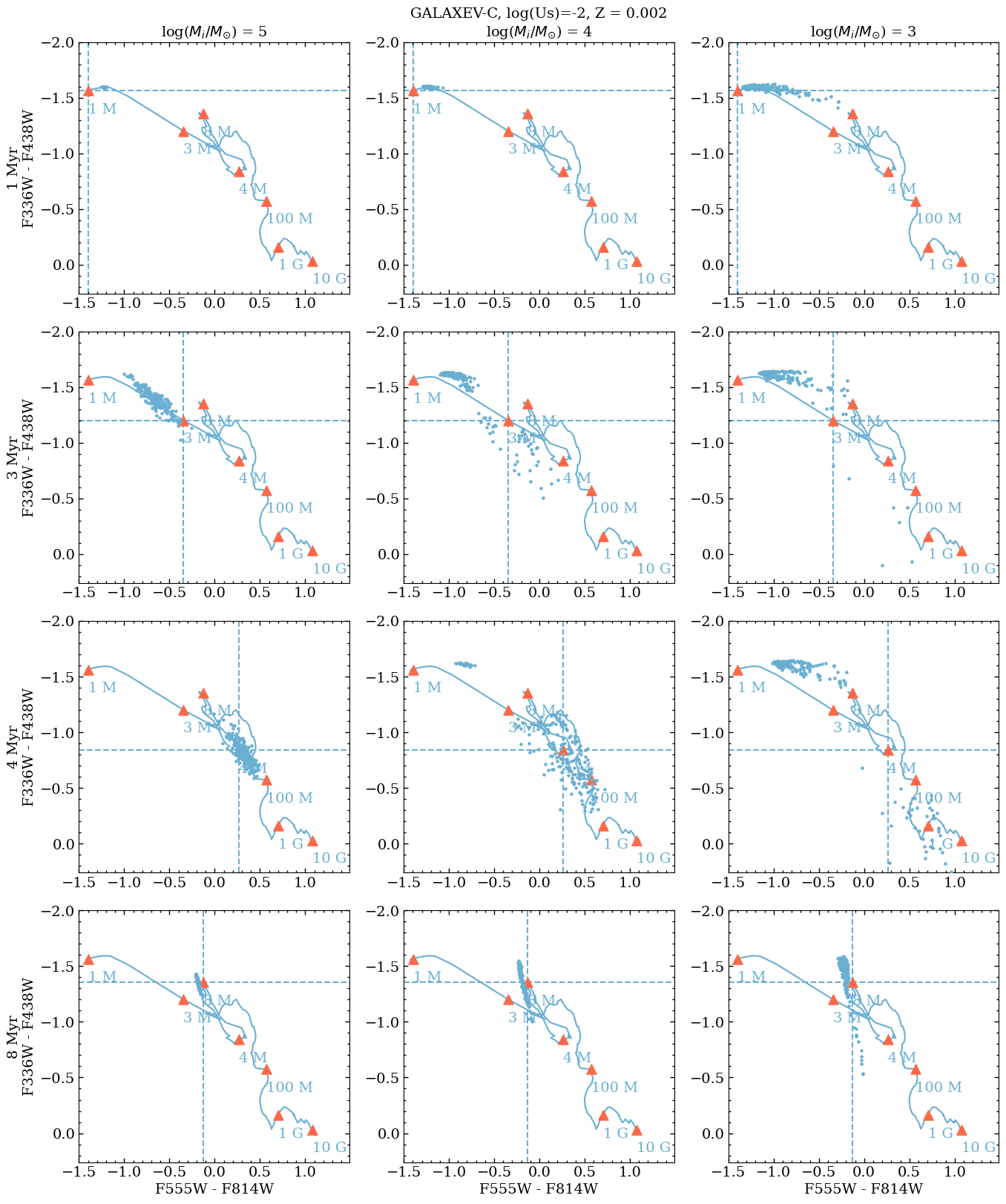}
    \caption{Similar to Figure~\ref{fig:ccd_z014_u3} but for $Z=0.002$.}
    \label{fig:ccd_z002_u2}
\end{figure*}

\begin{figure*}
    \centering
    \includegraphics[width=1.69\columnwidth]{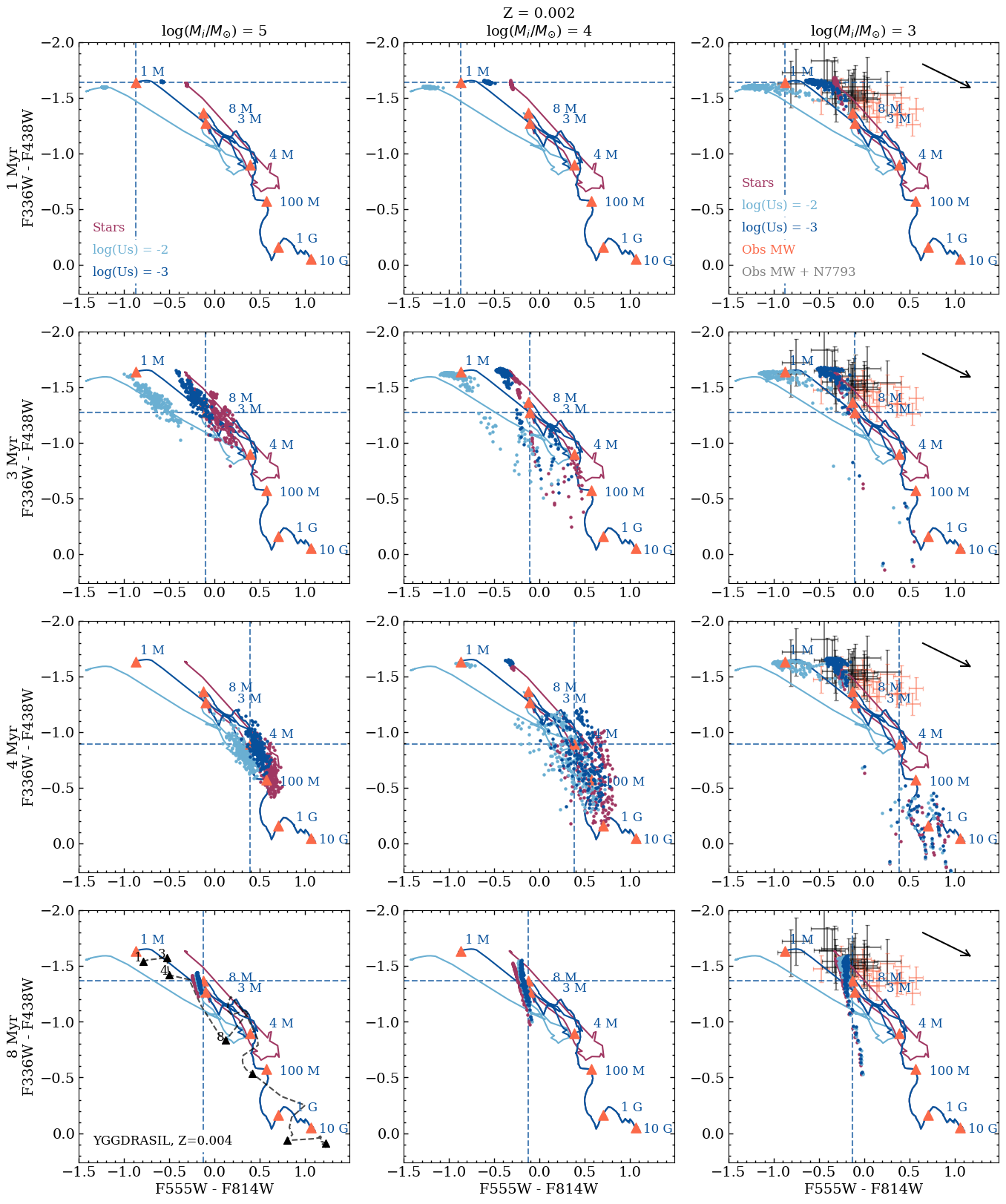}
    \caption{Similar to Figure~\ref{fig:ccd_z014} but for $Z=0.002$. In this case, the {\texttt Yggdrasil} predictions are for $Z=0.004$.}\label{fig:ccd_z002}
\end{figure*}

In Figures~\ref{fig:ccd_z014_u0} - \ref{fig:ccd_z014_u2} and Figure~\ref{fig:ccd_z014}, we compare Z=0.014 deterministic and stochastic predictions in the U - B vs. V - I diagram for cases: only-stars, log(Us)=-3, log(Us)=-2, and the three previous cases combined, respectively. Figures~\ref{fig:ccd_z002_u0} - \ref{fig:ccd_z002} are similar but for Z=0.002. Figures~\ref{fig:ccd_z002_u0} - \ref{fig:ccd_z002_u2} are discussed in Section~\ref{sec:AnalysisModels}, while Figure~\ref{fig:ccd_z002} is discussed in Section~\ref{sec:AnalysisObservations}.   


\section{Extinction and age PDF.}\label{app:cluster_pdfs}

\begin{figure*}
    \centering
    \includegraphics[width=0.99\columnwidth]{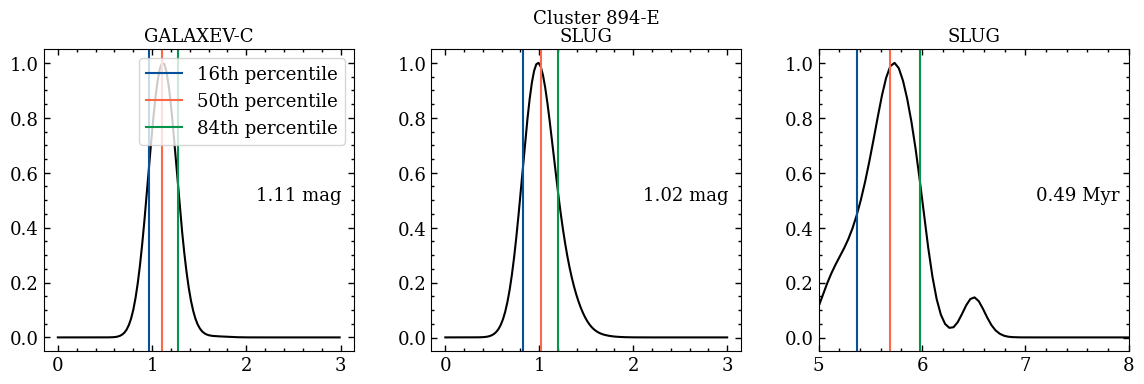}\includegraphics[width=0.99\columnwidth]{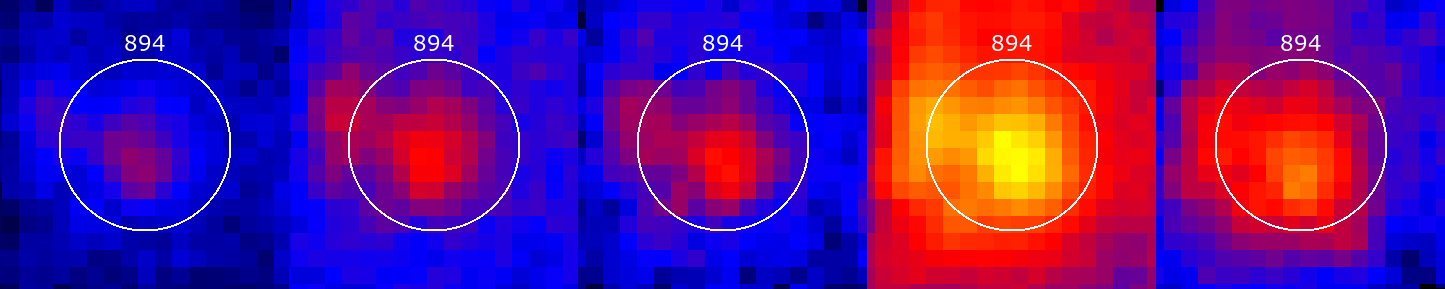}\\
    \includegraphics[width=0.99\columnwidth]{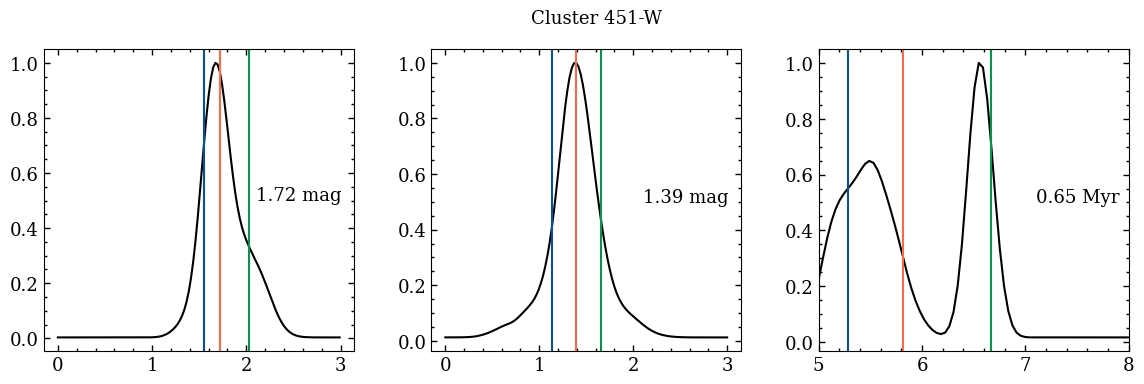}\includegraphics[width=0.99\columnwidth]{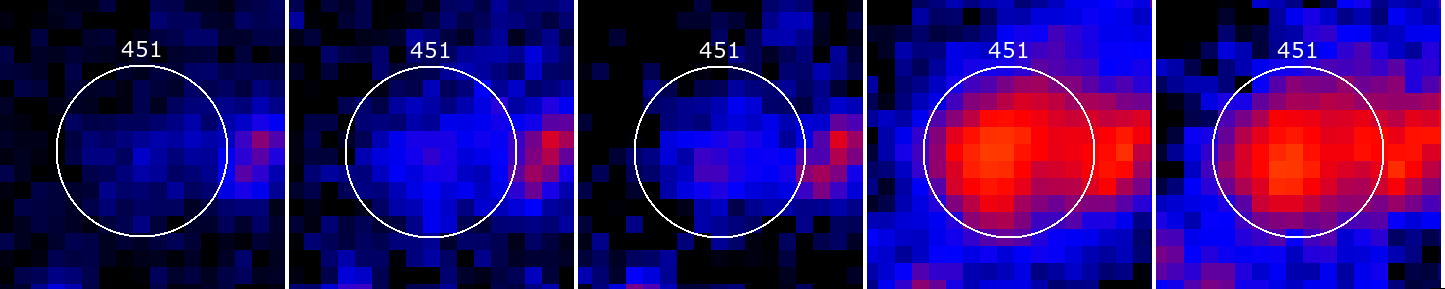}\\
    \includegraphics[width=0.99\columnwidth]{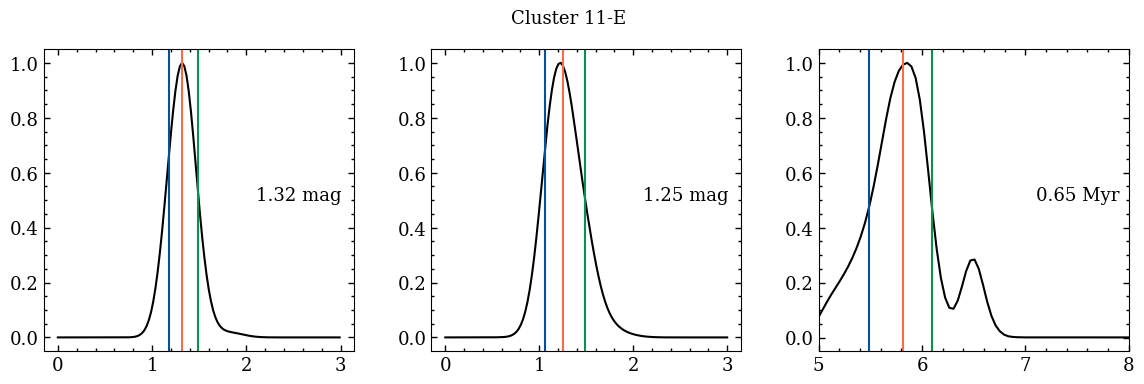}\includegraphics[width=0.99\columnwidth]{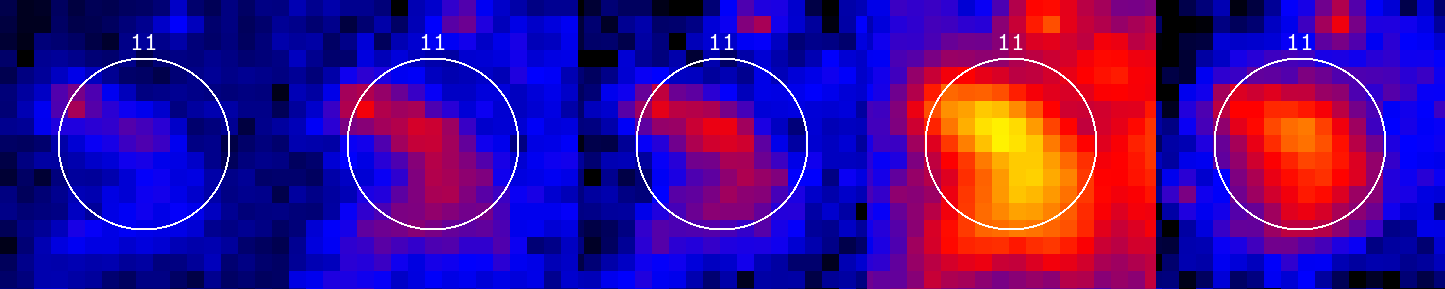}\\
    \includegraphics[width=0.99\columnwidth]{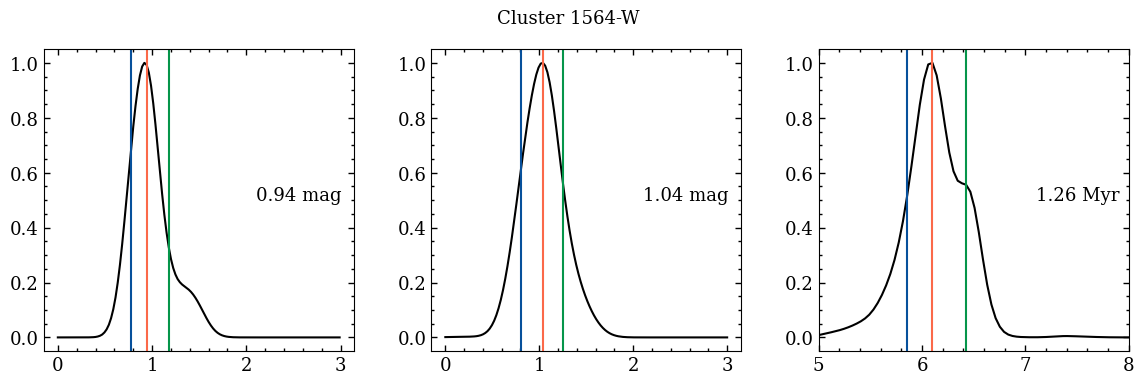}\includegraphics[width=0.99\columnwidth]{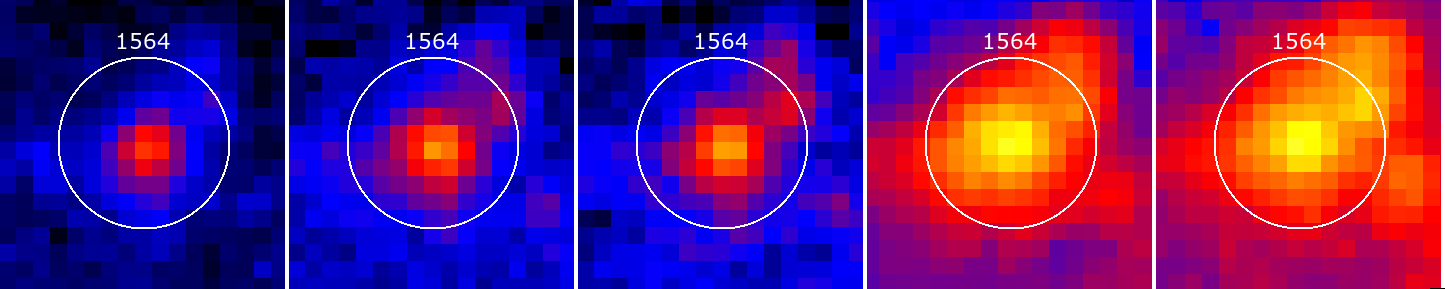}\\
    \includegraphics[width=0.99\columnwidth]{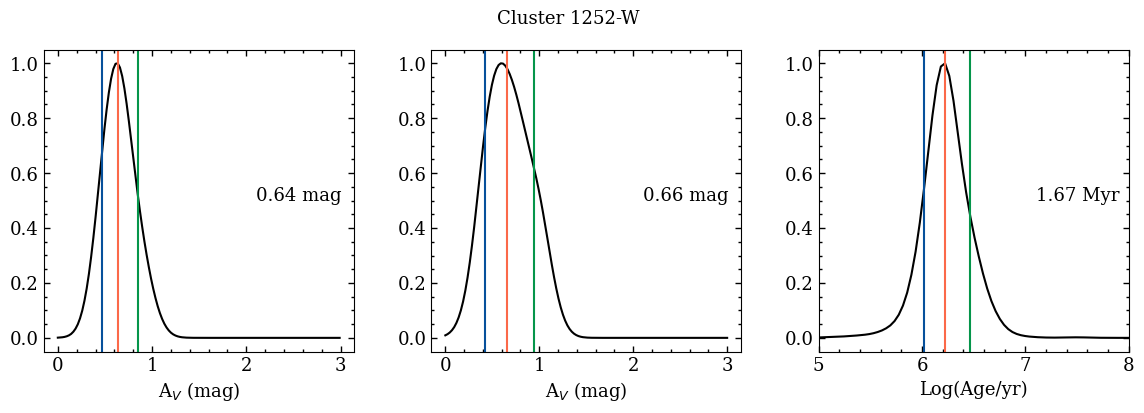}\includegraphics[width=0.99\columnwidth]{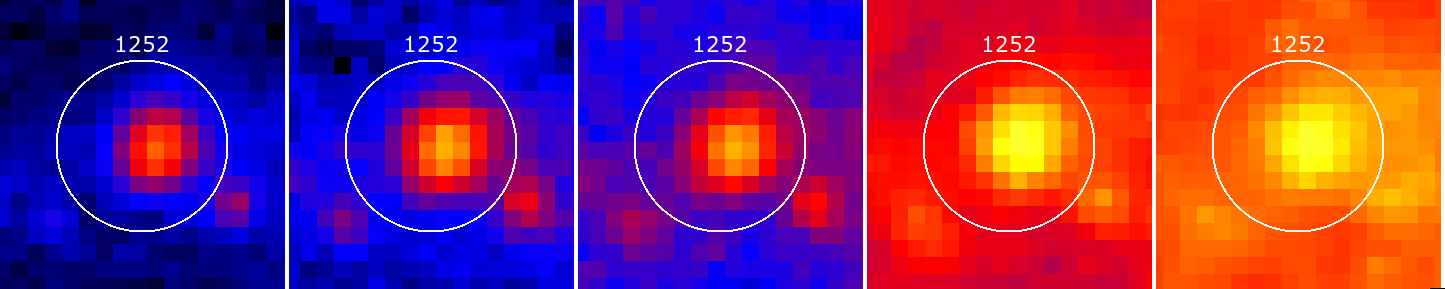}
    \caption{Left--. PDFs of V-band extinction (left and middle  columns) and age (right column) for five clusters in our sample. The left and middle columns show the \texttt{GALAXEV-C} and \texttt{SLUG}  PDFs, respectively, while the right column shows the \texttt{SLUG} age PDF. The clusters are arranged in order of increasing \texttt{SLUG} age. In each PDF sub-panel, we give the median value (50th percentile) of the extinction or age. Right--. Postage stamps of the clusters in the NUV, U, B, V and I LEGUS bands, from left to right. We use a logarithmic scale from 0 to 10 and SAO-ds9's colour scale "b", such that blue corresponds to the lowest number of counts and yellow corresponds to pixels with 10 counts or more.}
    \label{fig:pdf1}
\end{figure*}

\begin{figure*}
    \centering
    \includegraphics[width=0.99\columnwidth]{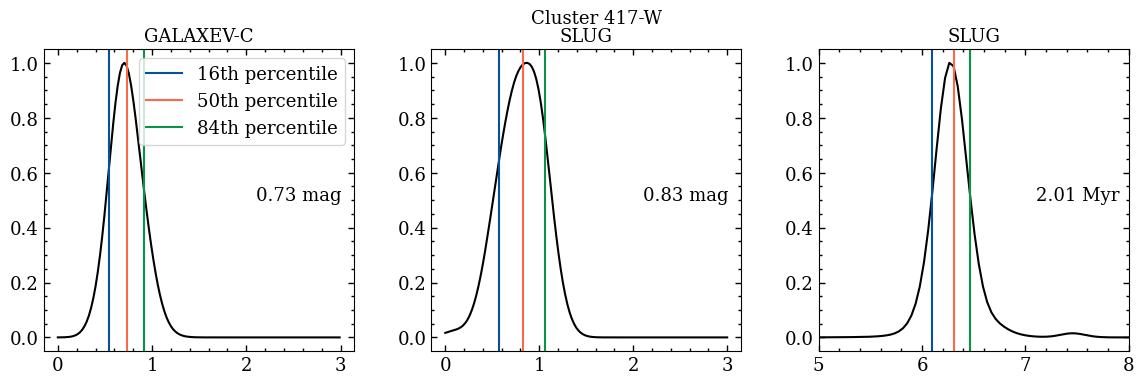}\includegraphics[width=0.99\columnwidth]{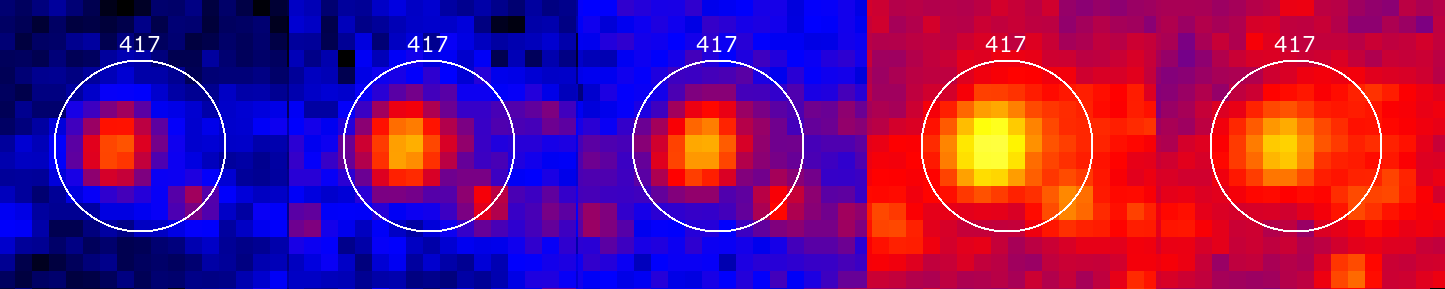}\\
    \includegraphics[width=0.99\columnwidth]{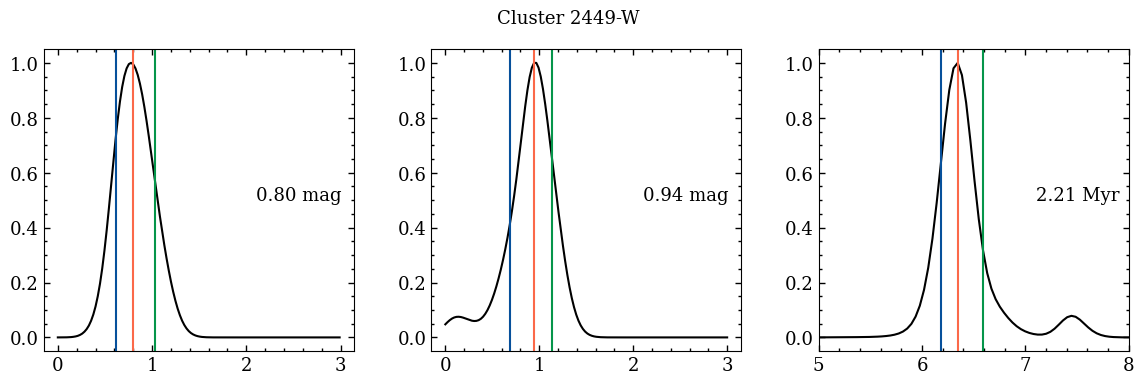}\includegraphics[width=0.99\columnwidth]{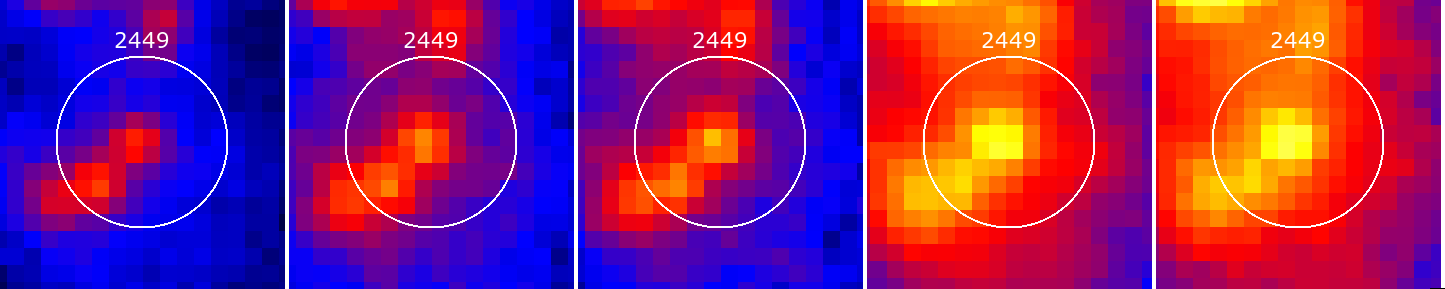}\\
    \includegraphics[width=0.99\columnwidth]{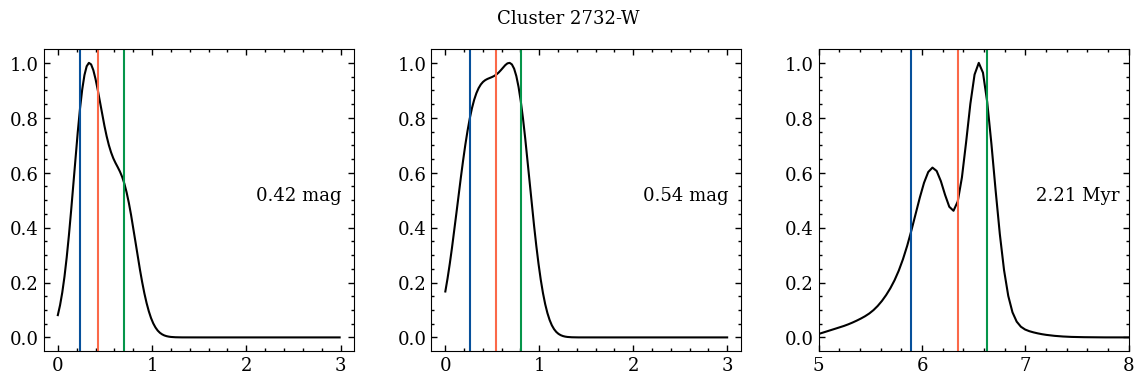}\includegraphics[width=0.99\columnwidth]{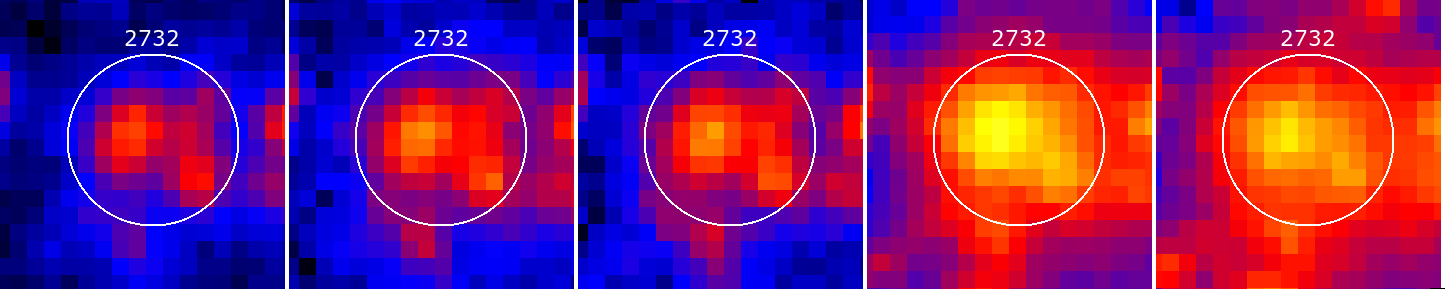}\\
    \includegraphics[width=0.99\columnwidth]{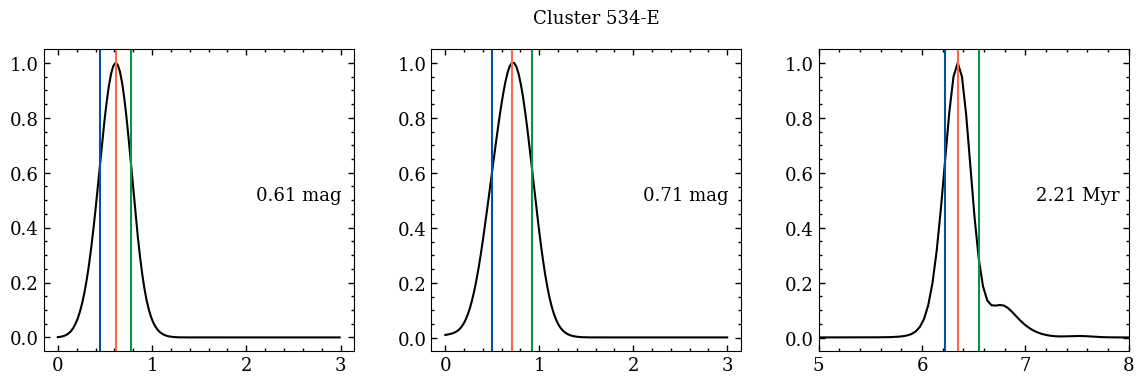}\includegraphics[width=0.99\columnwidth]{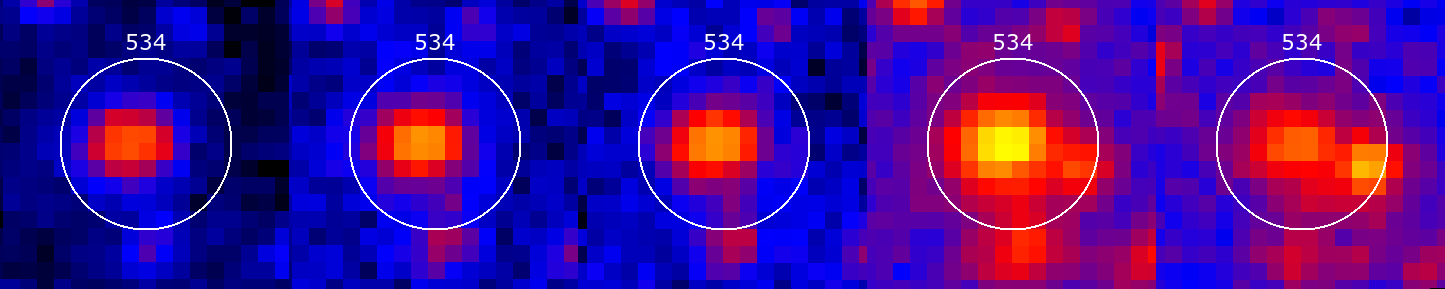}\\
    \includegraphics[width=0.99\columnwidth]{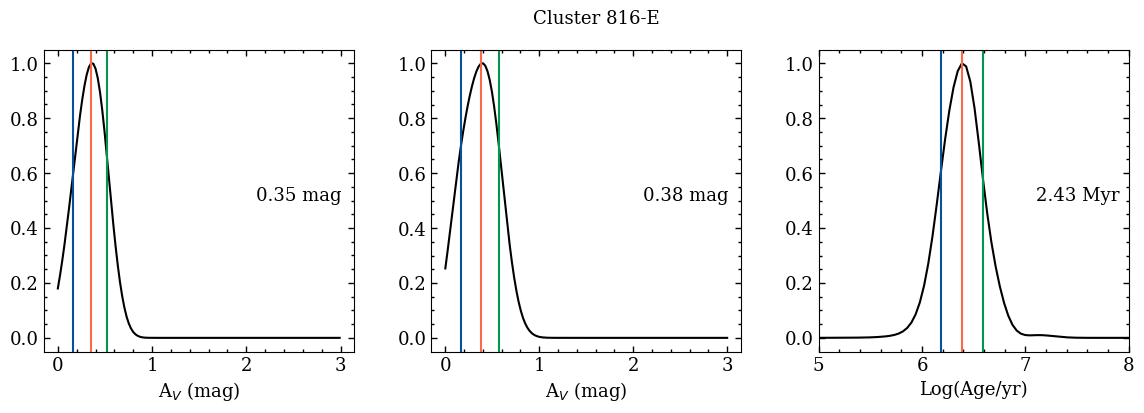}\includegraphics[width=0.99\columnwidth]{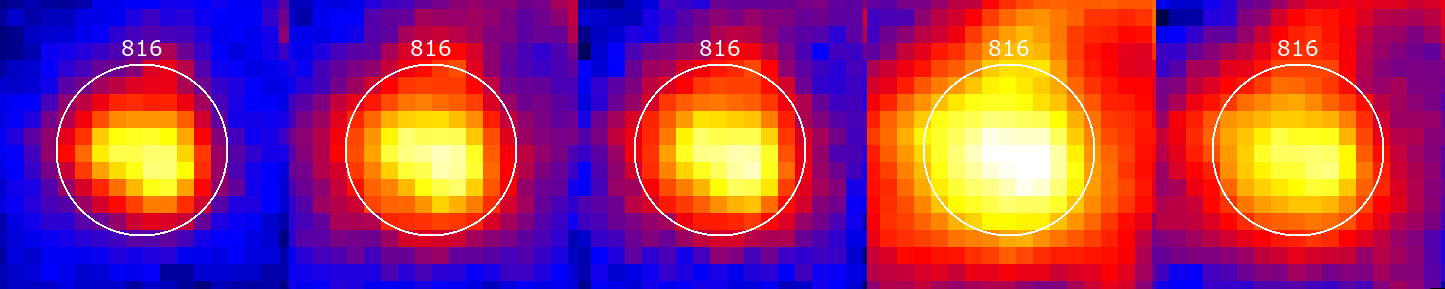}
    \caption{Similar to Figure~\ref{fig:pdf1} but for five different clusters.}
    \label{fig:pdf2}
\end{figure*}

\begin{figure*}
    \centering
    \includegraphics[width=0.99\columnwidth]{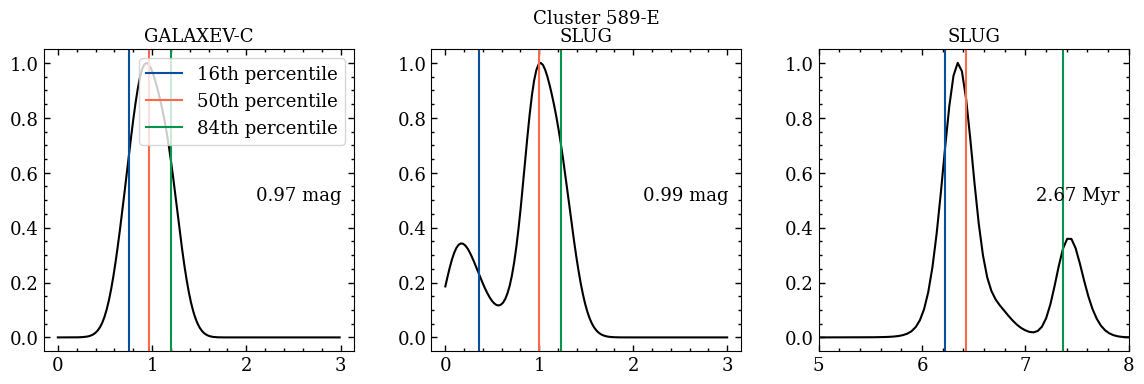}\includegraphics[width=0.99\columnwidth]{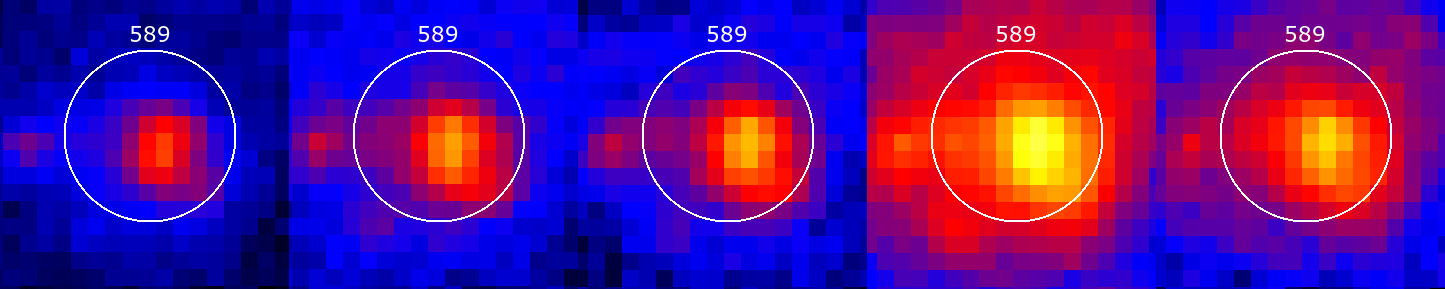}\\
    \includegraphics[width=0.99\columnwidth]{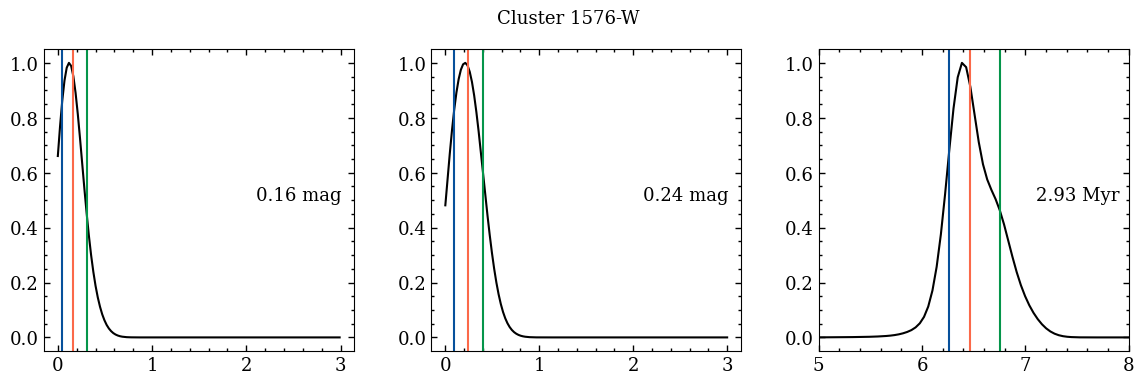}\includegraphics[width=0.99\columnwidth]{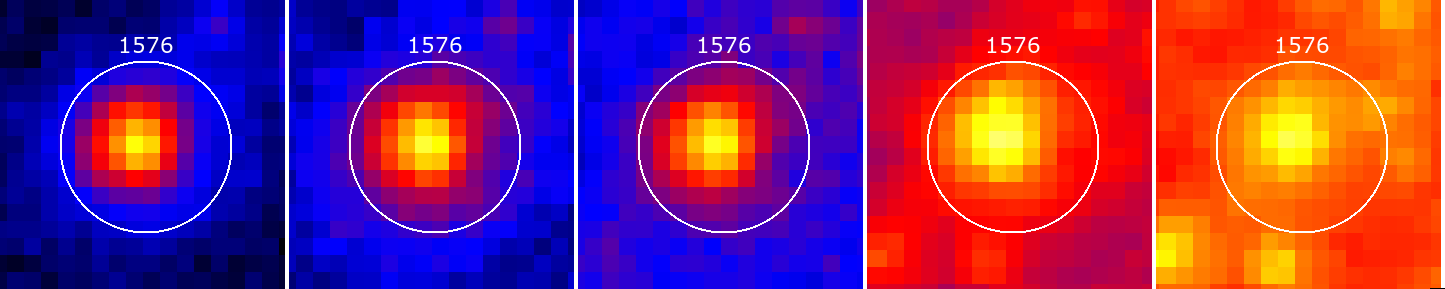}\\
    \includegraphics[width=0.99\columnwidth]{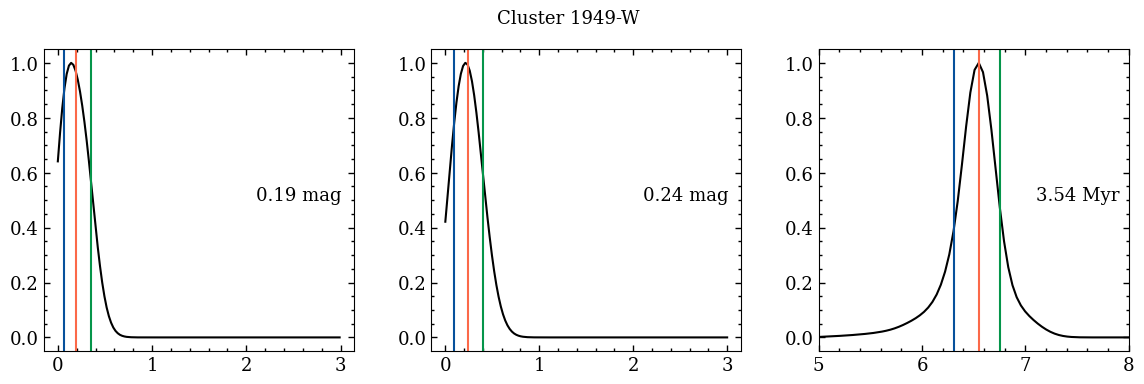}\includegraphics[width=0.99\columnwidth]{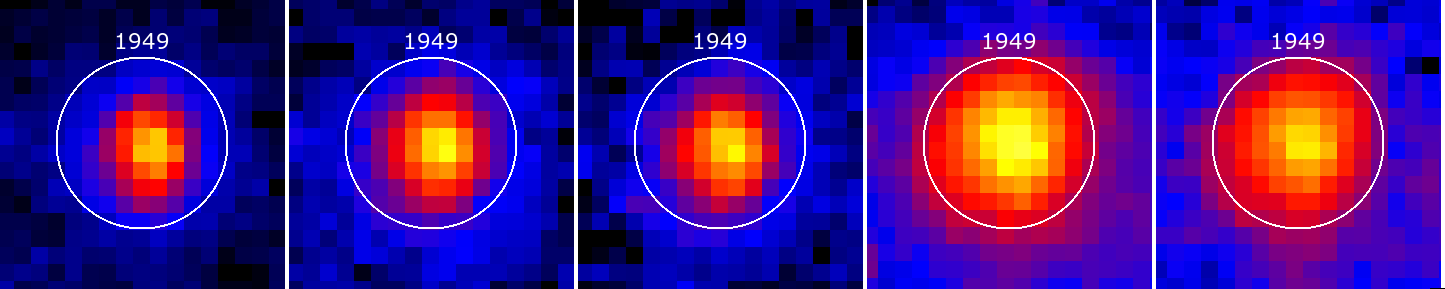}\\
    \includegraphics[width=0.99\columnwidth]{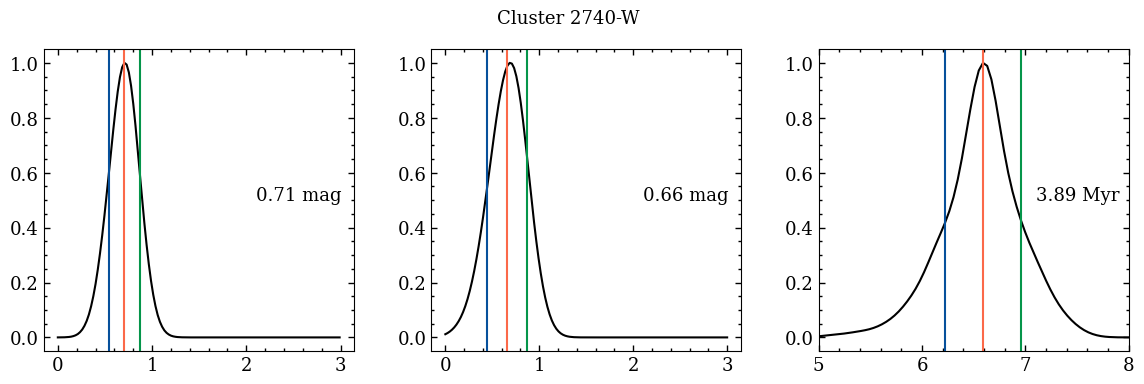}\includegraphics[width=0.99\columnwidth]{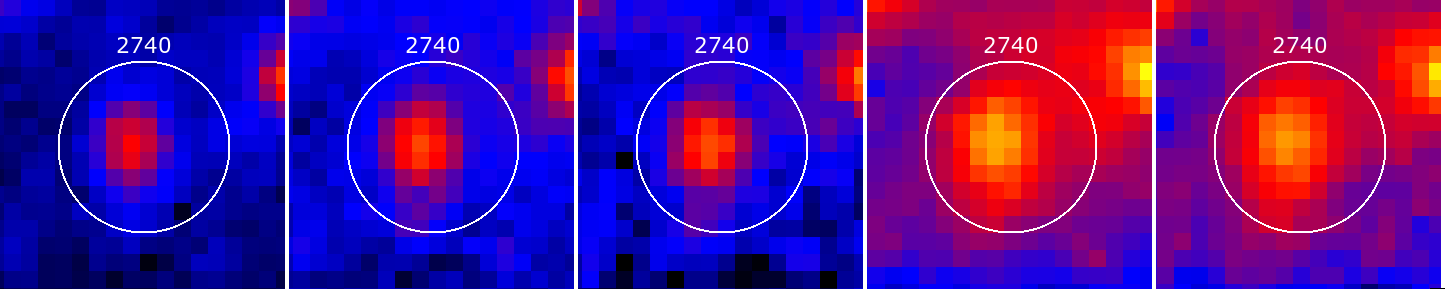}\\
    \includegraphics[width=0.99\columnwidth]{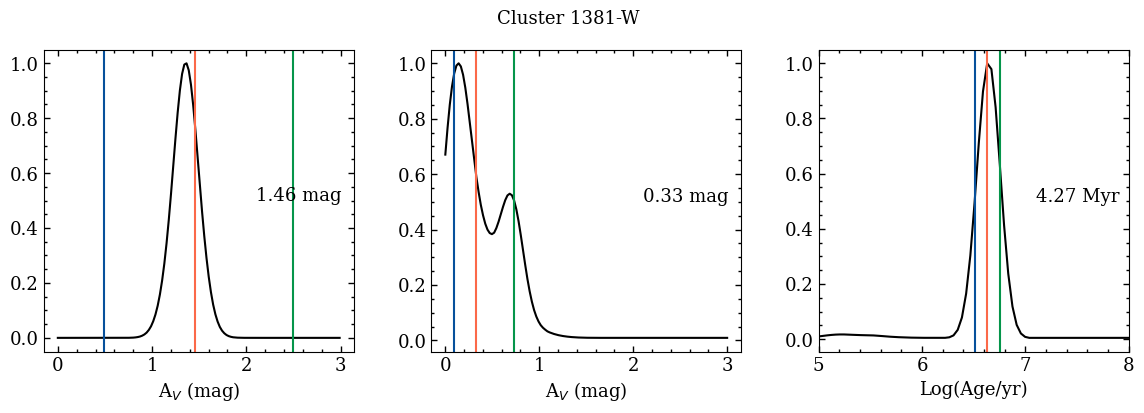}\includegraphics[width=0.99\columnwidth]{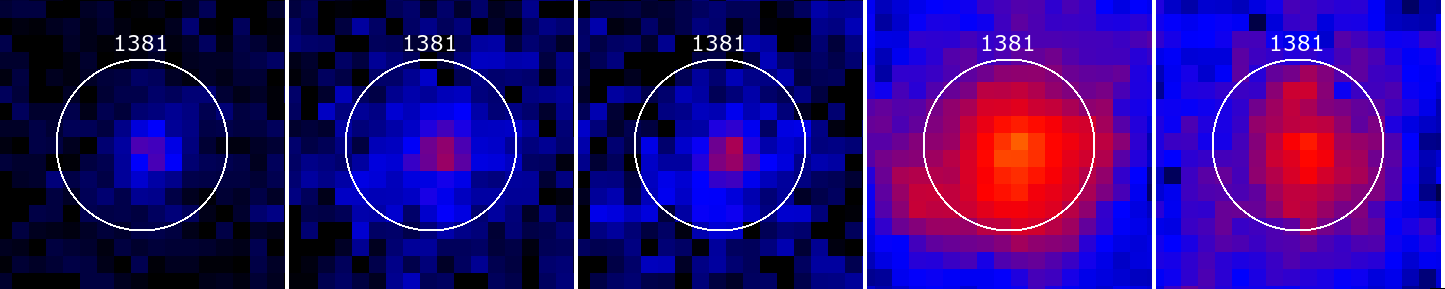}
    \caption{Similar to Figure~\ref{fig:pdf1} but for five different clusters. Note that cluster 1381-W, which is $\sim$4 Myr and has a high Av value (for the sample) according to \texttt{GALEX-C} and the second PDF peak of \texttt{SLUG}, has postage stamps that follow a similar pattern to that of the youngest ($<1$ Myr) high Av clusters of Figure~\ref{fig:pdf1}.}
    \label{fig:pdf3}
\end{figure*}

\begin{figure*}
    \centering
    \includegraphics[width=0.99\columnwidth]{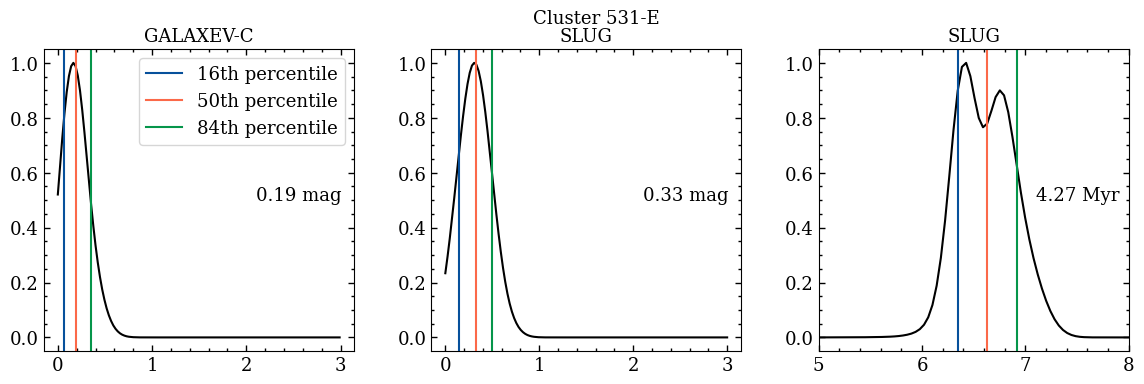}\includegraphics[width=0.99\columnwidth]{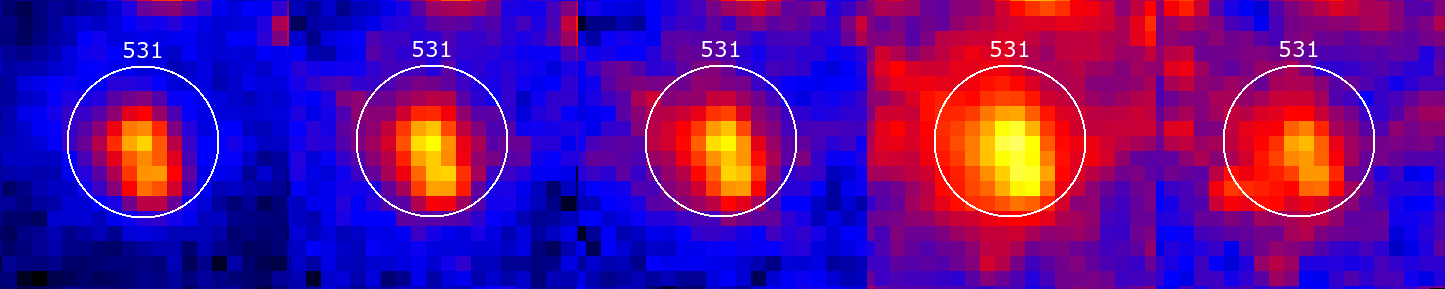}\\
    \includegraphics[width=0.99\columnwidth]{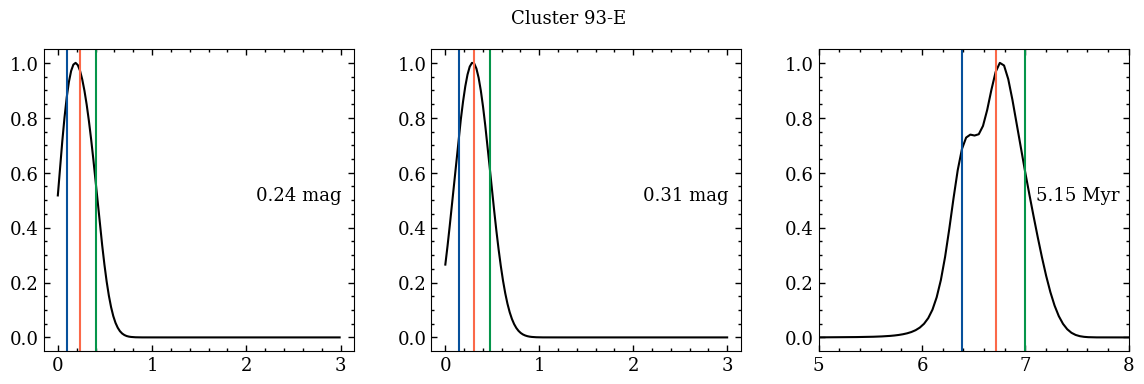}\includegraphics[width=0.99\columnwidth]{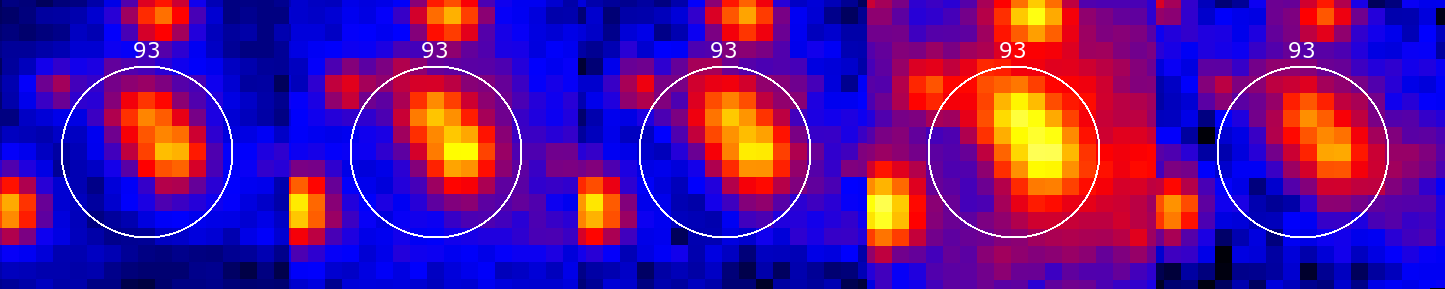}\\
    \includegraphics[width=0.99\columnwidth]{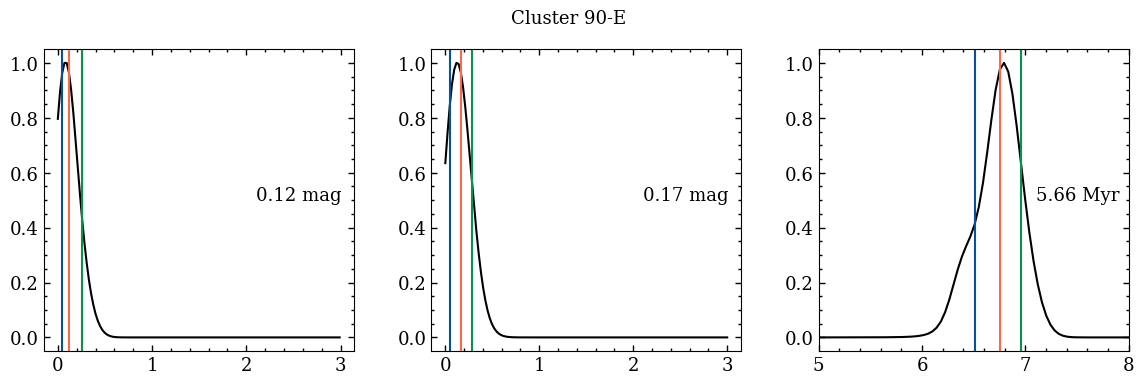}\includegraphics[width=0.99\columnwidth]{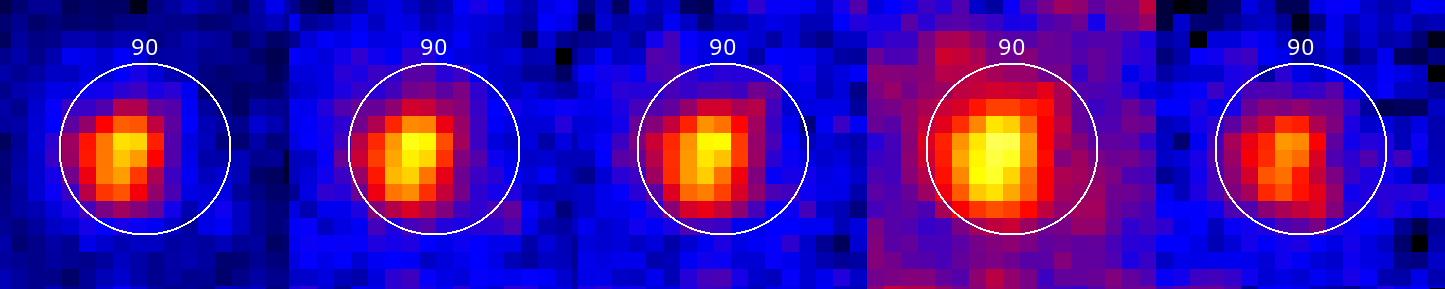}\\
    \includegraphics[width=0.99\columnwidth]{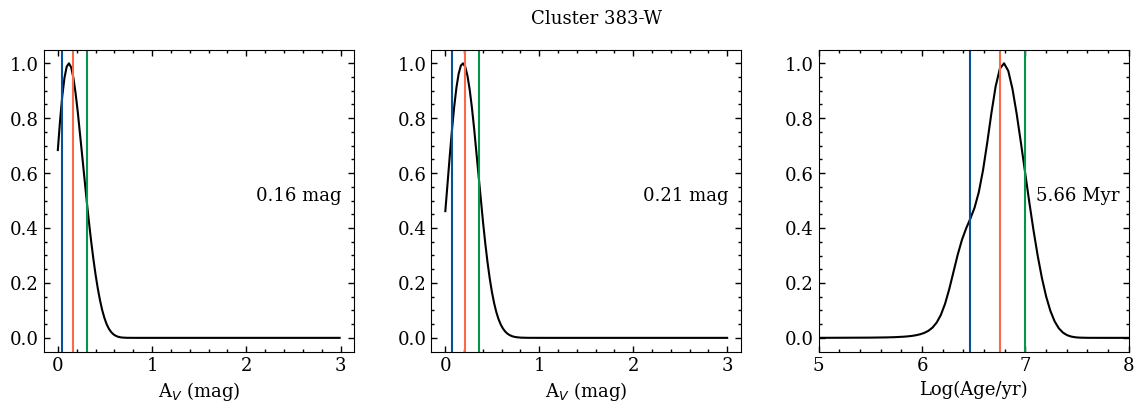}\includegraphics[width=0.99\columnwidth]{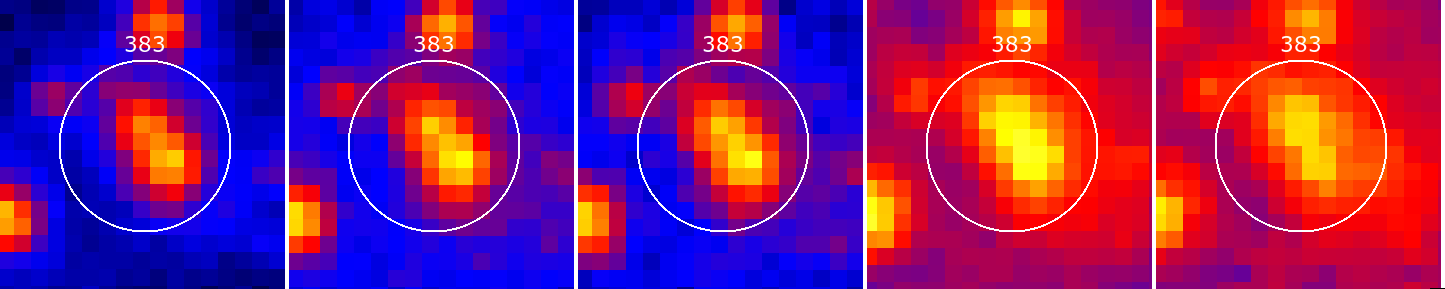}
    \caption{Similar to Figure~\ref{fig:pdf1} but for five different clusters.}
    \label{fig:pdf4}
\end{figure*}

For all clusters in our sample, the left-three panels of Figures~\ref{fig:pdf1} - \ref{fig:pdf4} show the V-band extinction PDFs from {\texttt{GALAXEV-C}} for models with log(U$_{\rm S})=-3$ and $Z = 0.014$ (left column) and from {\texttt{SLUG}} for models with log(U$_{\rm S})=-3$ and $Z = 0.020$ (middle column); and the age PDFs from {\texttt{SLUG}} (right column). The code and cluster ID is indicated in the column titles. The blue, red, and green lines give the 16th, 50th and 84th percentiles respectively. In each figure, the clusters are arranged in order of increasing age. The right-five panels of Figures~\ref{fig:pdf1} - \ref{fig:pdf4} show the NUV, U, B, V, and I postage stamps of the clusters, from left to right.

Although the top three clusters of Figure~\ref{fig:pdf1} have the youngest median {\texttt{SLUG}} ages, their age PDFs show a second peak at an older age. These three clusters also have high Av values (for the sample), as expected if the surrounding dust has not been as affected by the ionising photons from massive stars compared to other clusters. The high extinction of the youngest clusters leads to NUV/U/B-band postage stamps with low number of counts because reddening due to dust increases as wavelength decreases.

Figures~\ref{fig:pdf1} - \ref{fig:pdf4} also show that for some clusters the extinction PDF is multi-peaked according to one code but single-peaked according to the other. For instance, for cluster 451-W in Figure~\ref{fig:pdf1} {\texttt{GALAXEV-C}} yields two peaks while for cluster 589-E in Figure~\ref{fig:pdf3} it is the opposite. Cases where both codes yield single-peaked PDFs (e.g., cluster 534-E) or multiple-peaked PDFs (e.g., cluster 2732-W) also occur. The different shapes of the extinction PDFs is attributed to the different ways of modelling the ionising continuum in the {\texttt{GALAXEV-}} and {\texttt{SLUG}} libraries (see the introduction of this paper).

\section*{Acknowledgements}

We thank the jury of ROD's M.S. thesis (E. Terlevich, S. Sánchez, L. Aguilar, S. Srinivasan) as well as M. Cerviño and B. Elmegreen for comments and suggestions which have greatly improved the quality of this paper. ROD and AW acknowledge the support of UNAM via grant agreement PAPIIT no. 
IA-102120. AVG, SC and GB acknowledge support from the ERC via an Advanced Grant under grant agreement no. 321323-NEOGAL. AVG also aknowledges support from the ERC Advanced Grant MIST (FP7/2017-2022, No 742719). MRK acknowledges support from the Australian Research Council's Future Fellowship funding scheme, award FT180100375, and from resources and services provided by the National Computational Infrastructure (NCI), which is supported by the Australian Government.\\

\section{Data availability}

The \hst~data underlying this article are available in the Mikulski Archive for Space Telescopes at \url{https://mast.stsci.edu/portal/Mashup/Clients/Mast/Portal.html}, and can be accessed with the dataset identifiers 13364 and 13773. LEGUS high level science products can be found at \url{https://archive.stsci.edu/prepds/legus/dataproducts-public.html}




\bibliographystyle{mnras}
\bibliography{ref} 


\bsp	
\label{lastpage}
\end{document}